# In-liquid Plasma Catalysis for Nitrogen Reduction

P. Grosse[1*], J.L. Gembus[2], F. Landwehr[1], A.R. Silva Olaya[1], D. Escalera-López[1], N. Bibinov[2], A. R. Gibson[2,3], S.Z. Oener[1], B. Roldan Cuenya[1]*

[1]*Department of Interface Science, Fritz-Haber Institute of the Max Planck Society, Berlin, Germany*

[2]*Chair of Applied Electrodynamics and Plasma Technology, Ruhr University Bochum, 44801, Bochum, Germany*

[3]*York Plasma Institute, School of Physics, Engineering and Technology, University of York, Heslington, YO10 5DD, United Kingdom*

Corresponding authors: *roldan@fhi-berlin.mpg.de, grosspk7@fhi-berlin.mpg.de

**Abstract:**

Ammonia, a crucial component in fertilizers and fuels, is currently produced by the energy-intensive Haber-Bosch process. However, due to the high upfront investments required for large-scale centralized production, alternative routes for small-scale applications are being sought.

We integrate in-liquid plasma with electrocatalysis for ammonia generation via the nitrogen reduction reaction (NRR). Among materials tested, platinum emerged as the most stable and active catalyst—it evolves hydrogen without plasma but produces significant ammonia under cold or hot in-liquid plasma. Our system employs a dual mechanism: plasma activates $N_2$, while the elevated Pt electrode temperature drives water decomposition (thermally and via plasma pathways), releasing reactive hydrogen. This synergy stabilizes key NH intermediates, enabling ammonia production beyond conventional electrocatalysis and eliminating the need for added hydrogen. Under optimized plasma conditions, partial current densities up to 3 mmol h$^{-1}$ cm$^{-2}$ at 250 mA cm$^{-2}$ are achieved. Control measurements across various metals confirm a synergistic plasma-catalysis effect.

**TOC Graphic:**

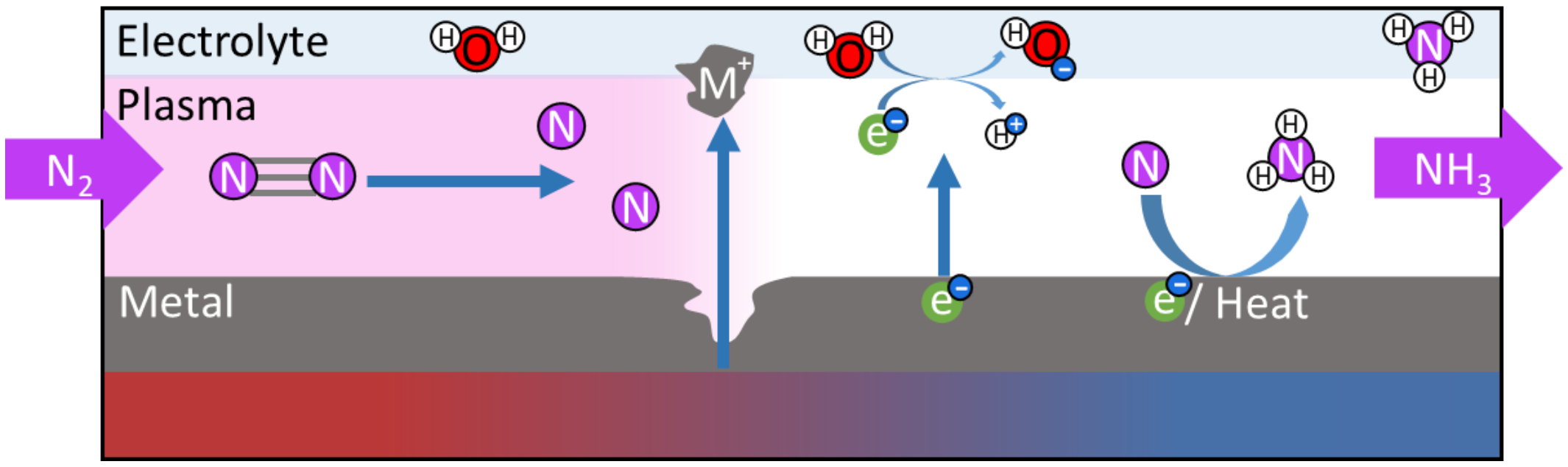

Ammonia ($NH_3$) production through the Haber–Bosch process has been a transformative achievement that enabled quadrupling global agricultural output and, thus, supporting a world population of over eight billion. Recently, ammonia has garnered attention as a potential green energy vector, where hydrogen produced from water electrolysis and renewable energies is used to generate green $NH_3$, which can then be thermally or electrochemically decomposed to release the green hydrogen on demand. However, the optimized Haber–Bosch process necessitates high operating temperatures (>400°C) and pressures (>200 atm) and requires cost-intensive upfront investments for large-scale centralized production.[1, 2] Additionally, the associated infrastructure requires substantial investments in $H_2$ feedstock generation, ammonia transportation, and storage.[3] Thus, small- to medium-sized, decentralized alternatives for ammonia production are highly sought after, especially the ones powered by renewable energy sources such as solar and wind for direct, on-demand $NH_3$ production.[4, 5]

Electrochemical ammonia synthesis, directly from nitrogen ($N_2$) and water, holds promise for small-scale, environmentally sustainable ammonia production, with potential economic and social benefits, particularly in developing countries and remote areas lacking sufficient infrastructure.[1, 6] Thus, the electrocatalytic nitrogen reduction reaction (NRR) has received interest as a promising avenue for small-scale $NH_3$ production. However, direct NRR faces intrinsic limitations due to the unreactive nature of triple-bonded $N_2$, its low solubility in water, and limited durability of non-aqueous electrolyte alternatives.[7, 8]

Conversion of the stable $N_2$ into a more reactive intermediary form is a critical step during this process. For example, for the electrocatalytic NRR, a lithium redox intermediate is thought to be critical for breaking the triple bond of $N_2$.[6, 9-13] Further, it has been recently postulated that inactive catalyst materials may be activated utilizing spin-mediated promotion by the addition of hetero metal atoms.[14, 15] Alternatively, a pathway utilizing nitrite and nitrates ($NO_3^-$ and $NO_2^-$) as highly soluble intermediates is under investigation.[16-18] NOx, derived from these species can be readily converted into $NH_3$.[19] However, industrially, nitrites and nitrates are typically produced from ammonia via the green-house gas intensive Ostwald process.[20] Historically, thermal plasma and arc furnaces were used for nitrogen fixation, which, in fact, predated the development of the Haber-Bosch process, but faced energy efficiency challenges.[21, 22]

Here, we propose a different ammonia synthesis route consisting in directly introducing gaseous $N_2$ into a plasma-electrocatalytic hybrid flow system with the goal of overcoming the low current densities and $NH_3$ production rates of electrocatalytic NRR. We study the electrochemical response of our reactor and various electrode materials and feed conditions to identify and understand key parameters influencing the efficiency and kinetics of ammonia formation. Additionally, optical emission spectroscopy (OES) provides key insights into the plasma properties and ammonia production rates. In our approach, we harness not only the energetic activation of $N_2$ by the plasma, but also exploit the high temperature of the electrode to drive thermal water decomposition. This produces a concentrated source of hydrogen radicals that, when combined with activated nitrogen species on the Pt surface, may form key intermediates (e.g., NH radicals), leading to efficient ammonia synthesis.

Finally, we discuss challenges and benefits of synthesizing ammonia from nitrogen ($N_2$) and water *via* this new pathway, keeping in mind that this is till now only a proof-of concept experiment and by no means a process that can in any way compete with the efficient Haber-Bosch process.

**Reactor Design and Plasma Ignition.**

To combine a plasma with electrocatalysis, we designed a single compartment, two-electrode reactor comprised of a platinum (Pt) counter electrode (anode) and a working electrode (cathode) (**Fig. 1a, Supplementary Figure 1**). A 1.0 mm diameter wire of variable length and material serves as active plasma and electrocatalytic working electrode (cathode). The wire is placed within a slightly larger inner diameter ceramic tube (alumina), which is exposed to a continuous $N_2$ gas flow. The counter electrode, consisting of a Pt-foil, surrounds the ceramic tube at a distance of approximately 5 mm from the working electrode. Additionally, an external circulating cooler is used to minimize the evaporation of the electrolyte (0.1 M KOH) and $NH_3$ outgassing during plasma operation. It was found that an alkaline pH (here 14) reduces the erosion of the plasma electrode. Various metal wires were tested as plasma electrodes. The selection was made based on their electrocatalytic properties in facilitating the hydrogen evolution reaction (HER), including Ni, Cu, Ir, and Pt. Additionally, Hf, Ti, Ta, Mo, and W, which are covered with amorphous or even non-conductive oxides under typical HER potentials, but become metallic at large negative potentials and were also tested.

With increasingly reducing potential, the HER in alkaline electrolyte ($2\ H_2O + 2\ e^- \leftrightarrow H_2 + 2\ OH^-$) starts to generate $H_2$. In combination with the nitrogen flow from the ceramic tube, a gas/vapor layer forms, which now comprises a mixture of nitrogen, hydrogen, and water vapor progressively enveloping the working electrode (cathode). At potentials exceeding 80 V, the gas/vapor sheath that forms around the cathode is not merely a barrier for charge transport; it becomes a reactive zone, where the elevated temperature and plasma trigger thermal decomposition of water. This process generates a rich supply of hydrogen radicals that should rapidly interact with activated $N_2$ species—produced by electron impact in the plasma—to form NH radicals, the critical intermediates for ammonia synthesis. In this state, free electrons, which are produced by ionization of gas molecules, drift towards the gas/electrolyte interface. Positively charged ions, produced during the ionization process, impact on the cathode surface and cause secondary electron emission. Because of the low mobility of the heavy ions, the electric field near the cathode surface is high and secondary electrons can be accelerated strongly. The thickness of this so-called plasma sheath region, or "cathode fall", is typically several micrometers under atmospheric pressure conditions. Outside of the cathode fall, the electrons lose kinetic energy by elastic and inelastic collisions with neutral and charged species in the gas-phase plasma bulk region. In addition, free electrons can collide with charged species and neutral species, leading to a variety of outcomes, such as (radiative) electron-positive ion recombination, electron-attachment to neutral species, electron impact excitation of rotational, vibrational and electronic states of electrons and ions, and electron impact dissociation of molecular species. The products of these reactions will also collide with one another, leading to a non-equilibrium gas-phase plasma chemistry. In addition, sputtering of metal atoms from the cathode may also occur. These sputtered atoms will also participate in the gas-phase chemistry described above. Some of these processes are shown schematically in **Fig. 1b.**

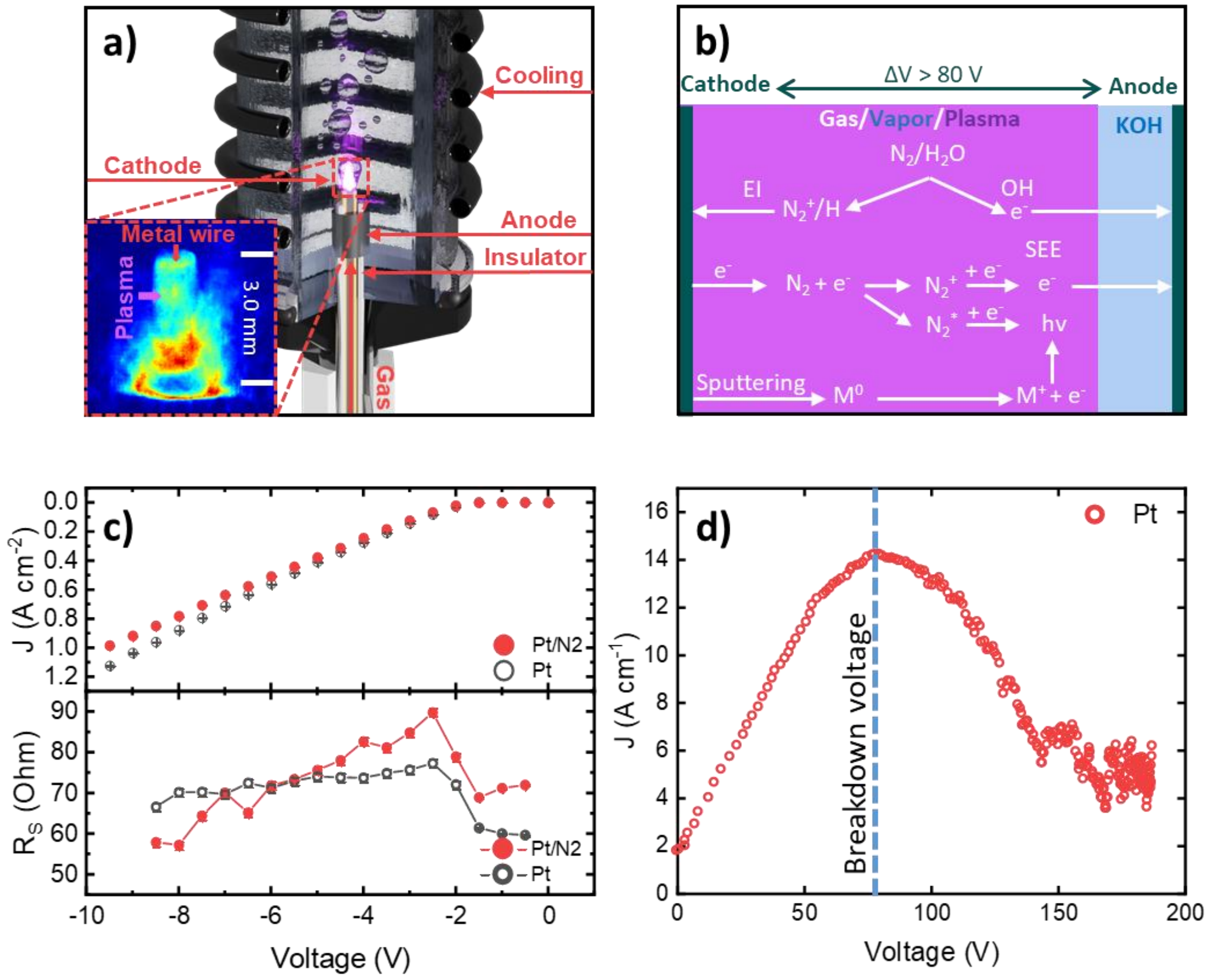

**Figure 1: Setup and electrochemical performance**. **(a)** Schematic representation of the reactor cell, with the cathode being enclosed within an insulating gas tube, and the Pt anode foil surrounding the external surface of the gas tube. **(b)** Schematic of relevant interfacial processes. The governing effects at the interface are the excitation-ionization (EI) of gas/vapor molecules, secondary electron emission (SEE) and radiative recombination. Sputtering is unlikely to occur at the applied voltages, but local hot spots in the plasma may eject metal atoms from the cathode. **(c)** Polarization curves of Pt acting as cathode material up to -10 V (top), with solution resistance (bottom). Current values are average values over 5-minute chronoamperometry at each potential. **(d)** Continuation of the diffusion-limited current-voltage curve, revealing a voltage breakdown and plasma ignition at approximately 80 V with 0.07 L $min^{-1}$ $N_2$ gas flow.

Chronoamperometry and impedance spectroscopy (EIS), both with and without nitrogen flow were used to study the electrochemical behavior of the system (**Figure 1c**). The series resistance was extracted from the high-frequency impedance at each potential step. **Figure 1d** shows an exemplary polarization curve for Pt until voltage breakdown and plasma ignition at ~ 80V with 0.07 L $min^{-1}$ $N_2$ gas flow. All polarization curves and corresponding EIS for the different metal wires are shown in **Supplementary Note 1**. For all materials, no significant differences in current and resistance are apparent once the potential enters the diffusion-limited regime. The sole exception is Hf, which features a natural passivation layer that only gets removed under harsh plasma conditions. The introduction of nitrogen gas resulted in a decreased current and increased resistance, by covering parts of the electrode in $N_2$ gas until eventually a full vapor layer forms at higher potentials. The $j\ vs.\ V$ plots in **Supplementary Figure 10** highlight differences between experiments conducted with and without nitrogen. These results indicate that the reactor is initially at low voltages mainly governed by HER, and no substantial NRR takes place with the tested materials.

**Optical emission spectroscopy and plasma characterization.**

To probe the influence of the cathode material (Pt and W wires) on the conditions of the plasma in the system and subsequent catalytic performance, optical emission spectroscopy was used (**Figure 2**). These two metals showed stable plasma generation, but distinctly different $NH_3$ production rates (see below). For both metals, we observe two distinctive plasma states, depending on the applied potential. The “cold” plasma, akin to a glow discharge, exhibits a characteristic neutral gas temperature (inferred from the rotational temperature of plasma-produced excited states i.e. the $N_2$(C-B) transition), ranging from 1000 K to 1600 K. Notably, for this “cold” plasma state, we observe already a substantial amount of excited nitrogen species emission for the Pt cathode, but lower contributions from the W electrode (**Figure 2a**). For the “hot” plasma, resembling an arc discharge, the gas temperatures exceed 3000 K. (See the Methods section for the determination of the neutral gas temperature in the plasma).

The temperature of the active plasma electrode (cathode) could be determined by background continuum fitting and showed an applied potential dependency. When 175 V were applied, the temperature of the Pt wire is close to room temperature. Increasing the applied voltage to 200 V leads to temperatures around 900 K for the “cold” state and 1300 K for the “hot” state. Lastly, for the highest ammonia formation at 230 V, the electrode is at 1600 K, where it starts to become unstable and ultimately melts after a few minutes. The electrode temperatures on W are similar at the respective applied voltages, though no “cold” states could be observed.

For the Pt cathode, transitioning from the “cold” to the “hot” state significantly amplifies the $N_2$ (C-B, v:0-0) (vibrational state transitions between the B $^3\Pi_g$ and the C $^3\Pi_u$ electronic states) peak intensities as well as the $H_\alpha$ spectral line, as shown in **Figure 2a**. Conversely, the ammonia production nearly quadruples for the hot plasma at around 200 V, suggesting energy-driven substantial dissociation of nitrogen. This observation supports our hypothesis that thermal water decomposition, occurring at electrode temperatures approaching 1000–1400 K on Pt[23], provides the necessary hydrogen radicals to stabilize the NH intermediate, thereby driving the formation of ammonia. When investigating the potential dependence for the Pt cathode, we observe that at lower potential (180-200V) the stronger emission from excited $N_2$ molecules is observed. With increasing applied potential (~ 230 V), the excited molecular $N_2$ peaks decrease in intensity, likely due to the breaking of the $N_2$ triple bond and decreasing electron temperatures. The behavior for Pt is in clear contrast to that of W, for which we only detect sharp metallic lines, which are indicative of a higher surface heating level in the "hot" plasma state, but not of ammonia formation.

Taken together, the results in **Figure 2** strongly indicate a real plasma-catalysis effect, mediated by the complex interaction between the plasma and the cathode material. Depending on the exact plasma parameters (hot vs. cold plasma and applied bias) the plasma switches between exciting $N_2$ molecules to strongly dissociating $N_2$ into atomic *N, which can then be bound on and converted to $NH_3$ on Pt (but not on W, probably due to tungsten nitriding[24]). Additionally, Pt possesses fast HER kinetics, which could enhance the catalytic conversion by providing more hydrogen for the reaction. A comparison between all materials at 10 V and 200 V can be found in **Supplementary Figure 2**. Furthermore, we note that the production rate of ammonia cannot be simply explained by changes in the gas temperature or electron density of the plasma, or $H_2$ production alone (see **Supplementary Note 2**), supporting our hypothesis of a real hybrid catalysis-plasma effect. At similar electron densities and neutral gas temperatures in the plasma, the production of $NH_3$ is far superior on Pt as compared to W.

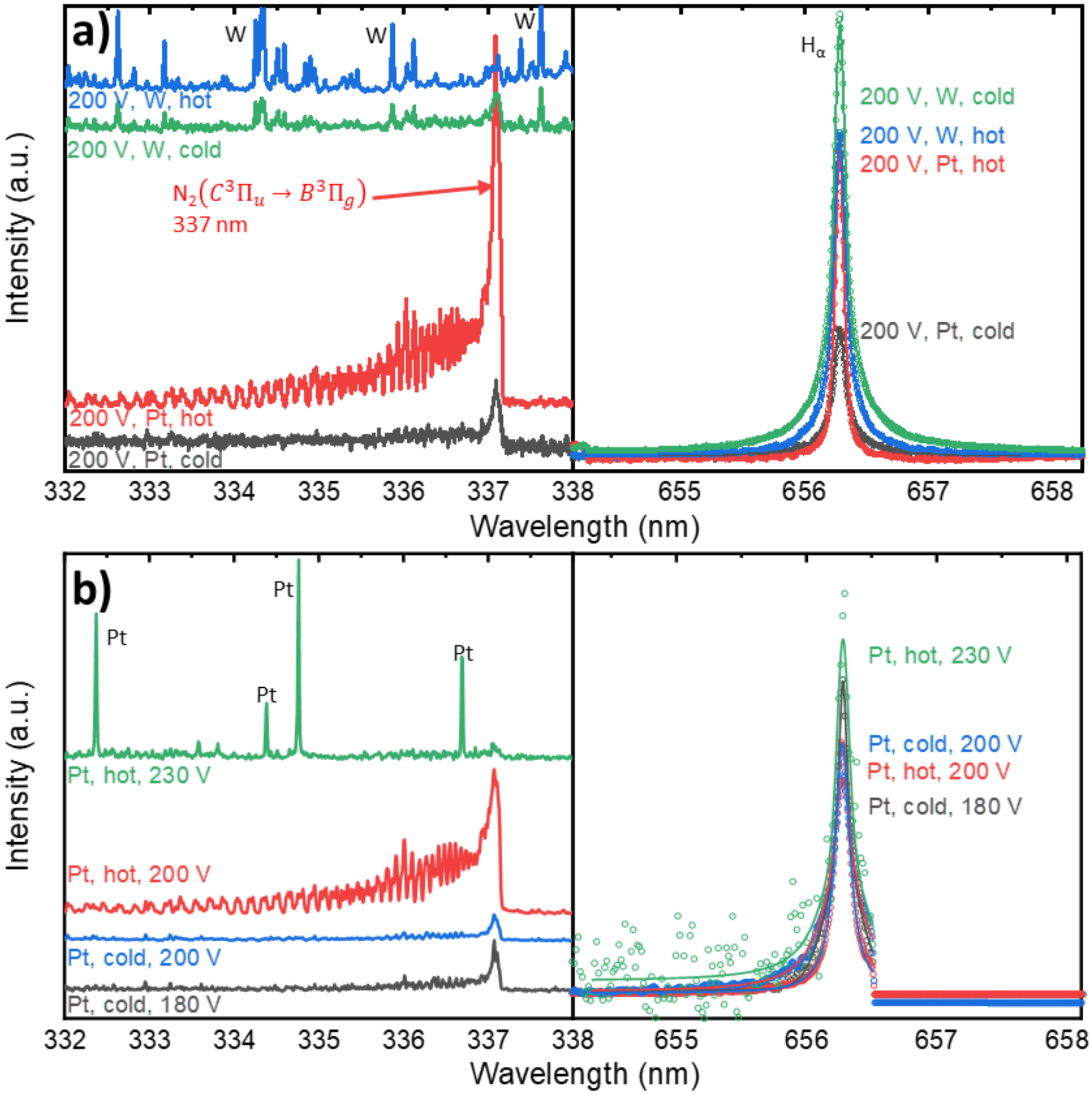

**Figure 2*:* Plasma characterization. (a)** OES data illustrating $N_2$ (0-0) vibrational transitions and rotational bands as well as the $H_\alpha$ line for both, hot and cold plasma states for Pt and W cathodes. **(b)** $N_2$ transition with a Pt cathode is depicted across different applied voltages ranging from 180-230 V. Vibrational excitation and ionization of nitrogen in relevant quantities were only observed on Pt. The peak height is directly proportional to the applied voltage, and at 230 V, dissociation occurs. Hydrogen was observed on W and Pt. All spectra are normalized by the OH radical signal, the right side of the $H_\alpha$ line in (b) is cut off by a sensitivity gap in the second spectrometer.

**Catalytic properties and parameter space.**

The ammonia produced could be quantified reliably using the indophenol blue method[10, 25], due to the fact that substantially larger amounts of $NH_3$ generated as compared to the electrocatalytic NRR, which is prone to misinterpretation of $NH_3$ trace amounts.[10]

**Figure 3a** illustrates the synthesis rate of ammonia plotted against the Faradaic efficiency, encompassing all tested Plasma cathode materials. Notably, materials such as Cu, Ni, and W exhibited significant reduction in synthesis rate and Faradaic efficiency with each consecutive measurement (despite stable plasma conditions ranging from 5 min up to one hour), while Pt, Hf, and Ir demonstrated greater stability. Despite their high activity, electrodes made of Ta and Ti endured only a few seconds under plasma conditions before their complete dissolution into the electrolyte. Hf, distinguished by its natural inert oxide layer, exhibited behavior akin to Ir. However, once the protective layer was compromised (indicated by the Hf** in **Figure 3a**), it mirrored the rapid degradation observed for Ti and Ta. As mentioned above, Pt is the best-performing and simultaneously the most plasma resistant material among those tested here, likely due to the combined plasma and catalytic properties that are needed for high $NH_3$ synthesis rates. **Supplementary Figures 3-6** display the XPS measurements on Cu, W, and Pt acquired before and after plasma operation. Whilst re-oxidation in air is a possibility (especially for Cu), an increase of oxide compared to the pristine state indicates changes originating from the experiment.

Compared to eNRR, the $NH_3$ production rate and current densities are here already superior, with only the efficiency partially behind (**Figure 3b)**. **Figure 3c** compares all experimental results of our study in terms of energy consumption vs. yield rate to the industrial standard (Haber Bosch process). Noteworthy, even small changes in the system, e.g. exact plasma potential, can lead to dramatic changes and improvements. See **Supplementary Note 3** for a discussion on the thermodynamics.

**Figure 3d** shows the Faradaic efficiency and $NH_3$ synthesis rate (log scale) in relation to the applied voltage used to generate the plasma. In the voltage range between the minimum threshold required for plasma ignition (80 V) and the onset of the hot plasma state (first observed at 200 V), only small amounts of ammonia are detected with Pt. However, as the potential reaches the range where the hot state becomes predominant (230 V), the efficiency of ammonia synthesis spikes dramatically. At even higher potentials the cathode and/or the gas sheath become unstable, likely due to excessive heating. **Figure 3e** shows that increasing the $N_2$ inflow is beneficial up to a certain threshold, beyond which it disrupts the formation of the gas sheath. **Figure 3f** shows the $NH_3$ production over time, with aliquots taken every 5 min with a continuous "hot" Plasma state for 30 min and 0.07 L min$^{-1}$ of nitrogen inflow.

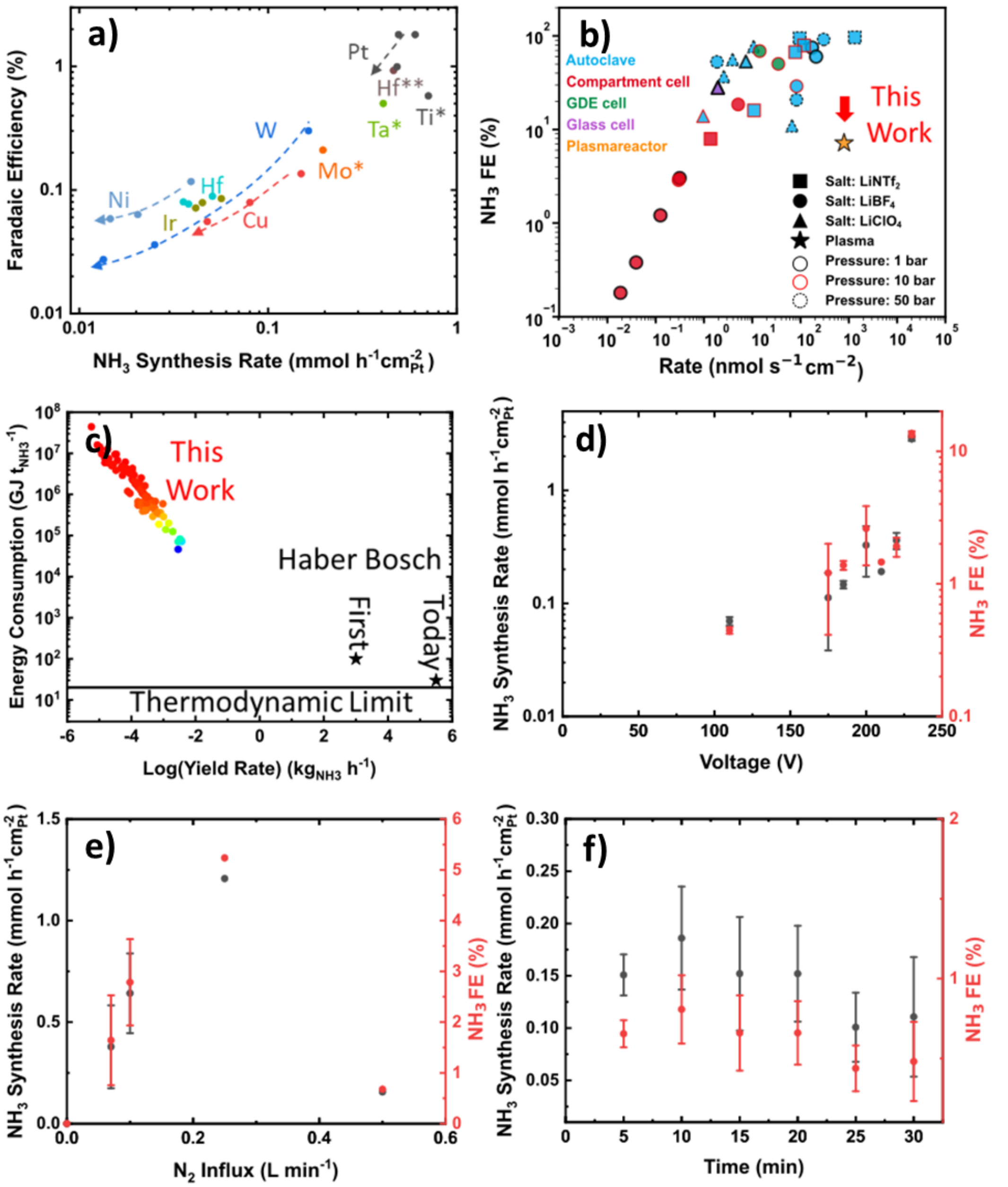

**Figure 3: Faradaic Efficiency and $NH_3$ Synthesis Rate. (a)** Faradaic efficiency vs. $NH_3$ Synthesis rate measured with 0.07 L min$^{-1}$ of $N_2$ flow and 200 V plasma voltage. Three successive measurements were performed for each material, with arrows in (a) indicating changes for the same wire between successive experiments. A general deactivation with each consecutive

measurement was observed for the stable materials. Pt had the best reproducibility and was used as the standard for the following measurements. **(b)** Comparison of results from various Li-mediated NRR studies to the findings of this work in terms of faradaic efficiency and ammonia production rate (figure based on Ref. [26] see **Supplementary Note 4**) **(c)** Comparison of performance metrics to industrial Haber-Bosch thermal catalysis process in terms of energy efficiency. The color coding represents the Faradaic efficiency from 0% (red) to 8% (blue) over all experiments. **(d-e)** $NH_3$ synthesis rate and faradaic efficiency with bias, $N_2$ inflow, and time, respectively. Materials labeled with '*' are deemed unstable due to fast dissolution of the cathode, while '**' denotes an initially inefficient but stable catalyst (e.g., Hf) after the natural passivation layer has worn off, and subsequently becoming unstable. If not stated otherwise, the measurements were conducted at 200 V bias, 0.07 L min$^{-1}$ of $N_2$ inflow on a Pt cathode with 5 min of plasma.

Taken together, the results in Figure 3 clearly show that the Pt plasma cathode leads to the highest ammonia production among all tested materials, with almost two orders of magnitude higher yields compared to W. Furthermore, doubling the $N_2$ flow leads to almost double the amount of $NH_3$ produced, which strongly indicates that the exact flow and plasma conditions, which are not yet optimized in this proof of concept study, are critical for a high $NH_3$ rate, and that there is a great potential for improved yields with new reactor designs. In particular, we hypothesized that a larger exposed plasma area on the Pt wire should lead to increased yields.

**Wire length & plasma correlation.**

To assess the effect of the exposed wire surface, we systematically varied the length of the plasma cathode wire in our reactor 1.0 mm to 4.0 mm. We observed a distinct transition in the catalytic conversion of nitrogen between 3.0 mm and 3.5 mm **(Figure 4 a)**. For wire lengths > 3 mm, a substantial enhancement in faradaic efficiency is observed, paralleled by a drop in current density and a 40 % reduction of the temperature change at 4 mm. We ascribe this behavior to the formation of a stable contact glow discharge (CGDC) plasma, a condition hindered by the emergence of bubbles from the $N_2$ inlet at smaller wire lengths. We observed that increasing the plasma wire length not only enhances the Faradaic efficiency but also creates spatial regions with distinct thermal profiles. The longer wire exposes a larger area to the hot plasma conditions required for water decomposition, thereby generating abundant hydrogen radicals. Simultaneously, cooler regions along the wire allow the Pt surface to effectively bind and hydrogenate activated nitrogen species, culminating in a robust and synergistic pathway for ammonia synthesis. The reduction in current density (at constant potential of 200 V) and temperature increase over the 5 min of reaction is attributed to the increased resistance at the interface, stemming from the presence of the gas sheath. Moreover, the transition of liquid water to vapor during the process reduces the energy requirement for water electrolysis and thermal decomposition, increasing thus the available hydrogen for ammonia formation. In the plasma, the generation of ions and radicals might enable a non-thermal Eley-Rideal (ER) reaction mechanism alongside the thermal Langmuir-Hinshelwood (LH) mechanism. Short-lived atomic nitrogen species produced in the plasma stabilize upon adsorption onto the cathode surface, particularly favored by Pt, which exhibits strong nitrogen binding. Possibly following a reverse Gerischer-Maurer Mechanism[27], the Pt surface can bind nitrogen, and partially hydrated intermediates, which eventually form ammonia that might detach in the cooler regions of the cathode. A crucial intermediate seems to be the NH radical. Partial hydration or direct reaction with hydrogen are possible. ER/LH pathways on the catalyst surface are the main contributors here. Using materials other than Pt as cathode yields little to no ammonia, indicating that the pure gas/vapor/plasma reaction is only contributing negligible amounts of ammonia. This process is schematically depicted in **Figure 4b** and summarized in **Supplementary Note 5**. This could also explain the enhanced activity with increasing wire length, where the hotter regions excite and dissociate the nitrogen and form hydrogen radicals that might then react at the cooler sites. **Supplementary Note 2** contains high speed camera measurements of the stochastic discharge distribution, corroborating this hypothesis of an inhomogeneous plasma distribution.

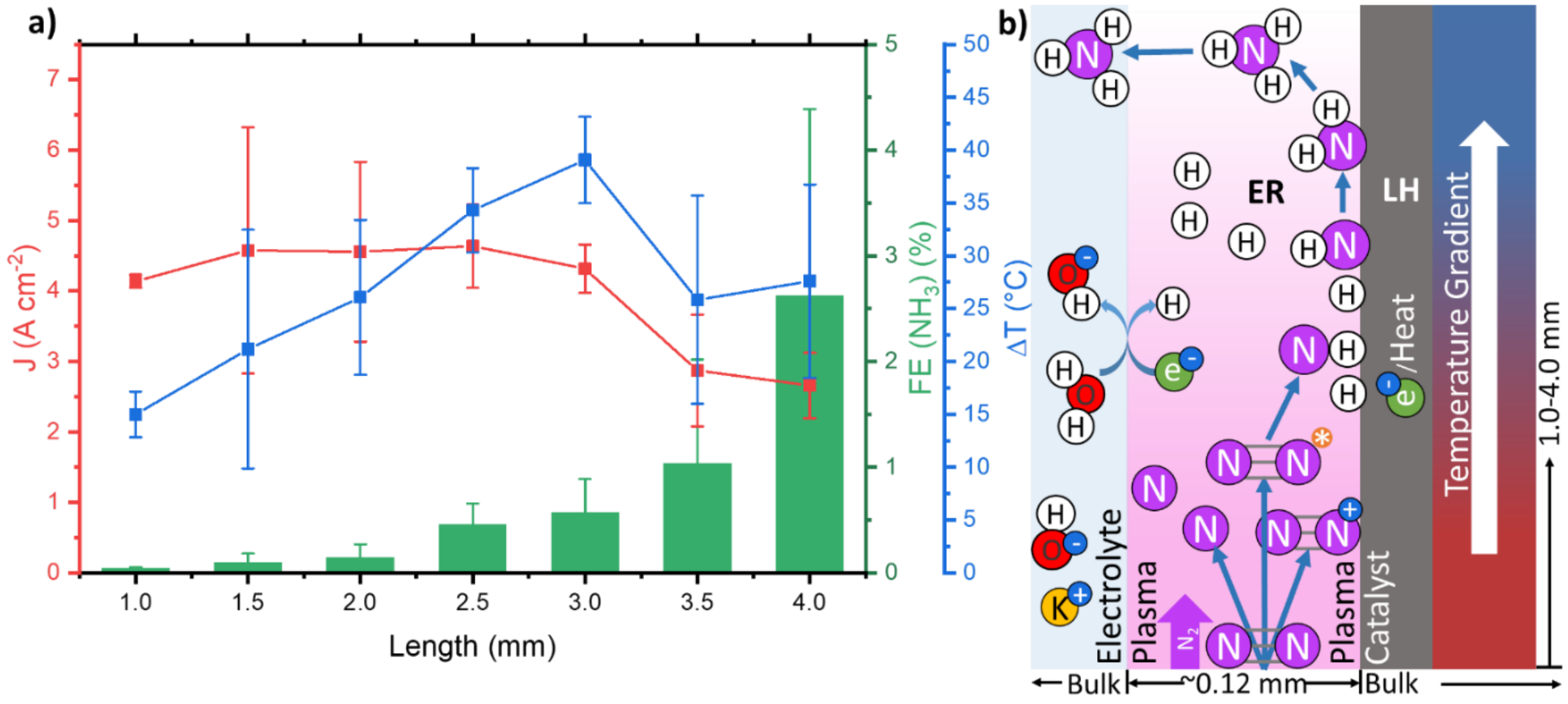

**Figure 4: Impact of wire length on current density, temperature increase and Faradaic efficiency. (a)** Relationship between wire length of the active plasma electrode, elevation in the electrolyte temperature during the experiment (blue), the current density (red), and the corresponding Faradaic efficiency (green). Between wire lengths of 3.0 mm and 3.5 mm, a transition can be observed, where the FE% drastically increases. **(b)** Schematic of the interface and possible mechanistic pathways for the hydrogenation of nitrogen species, starting from $N_2$*, $N_2^+$, NH, and atomic N.

To probe any remaining species on the electrode surface, cyclic voltammetry was conducted immediately after 5 min of plasma. The surface of the electrode becomes roughened, which can be seen from the pseudo-capacitive increase of the CVs of about a factor of three. Additionally, it can be observed that the hydrogen adsorption/desorption regime is shifted down due to the presence of oxygen. At 0.5 $V_{RHE}$, a new peak emerges but disappears with consecutive cycles. This could be the onset of the ammonia oxidation reaction (AOR), which disappears once the adsorbed ammonia is fully oxidized. A clear distinction between Pt oxidation and AOR around 0.9 $V_{RHE}$ is not possible. A shoulder peak at 0.7 to 0.8 $V_{RHE}$ is associated with Pt-hydroxide formation and gets more pronounced with the exposure to hydroxyl radicals during plasma. Directly after the plasma, the Pt-O reduction peak at 0.57 $V_{RHE}$ increases and the $H_{upd}$ adsoption and desorption become more pronounced, as seen in Figure 5.

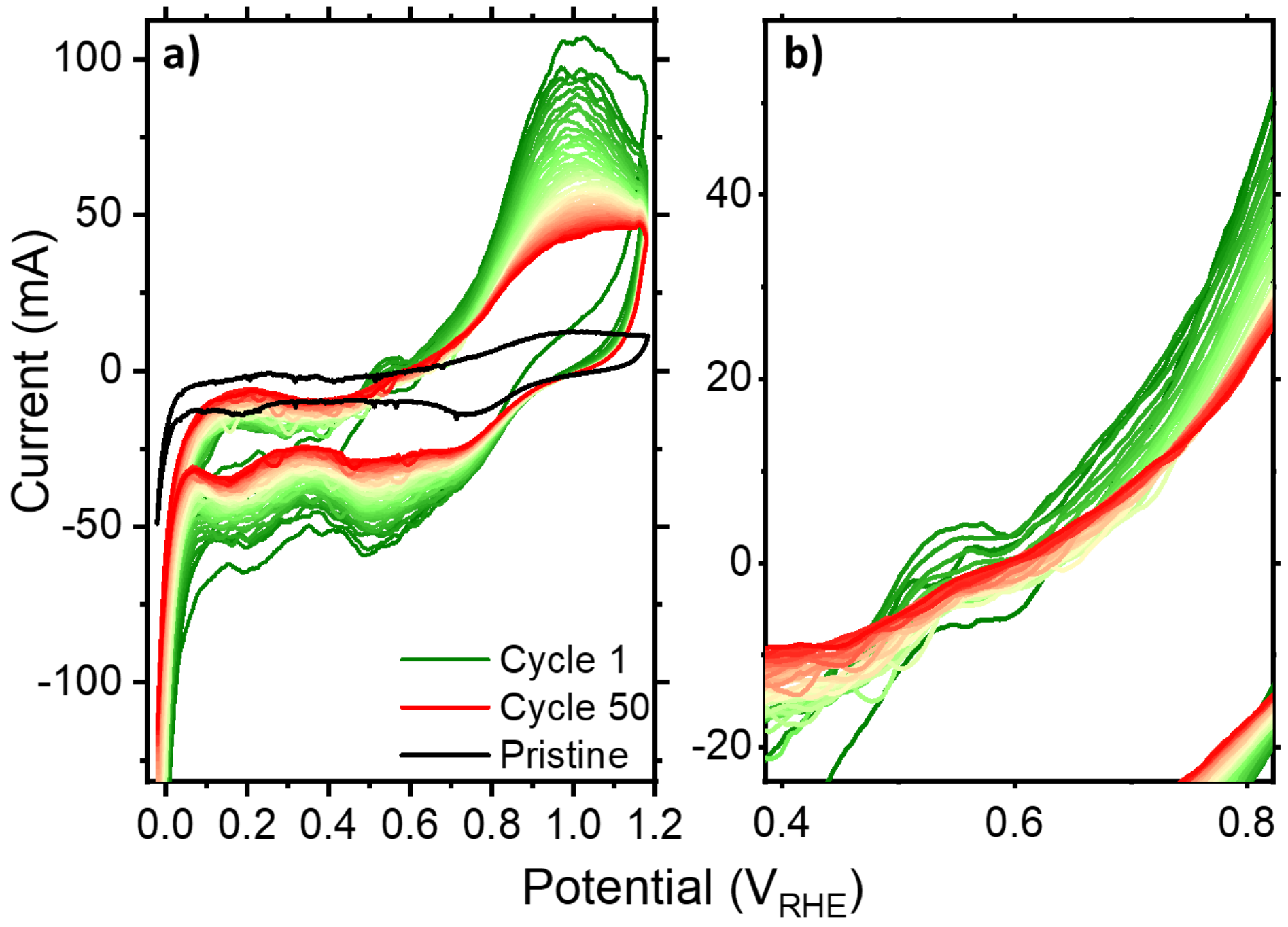

**Figure 5: Cyclic voltammetry before and after plasma. (a)** Cyclic voltammetry on a fresh 3.0 mm Pt wire before (black) and after plasma (green to red). The first 50 cycles after plasma were collected between 45°C to 30°C (constant cooling). **(b)** is a zoom in the AOR region. All CVs were collected at 50 mV $s^{-1}$ in 0.1 M KOH and 0.07 L $min^{-1}$ of $N_2$ inflow around the electrode.

In conclusion, we have carried out proof-of-concept experiments and demonstrated the ability of generating $NH_3$ from $N_2$ with an in-liquid plasma-electrocatalytic reactor. Among the tested electrode materials, platinum exhibited optimal stability and catalytic conversion, in clear contrast to W, which also formed a stable plasma, but produced only minor amounts of $NH_3$. The attainment of the hot plasma state at potentials exceeding 200 V resulted in a fourfold improvement in faradaic efficiency for ammonia compared to the cold plasma state. The optimal nitrogen flow was observed at 0.07 L $min^{-1}$, striking a balance between plasma gas sheath stability and activity. While the system demonstrated stability over 30 min, the operation conditions are not yet fully optimized, and challenges persisted in preventing the potential outgassing or thermal decomposition of the formed ammonia, even with intermittent cooling intervals every five minutes. Addressing this issue necessitates an improved solution for ammonia collection or an enhanced cooling mechanism. Distinct hot and cold states resembling those reported in established plasma systems such as arc and glow discharge were monitored via Optical emission spectroscopy. In the arc discharge state, molecular excited nitrogen species were absent, with nitrogen undergoing atomic splitting and recombination with hydrogen from water splitting at temperatures exceeding 3000 K. Thermal decomposition likely also contributes to the process, potentially resulting in the complete dissociation of $N_2$ in the hot state. Thus, we propose here a reaction pathway from a combination of Gerischer-Maurer mechanism to bind atomic nitrogen to the catalyst surface followed by Langmuir-Hinshelwood and Eley-Rideal mechanisms to form ammonia. Our findings collectively contribute to the understanding of plasma-catalyzed nitrogen reduction in liquid environments, laying the foundation for further advancements in this field.

The presented system is still in its early development, and the full understanding of the intriguing findings observed will require more in-depth insight into the thermal catalysis, electrochemical catalysis and plasma catalysis processes involved. In addition, further optimization of the reactor configuration and electrode parameters should be undertaken in order to enhance the currently low energy efficiency. The first attempts were illustrated here, including varying the plasma wire length, the plasma electrode geometries and materials as well as implementing a new reactor design with improved $N_2$ gas flow.

Finally, beyond $NH_3$ production, we envision that such a plasma-catalytic reactor might also be used to attain high rate $CO_2$ reduction and other critical chemical energy conversions, especially when green electricity is cheap and abundant and a high conversion rate is needed.

**Experimental Methods:**

**Sample preparation**

Prior to utilization, each cathode wire was mechanically and electrochemically polished. Furthermore, the materials underwent flame annealing using a propane burner positioned at the apex of the flame to mitigate the risk of contamination arising from impure combustion.

**Plasma Generation**

For the generation of a contact glow discharge, a potential was applied between two electrodes that are submerged in an aqueous electrolyte. The plasma electrode is defined as the active working electrode. The counter electrode has a > 10 times larger surface area than the active electrode. The distance of both electrodes was intentionally kept small, but far enough (> 5mm) that no arc over can occur. The configuration is depicted in in **Supplementary Figure 1**, the shortest distance between the active electrode and the counter electrode is roughly 5 mm in all directions.

**Determination of $NH_3$ concentration *via* UV-VIS spectroscopy and Indophenol Blue**

The concentration of $NH_3$ in the liquid phase was determined through UV-Vis spectroscopy and the indophenol blue method. A 5 mL aliquot of the solution was extracted from the plasma reactor. Subsequently, 2 mL of the cooled solution were mixed with 0.5 mL of phenol nitroprusside solution (50 g/L phenol in a sodium nitroprusside solution containing a stabilizer, MERCK) and 0.5 mL of alkaline hypochlorite solution (0.2% Sodium Hypochlorite in 0.625 M Sodium Hydroxide, MERCK). The resulting mixture was agitated and then kept in the darkness for 30 minutes. Ammonia selectively mediates the linkage of two phenol rings in the presence of a nitroprusside catalyst, thereby engendering the formation of the indophenol molecule, which exhibits light absorption in the red spectrum, resulting in a blue color perceivable by eye. To quantify the amount of $NH_3$, 3 mL of the prepared solution were transferred to a cuvette for UV-VIS spectroscopy (Agilent Cary 60 UV-Vis). With increasing $NH_3$ concentration, the absorption peak at 633 nm increases. The peak intensity was calibrated with a solution of known $NH_3$ concentration (**Supplementary Figure 7**). Additionally, UV-Vis spectroscopy was performed to test for NOx and hydrazine. However, only trace quantities of these compounds were detected.

**Optical emission spectroscopy (OES)**

The gas temperature was determined through two methods: fitting transitions of $N_2$(C-B) and $N_2^+$(B-X), and fitting the OH(A-X) transition with the LIFBASE[28] software. Based on the in-house software analysis, the neutral gas temperature was estimated to range between 1000 K (180 V) and 3300 K (230 V), with a reduced electric field of 36 to 38 Td at 200 V in the hot state. Additionally, by fitting spectral lines of the Balmer series ($H_\alpha$ and $H_\beta$), electron density ($n_e$) was determined even when no nitrogen lines were observed, showing a linear correlation with the applied voltage and an increase from the cold to hot states. Electron densities ranged from $10^{14}$ to $10^{16}$ $cm^{-3}$ across all experiments, as detailed in **Supplementary Note 2**. The electrode temperatures were determined by background continuum fitting and ranged from room temperature at 175 V (Pt-electrode) to 1600 K at 230 V.

Examining the electrical potential's influence on the emission intensity of the $N_2$(C-B) transition revealed that the hot plasma state was only achieved at or above 200 V. Beyond 230 V, stabilizing the cold state became challenging, leading to an immediate transition to the hot state. The difference in emission lines between 180 V and 200 V (cold state) was negligible, but upon transitioning to the hot state, the $N_2$(C-B) band intensified significantly, correlating with increased ammonia production.

**Determination of gaseous product concentration via GC**

Gas products such as $H_2$ and $NO_x$ were detected and quantified from the head space by online gas chromatography (GC, Agilent 8860 GC System). The faradaic efficiency (FE%) was determined *via* the equation:

$$FE\%_x = \frac{\dot{V} * C_x * n_x * F}{A * V_M * j_{total} * t} * 100\% \quad (1)$$

the flow rate $\dot{V}$ of $N_2$ (mL $min^{-1}$), the volume fraction $C_x$ of the gas x determined by GC, the number of electrons $n_x$ involved in the reaction, the Faraday constant F ($9.648x10^4$ C $mol^{-1}$), geometric surface area A of the electrode, the molar volume $V_m$ of an ideal gas (24.0 L $mol^{-1}$ at 25°C and 1 atm) and the total current density $j_{total}$ (A $cm^{-2}$).

**$NH_3$ Faradaic Efficiency**

The Faradaic efficiency was calculated from the concentration of ammonia in the electrolyte according to the equation below. In the gas-phase, no ammonia was detected via GC, MS or in the wash-bottle.

$$Q_{NH_3} = 3 \cdot F \cdot C_{NH_3} \cdot V \tag{2}$$

$$FE_{NH_3} = \frac{Q_{NH_3}}{Q} \cdot 100 = \frac{n \cdot F \cdot C_{NH_3} \cdot V}{Q} \cdot 100 \tag{3}$$

With the charge $Q_{NH_3}$, Faraday constant $F$, the Volume V, $NH_3$ concentration in the electrolyte $C_{NH_3}$ and the number of electrons n transferred during the reaction for each mole of $NH_3$ ($6H_2O + 6e^- + N_2 \rightarrow 2NH_3 + 6OH^-$), i.e. n=3.

**Erroneous Sources of Ammonia**

Previous studies have highlighted the significant number of potential trace $NH_3$ sources that can contribute to erroneous quantification.[29, 30] The accurate detection of ammonia at (sub-) parts per million (ppm) levels presents a non-trivial challenge, necessitating a meticulous testing approach that systematically eliminates all alternative factors. Here, however thanks to the introduction of the plasma the overall amounts of $NH_3$ produced are substantially larger than in previous electrocatalytic studies (see **Figure 3b**), which lead to the observation of a clear bias dependent $NH_3$ production rate. Nonetheless, in pursuit of a rigorous assessment, the experiments were replicated using argon (Ar) in lieu of nitrogen ($N_2$). Furthermore, the investigation encompassed the examination of both, the pristine electrolyte and the electrolyte subjected to $N_2$ saturation for a duration of 10 minutes. All involved materials and equipment that was in contact with the electrolyte has been tested for ammonia impurities. Only in the presence of $N_2$ in combination with plasma ignition could we detected non-trace amounts of $NH_3$. Without plasma or without $N_2$, no $NH_3$ is detected beyond trace amounts. The background levels of ammonia for all measurements varied between 0.01 ppm and 0.03 ppm. Ammonia concentrations below 0.01 ppm could not be detected.

**Current-Voltage measurements**

For voltages between 0-10 V and electrochemical impedance spectroscopy, a commercial potentiostat (Gamry Instruments, Reference 3000) was used. The resistance of the electrolyte was determined via stepped potentiostatic electrochemical impedance spectroscopy (SPEIS) for the ±10 V window. The Nyquist plots were fitted *via* the ZFit software (Biologic). Above 10 V, the current-voltage curves were collected *via* a self-built high voltage probe and recorded with an oscilloscope (Rhode & Schwarz RTC1002). These voltages were generated by an external power supply (FUG NTN 700-350).

**Energy Efficiency**

The energy efficiency was calculated based on the total electric energy that went into the system, measured with a high voltage probe and current probe (self-built). The specific energy for the energy efficiency was based on the formation enthalpy and the respective calculated energy for the ammonia formation (see **Supplementary Note 3 & 5**). The calculations are summarized in **Supplementary Note 6**.

**Ex-situ ICP-MS Measurements**

ICP-MS measurements were conducted on a Thermo Scientific iCAQ RQ ICP-MS in KED mode (He flow rate: 0.65 L $min^{-1}$) using an ESI 2DX autosampler for sample uptake. The examined electrolyte aliquots (5mL) before/after plasma catalysis experiments (total initial volume: 95 mL) were diluted in a 1:10 ratio to ensure total solid contents of below 1 at. %. Standard solutions with known concentrations of Pt (from 10 to 500 µg $L^{-1}$, diluted from 10mg $L^{-1}$ ICP-MS standard, Honeywell Flucka) were freshly prepared to calibrate the ICP-MS prior to quantification, along with $^{186}Re$ as internal standard (200 µg $L^{-1}$, diluted from 10mg $L^{-1}$ ICP-MS standard, Honeywell Flucka).

The determination of dissolved metal ion concentrations was accomplished through inductively coupled plasma mass spectrometry (ICP-MS) analysis of the electrolyte. Sequential samples were extracted from the reactor at 5 min intervals during each operational increment with the plasma activated. Subsequent to each increment, the reactor was cooled down to room temperature before initiating the subsequent stage. A deliberate alternation between the hot and cold plasma states was implemented to discern potential variations in metal ion dissolution contingent upon temperature fluctuations. This methodology aimed to unveil any changes in metal ion dissolution associated with the varying thermal conditions. The combined data, showing the concentration profiles of all feasible metal ions originating from ceramics, reactor materials, and the catalyst, is presented in **Supplementary Figure 8**. This dataset is concurrently compared with the ammonia concentration determined using the previously mentioned indophenol blue method. To also compare the changes in ammonia concentration to the current and temperature increase, they were plotted for one individual measurement. This is the same measurement that was used for the set of ICP-MS measurements. A clear distinguishable difference in current, temperature increase, and ammonia concentration increase between the hot and the cold state was observed, while the metal ion concentration was not significantly influenced by the different states. The resulting changes in temperature increase, current, and ammonia concentration are plotted together in **Supplementary Figure 9.**

**Ex-situ XPS measurements:**

XPS data were collected in an ultrahigh vacuum chamber equipped with a Specs XR50 Al/Mg X-ray source (Al $K_\alpha$, 1486.61 eV) and a Specs Phoibos 100 MCD5 electron energy analyzer. For peak fitting, Shirley backgrounds were used in combination with Gaussian-Lorentzian (GL) line shapes for the peaks deconvoluted with the CasaXPS[31] software. The quantitative information of the ratio between different elements and for one element the relative contribution of different oxidation states were extracted from the analysis of the respective peak areas in the XPS spectra. The area of each element was previously corrected with the relative sensitivity factors RSF (tabulated values specific for each XPS instrument). In the case of Cu, differentiating between Cu(0) and Cu(I) is inaccurate with only the Cu2p peak, therefore, a linear deconvolution of the Cu LMM Auger peak was conducted based on reference spectra acquired from the pure species or oxide compounds (Cu, CuO and $Cu_2O$). All XPS measurements were conducted ex-situ, therefore some re-oxidation due to the transport in air is expected.

By comparison of the Pt 4f regions before and after 5 min of plasma at 200 V and 0.07 L $min^{-1}$ of $N_2$ flow, it can be observed that the elemental composition of the Pt cathode does not significantly change. The peaks were fitted with an asymmetric line shape and appear fully metallic in the pristine as well as the post plasma state, as can be seen in **Supplementary Figure 3**. From the XPS on other cathode materials, a clear change in oxidation state between the pristine and the post plasma state can be observed as for example in case of Cu. In case of Cu, an initial ratio of 6:1 was found in the Cu2p region for the Cu(0/I) to Cu(II) ratio. After 5 min of plasma, this ratio is changed to 2:1 due to the strong oxidation of the Cu. A similar picture can be seen from the Cu LMM Auger spectrum. The spectra with the respective fits can be seen in **Supplementary Figure**

**4 & 5**. In the W 4f, no reduction was observed. In the pristine state, $WO_3$ is the only species found, coming from a thick natural oxide layer. Post plasma, the amount of oxide does not change, likely due to reoxidation by exposure to air or the hot electrolyte when the plasma is switched off. The spectra and respective fits can be seen in **Supplementary Figure 6**. The fit results and elemental composition can be found in **Supplementary Note 7** in the **Supplementary Table 7**. These XPS measurements corroborates, that the change in oxidation state during plasma leads to substantially worse catalytic conversion and explains the trends observed in **Figure 3 (a)** nicely. This means, that finding a better NRR catalyst that does not oxidize under these conditions could significantly boost the energy efficiency of this system.

**Acknowledgments:** We acknowledge funding by the Deutsche Forschungsgemeinschaft (DFG, German Research Foundation) – project no. 327886311 – SFB 1316, subproject B1 and B5. We also appreciate the insightful discussions with Prof. Robert Schlögl.

**Author Contributions:** P.G., S.Z.O. and B.R.C conceived of the project and planned the experiments. P.G. built the reactor and conducted the experiments. P.G., J.L.G., N.B., and A.R. G. conducted and analyzed the spectroscopic plasma experiments. D.E.L. conducted the ICP-MS experiments. A. R.O.S. and P.G. conducted and analyzed the GC measurements. P.G., S.Z.O., N.B. and J.L.G. analyzed and interpreted the data. P.G., S.Z.O. and B.R.C. wrote the manuscript with input from all authors.

**Competing Interests**

**Data and materials availability**

**License information**

**Supplementary Table of Content:**

**References:**

1. Humphreys, J.; Lan, R.; Tao, S. J. A. E.; Research, S., Development and recent progress on ammonia synthesis catalysts for Haber–Bosch process. **2021,** *2* (1), 2000043.
2. Kyriakou, V.; Garagounis, I.; Vourros, A.; Vasileiou, E.; Stoukides, M. J. J., An electrochemical haber-bosch process. **2020,** *4* (1), 142-158.
3. MacFarlane, D. R.; Cherepanov, P. V.; Choi, J.; Suryanto, B. H.; Hodgetts, R. Y.; Bakker, J. M.; Vallana, F. M. F.; Simonov, A. N. J. J., A roadmap to the ammonia economy. **2020,** *4* (6), 1186-1205.
4. Smith, C.; Hill, A. K.; Torrente-Murciano, L. J. E.; Science, E., Current and future role of Haber–Bosch ammonia in a carbon-free energy landscape. **2020,** *13* (2), 331-344.
5. Wang, M.; Khan, M. A.; Mohsin, I.; Wicks, J.; Ip, A. H.; Sumon, K. Z.; Dinh, C.-T.; Sargent, E. H.; Gates, I. D.; Kibria, M. G. J. E.; Science, E., Can sustainable ammonia synthesis pathways compete with fossil-fuel based Haber–Bosch processes? **2021,** *14* (5), 2535-2548.

6. Li, K.; Andersen, S. Z.; Statt, M. J.; Saccoccio, M.; Bukas, V. J.; Krempl, K.; Sažinas, R.; Pedersen, J. B.; Shadravan, V.; Zhou, Y.; Chakraborty, D.; Kibsgaard, J.; Vesborg, P. C. K.; Nørskov, J. K.; Chorkendorff, I., Enhancement of lithium-mediated ammonia synthesis by addition of oxygen. *Science* **2021,** *374* (6575), 1593-1597.
7. Fu, X.; Pedersen, J. B.; Zhou, Y.; Saccoccio, M.; Li, S.; Sažinas, R.; Li, K.; Andersen, S. Z.; Xu, A.; Deissler, N. H.; Mygind, J. B. V.; Wei, C.; Kibsgaard, J.; Vesborg, P. C. K.; Nørskov, J. K.; Chorkendorff, I., Continuous-flow electrosynthesis of ammonia by nitrogen reduction and hydrogen oxidation. **2023,** *379* (6633), 707-712.
8. Mushtaq, M. A.; Arif, M.; Yasin, G.; Tabish, M.; Kumar, A.; Ibraheem, S.; Ye, W.; Ajmal, S.; Zhao, J.; Li, P. J. R.; Reviews, S. E., Recent developments in heterogeneous electrocatalysts for ambient nitrogen reduction to ammonia: Activity, challenges, and future perspectives. **2023,** *176*, 113197.
9. Fichter, F.; Girard, P.; Erlenmeyer, H. J. H. C. A., Elektrolytische Bindung von komprimiertem Stickstoff bei gewöhnlicher Temperatur. **1930,** *13* (6), 1228-1236.
10. Andersen, S. Z.; Čolić, V.; Yang, S.; Schwalbe, J. A.; Nielander, A. C.; McEnaney, J. M.; Enemark-Rasmussen, K.; Baker, J. G.; Singh, A. R.; Rohr, B. A. J. N., A rigorous electrochemical ammonia synthesis protocol with quantitative isotope measurements. **2019,** *570* (7762), 504-508.
11. Nielander, A. C.; McEnaney, J. M.; Schwalbe, J. A.; Baker, J. G.; Blair, S. J.; Wang, L.; Pelton, J. G.; Andersen, S. Z.; Enemark-Rasmussen, K.; Colic, V. J. A. C., A versatile method for ammonia detection in a range of relevant electrolytes via direct nuclear magnetic resonance techniques. **2019,** *9* (7), 5797-5802.
12. Suryanto, B. H. R.; Matuszek, K.; Choi, J.; Hodgetts, R. Y.; Du, H.-L.; Bakker, J. M.; Kang, C. S. M.; Cherepanov, P. V.; Simonov, A. N.; MacFarlane, D. R., Nitrogen reduction to ammonia at high efficiency and rates based on a phosphonium proton shuttle. **2021,** *372* (6547), 1187-1191.
13. Li, S.; Zhou, Y.; Fu, X.; Pedersen, J. B.; Saccoccio, M.; Andersen, S. Z.; Enemark-Rasmussen, K.; Kempen, P. J.; Damsgaard, C. D.; Xu, A.; Sažinas, R.; Mygind, J. B. V.; Deissler, N. H.; Kibsgaard, J.; Vesborg, P. C. K.; Nørskov, J. K.; Chorkendorff, I., Long-term continuous ammonia electrosynthesis. *Nature* **2024,** *629* (8010), 92-97.
14. Zhang, K.; Cao, A.; Wandall, L. H.; Vernieres, J.; Kibsgaard, J.; Nørskov, J. K.; Chorkendorff, I., Spin-mediated promotion of Co catalysts for ammonia synthesis. *Science* **2024,** *383* (6689), 1357-1363.
15. Cao, A.; Bukas, V. J.; Shadravan, V.; Wang, Z.; Li, H.; Kibsgaard, J.; Chorkendorff, I.; Nørskov, J. K., A spin promotion effect in catalytic ammonia synthesis. *Nature Communications* **2022,** *13* (1), 2382.
16. Liang, J.; Li, Z.; Zhang, L.; He, X.; Luo, Y.; Zheng, D.; Wang, Y.; Li, T.; Yan, H.; Ying, B.; Sun, S.; Liu, Q.; Hamdy, M. S.; Tang, B.; Sun, X., Advances in ammonia electrosynthesis from ambient nitrate/nitrite reduction. *Chem* **2023,** *9* (7), 1768-1827.
17. Bai, L.; Franco, F.; Timoshenko, J.; Rettenmaier, C.; Scholten, F.; Jeon, H. S.; Yoon, A.; Rüscher, M.; Herzog, A.; Haase, F. T.; Kühl, S.; Chee, S. W.; Bergmann, A.; Beatriz, R. C., Electrocatalytic Nitrate and Nitrite Reduction toward Ammonia Using Cu2O Nanocubes: Active Species and Reaction Mechanisms. *Journal of the American Chemical Society* **2024,** *146* (14), 9665-9678.

18. Yoon, A.; Bai, L.; Yang, F.; Franco, F.; Zhan, C.; Rüscher, M.; Timoshenko, J.; Pratsch, C.; Werner, S.; Jeon, H. S.; Monteiro, M. C. d. O.; Chee, S. W.; Roldan Cuenya, B., Revealing catalyst restructuring and composition during nitrate electroreduction through correlated operando microscopy and spectroscopy. *Nature Materials* **2025,** *24* (5), 762-769.
19. Liu, H.; Lang, X.; Zhu, C.; Timoshenko, J.; Rüscher, M.; Bai, L.; Guijarro, N.; Yin, H.; Peng, Y.; Li, J.; Liu, Z.; Wang, W.; Cuenya, B. R.; Luo, J., Efficient Electrochemical Nitrate Reduction to Ammonia with Copper-Supported Rhodium Cluster and Single-Atom Catalysts. **2022,** *61* (23), e202202556.
20. Ma, H.; Schneider, W. F. J. A. C., Structure-and temperature-dependence of Pt-catalyzed ammonia oxidation rates and selectivities. **2019,** *9* (3), 2407-2414.
21. Hetherington, H. J. J. o. C. E., The fixation of atmospheric nitrogen. **1926,** *3* (2), 170.
22. Wang, W.; Patil, B.; Heijkers, S.; Hessel, V.; Bogaerts, A. J. C., Nitrogen fixation by gliding arc plasma: better insight by chemical kinetics modelling. **2017,** *10* (10), 2145-2157.
23. Jellinek, H. H. G.; Kachi, H., The catalytic thermal decomposition of water and the production of hydrogen. *International Journal of Hydrogen Energy* **1984,** *9* (8), 677-688.
24. Tulenbergenov, T.; Skakov, M.; Kolodeshnikov, A.; Zuev, V.; Rakhadilov, B.; Sokolov, I.; Ganovichev, D.; Miniyazov, A.; Bukina, O., Interaction between nitrogen plasma and tungsten. *Nuclear Materials and Energy* **2017,** *13*, 63-67.
25. Bolleter, W.; Bushman, C.; Tidwell, P. W. J. A. C., Spectrophotometric determination of ammonia as indophenol. **1961,** *33* (4), 592-594.
26. Chang, W.; Jain, A.; Rezaie, F.; Manthiram, K., Lithium-mediated nitrogen reduction to ammonia via the catalytic solid–electrolyte interphase. *Nature Catalysis* **2024,** *7* (3), 231-241.
27. Gerischer, H.; Mauerer, A., Untersuchungen Zur anodischen Oxidation von Ammoniak an Platin-Elektroden. *Journal of Electroanalytical Chemistry and Interfacial Electrochemistry* **1970,** *25* (3), 421-433.
28. Luque, J.; Crosley, D. R. J. S. i. r. M., LIFBASE: Database and spectral simulation program (version 1.5). **1999,** *99* (009).
29. Choi, J.; Suryanto, B. H. R.; Wang, D.; Du, H. L.; Hodgetts, R. Y.; Ferrero Vallana, F. M.; MacFarlane, D. R.; Simonov, A. N., Identification and elimination of false positives in electrochemical nitrogen reduction studies. *Nat Commun* **2020,** *11* (1), 5546.
30. Andersen, S. Z.; Colic, V.; Yang, S.; Schwalbe, J. A.; Nielander, A. C.; McEnaney, J. M.; Enemark-Rasmussen, K.; Baker, J. G.; Singh, A. R.; Rohr, B. A.; Statt, M. J.; Blair, S. J.; Mezzavilla, S.; Kibsgaard, J.; Vesborg, P. C. K.; Cargnello, M.; Bent, S. F.; Jaramillo, T. F.; Stephens, I. E. L.; Norskov, J. K.; Chorkendorff, I., A rigorous electrochemical ammonia synthesis protocol with quantitative isotope measurements. *Nature* **2019,** *570* (7762), 504-508.
31. Fairley, N.; Fernandez, V.; Richard-Plouet, M.; Guillot-Deudon, C.; Walton, J.; Smith, E.; Flahaut, D.; Greiner, M.; Biesinger, M.; Tougaard, S.; Morgan, D.; Baltrusaitis, J., Systematic and collaborative approach to problem solving using X-ray photoelectron spectroscopy. *Applied Surface Science Advances* **2021,** *5*, 100112.

Supporting Information for

# In-liquid Plasma Catalysis for Nitrogen Reduction

P. Grosse[1*], J.L. Gembus[2], F. Landwehr[1], A.R. Silva Olaya[1], D. Escalera-López[1], N. Bibinov[2], A. R. Gibson[2,3], S.Z. Oener[1], B. Roldan Cuenya[1]*

[1]*Department of Interface Science, Fritz-Haber Institute of the Max Planck Society, 14195, Berlin, Germany*

[2]*Chair of Applied Electrodynamics and Plasma Technology, Ruhr University Bochum, 44801, Bochum, Germany*

[3]*York Plasma Institute, School of Physics, Engineering and Technology, University of York, Heslington, YO10 5DD, United Kingdom*

Corresponding authors: *roldan@fhi-berlin.mpg.de, grosspk7@fhi-berlin.mpg.de

## Contents

**Supplementary Note 1:**

**Staircase Potentio-Electrochemical Impedance Spectroscopy (SPEIS)**

To determine the electrocatalytic NRR activity of the different cathode materials, staircase potentio- electrochemical impedance spectroscopy (SPEIS) was conducted in a range from 0 V to -10 V. The materials were tested with and without 0.07 L $min^{-1}$ of $N_2$ flow. Each potential step was kept for 5 min and the last 70% of the current was used to calculate the average current in order to decouple capacitive effects from the polarization curve. From the impedance spectroscopy at each voltage step, the electrolyte resistance was calculated by fitting a Nyquist plot with an equivalent circuit. The resulting polarization curves and resistances are depicted in **Supplementary Figure 10** for all tested cathode materials. The only exception here is Hf, which has a natural passivation layer that only breaks down at substantially higher voltages during plasma.

The linear parts of the polarization curves were fitted to identify changes in the slopes. From the fitted slopes with and without $N_2$ flow, it can be seen that the effect of reduced surface area (covered by the gas) on the change of the resistance (according to Ohm's law) is approximately similar over all materials at -19 ± 5 µS $cm^{-2}$. The average change in onset (derived from the x0 of the linear fit) has a larger error at -71.49 ± 36.70 mV.

**High Speed IV-curves**

To determine the changes in the plasma ignition voltage, the current and voltage were measured with an oscilloscope (RTC1002, Rohde & Schwarz) and a self-built high voltage probe. The ramp-up time of the voltage is determined by a combination of electrochemical effects and the electronics of the system. **Supplementary Figure 11** depicts the IV-curves for all investigated cathode materials.

It can be seen, that the beginning of the ramp-up is fully governed by Ohm's law. Around 80 V, the gas evolution reduces the amount of current that can be transferred over the interface. This proceeds until the interface is fully covered in a gas-vapor layer. At a sufficiently high voltage, a gas breakdown occurs and a plasma is ignited between the electrolyte|gas and the gas|electrode interfaces. This effect is slightly influenced by the speed at which gas bubbles can form. With the addition of gas, the breakdown voltage slightly shifts to higher voltages, probably due to the higher gas coverage of the electrode material.

The current densities at a fixed potential are compared across materials in **Supplementary Figure 2**. Assuming 99% FE for HER, the current density can be used as qualitative indicator for the amount of $H_2$ produced.

**Supplementary Note 2:**

**Optical emission spectroscopy & high-speed camera imaging:**

The plasma generated in the liquid within our experimental setup was characterized using optical emission spectroscopy (OES). The light emitted by the plasma was collected using an optical fiber, incorporating a lens for both focusing and capturing the emission from the center of the plasma. Simultaneous utilization of multiple spectrometers enabled the acquisition of spectra spanning the UV-VIS and IR ranges with high temporal resolution.

All discharges were characterized by using echelle spectrometers. For the full range spectra that cover UV and Vis, an echelle spectrometer (ESA 4000, LLA Instruments, Germany) with a spectral resolution of $0.015\ \text{nm} < \Delta\lambda < 0.06\ \text{nm}$ in the spectral range of $200\ \text{nm} < \lambda < 800\ \text{nm}$ was used. The second spectrometer was also an echelle spectrometer (Aryelle Butterfly, LTB, Germany) capable of measuring two separate spectral ranges at high resolution $0.006\ \text{nm} < \Delta\lambda < 0.011\ \text{nm}$ in the range of $190\ \text{nm} < \lambda < 330\ \text{nm}$ and $0.022\ \text{nm} < \Delta\lambda < 0.057\ \text{nm}$ in the spectral range of $330\ \text{nm} < \lambda < 850\ \text{nm}$. The attached optical fibers used to guide light into each spectrometer were placed on opposite sides of the wire. The optical fibers were focused with a lens to collect only emission from the plasma. The setup is depicted in **Supplementary Figure 12**.

Within the spectral range and under consideration of the absorption from the reactor material, there are three transitions observable, $N_2$(C-B), $N_2^+$(B-X) and OH(A-X). A collisional-radiative model of excited state formation of $N_2$ and OH molecules based on values from NIST[1] is shown in **Supplementary Figure 13**. The red arrows indicate the observed (spontaneous) transitions, the blue arrows indicate the electron impact excitation processes. Lastly, the purple arrow indicates the desired nitrogen dissociation pathway. Next to the arrows, the respective rate constants of the electron impact excitations are shown. From this model, the reduced electric field (E/N in Td) and electron density ($n_e$) can be determined from the spectra according to Gröger et al.[2]

**Supplementary Figure 14 a & b** display the extended range of the molecular nitrogen regions for the Pt and W cathodes in hot and cold plasma states. For the W, no molecular ammonia was observed at 200 V, 0.07 L min$^{-1}$ of $N_2$ flow in either the hot or the cold state. However, atomic lines of W were found, indicating a much higher surface temperature. The reduced electric field can be calculated from the intensity ratio of the $N_2(C^3\Pi_u)$ and the $N_2^+(B^2\Sigma_u^+)$ states according to the equation 1 by Gröger et al.[2]

$$\frac{\left(I_{N_2(C)}\right)^2 \cdot Q_{N_2^+(B)}}{I_{N_2^+(B)} \cdot Q_{N_2(C)}^2 \cdot [N_2]^2} = \frac{\left(k_{N_2(C)}^{N_2(X)} + k_{N_2(A)}^{N_2(X)} \cdot \frac{B1 \cdot B2 \cdot k_{N_2(C)}^{N_2(A)}}{k_{N_2(B)}^{N_2(A)} + k_{N_2(C)}^{N_2(A)} + k_{N_2^+(X)}^{N_2(A)} + k_{diss.}^{N_2(A)}}\right)^2}{k_{N_2^+(B)}^{N_2^+(X)}} = \frac{\left(k_{N_2(C)}^{exc}\right)^2}{k_{N_2^+(B)}^{N_2(X)}} = F_2\left(\frac{E}{N}\right) \quad (1)$$

Here, I describes the intensity, Q is the collisional quenching of the respective state, and k is the rate constant for the transition between two states (superscript: lower state, subscript: upper state). The factor $B_1$=0.5 represents the relative population of the $N_2$(C) vibrational level (v'=0) via electron impact excitation of the $N_2$(A) state. $B_2$ = 0.5 is the branching fraction of the Einstein coefficient of the $N_2$(C-B; v'=0, v''=0) transition with respect to the total probability of photoemission from the upper level. Lastly, [$N_2$] describes the density of nitrogen, assuming a pressure of 1 atm and the respective temperatures.

The electron density can be calculated the same way via the intensity, collisional quenching and rate constants, according to equation 2.

$$n_e = \frac{I_{N_2(C)}}{Q_{N_2(C)} \cdot [N_2] \cdot k^{exc}_{N_2(C)}} \quad (2)$$

The obtained spectra were systematically normalized based on the $H_\alpha$ peak and OH(A-X) transition, ensuring robust comparability across various measurements. This normalization procedure enhances the reliability of comparative analysis across different experimental conditions, providing a standardized basis for interpreting the resulting OES spectra. For the same voltage, similar OH(A-X) transitions are observed across different materials.

The OH (A-X) and $N_2^+$(B-X) rovibrational and rotational transition bands were fitted by using the LIFBASE[3] spectroscopy tool. The Boltzmann equation was solved numerically to determine the electron distribution function for nitrogen and different reduced electric fields. For this calculations, the 'EEDF' code, developed by Napartovich et al.[4] was used. For the $N_2$(C-B) transitions, an in house developed code by Behringer et al. [5] was used to estimate the temperatures, analogous to the work of Gröger et al.[2]

**Estimating the Plasma Volume:**

To further characterize the plasma and associated electron density, high-speed videos were captured. This is required in order to estimate the plasma volume, which is used as an input to the reduced electric field and electron density calculations. The determination of the plasma volume necessitates two key parameters: a statistical analysis of the emission distribution on the wire surface and the thickness of the gas sheath enveloping the wire.

For the statistical distribution, emission emanating from the plasma was collected over a 250 µs duration at a frame rate of 20000 frames per second (fps) using a high-speed camera (Phantom VEO 410L). By confining the analysis to the wire area, reflective interferences were mitigated. Normalization of each frame to the video's maximum intensity, followed by selective extraction of the lower 70% of intensities, facilitated the isolation of emissions predominantly originating from the contact glow discharge on the surface. The processed results are depicted in **Supplementary Figure 15**.

**Gas Sheath Thickness:**

To ascertain the gas sheath thickness, a binary profile section of the back-illuminated video was extracted. The image section was filtered, so that no disconnected bubble and only objects directly connected to the cathode are considered. The average thickness of the profile was computed for each frame, and a Gaussian curve was fitted to the resulting histogram. Three distinct sections were delineated: the formation and detachment of a larger gas bubble forming due to the nitrogen flow, the electrode completely enveloped by the bubble, and the region corresponding to the gas-vapor layer between emerging bubbles. An average gas-vapor layer thickness (d) of 0.244 ± 0.046 mm for the cold and hot plasma state was determined at 200 V and 0.07 L min$^{-1}$ $N_2$ flow.

**Plasma Volume:**

To determine the plasma volume, we utilized the statistical distribution of plasma emissions (**Supplementary Figure 15**) along the wire. We hypothesize that each emission corresponds to a discharge occurring between the electrode-gas/vapor interface and the gas/vapor-electrolyte interface. Statistically, the distance of these discharges should align with the average gas sheath length. Additionally, we assume that the number and intensity of plasma discharges on the cathode side facing the camera are equivalent to those on the side facing away from the camera. On average, we determined the plasma volume to be 0.014 ± 0.001 mm$^3$ in the cold state and 0.312 ± 0.005 mm$^3$ in the hot state (at 95% confidence interval). For the calculation of the reduced electric field and electron density, this value has to be divided by two, since the spectrometer only catches the emission form the side facing the optical fiber.

**Reduced Electric Field:**

To determine the reduced electric field, the geometry factor for the fiber needs to be determined.

$$g = \frac{\pi \cdot r^2}{4 \cdot \pi \cdot d^2} = 2.34 \cdot 10^{-5} \quad (3)$$

With the radius of the fiber r (r = 0.3 mm) and the distance d between the plasma electrode and the optical fiber (d = 31 mm). The intensity of the respective transition is calculated according to equation 4:

$$I_{state} = \frac{Photons}{V_{plasma} \cdot g} \quad (4)$$

With $V_{plasma}$ being the volume of the plasma seen by the fiber and *g* being the geometric factor from equation 3. The intensity for the $N_2$(C-B,v'=0, v''=0) transition at 200 V has been determined as $I_{CB} = 3.87 \cdot 10^{17}$ Photons cm$^{-3}$ s$^{-1}$ and for the $N_2^+$(B-X,v'=0, v''=0) as $I_{BX} = 1.95 \cdot 10^{17}$ Photons cm$^{-3}$ s$^{-1}$. From this, the collisional quenching factor can also be determined according to equations 5-6:

$$Q_{(CB)} = \frac{A_{N_2(CB)}}{A_{N_2(CB)} + [N_2] \cdot k_q^{N_2(CB)} + [H_2O]} = \frac{1}{5} = 0.2 \quad (5)$$

$$Q_{(BX)} = \frac{A_{N_2^+(BX)}}{A_{N_2^+(BX)} + [N_2] \cdot k_q^{N_2^+(BX)} + [H_2O]} = \frac{1}{128} = 7.81 \cdot 10^{-3} \quad (6)$$

Here, *A* are the Einstein coefficients with $A_{N2(CB)} = 2.38 \cdot 10^7$ and $A_{N2+(BX)} = 1.61 \cdot 10^7$. With these values, it is now possible to determine the reduced electric field according to equation 7:

$$\frac{\left(I_{N_2(C)}\right)^2 \cdot Q_{N_2^+(B)}}{I_{N_2^+(B)} \cdot Q_{N_2(C)}^2 \cdot [N_2]^2} = 8.11 \cdot 10^{-20} \quad (7)$$

For the calculation, the partial pressure of gaseous species in the gas sheath needs to be determined. From gas-chromatography measurements, we have determined, that the partial pressure of hydrogen is about 2.41 kPa and the peak of the $NO_x$ species is too low to distinguish from the background noise. We assume that these two species as well as the ammonia are negligible in comparison to the $N_2$ and $H_2O$ partial pressures. The vapor pressure was approximated by assuming steady state conditions for bulk electrolyte temperatures ranging from 35°C to 50°C (measured). The reactor is open, so the total pressure does not exceed 1 atm, meaning that:

$$P_{total} = P_{N_2} + P_{H_2O,vapor} = 101.32\ kPa \quad (8)$$

This method is only applicable, if the $N_2(C^3\Pi_u)$ and the $N_2^+(B^2\Sigma_u^+)$ states are sufficiently high in intensity. This was only possible for the 200 V Pt hot state measurements. The results are collectively shown in supplementary table 1:

*Supplementary Table 1:* Results for the reduced electric field and the electron density.

| **Temp.** | **E/N** | **$n_e$** |
|---|---|---|
| 35°C | 36.08 ± 0.24 Td | $(5.96 \pm 0.24)\cdot 10^{14}$ cm$^{-3}$ |
| 50°C | 37.62 ± 0.20 Td | $(4.66 \pm 0.11)\cdot 10^{14}$ cm$^{-3}$ |

Due to the lack of $N_2^+(B^2\Sigma_u^+)$ states in most of the other spectra, the electron density can be derived with a different method from the spectral line Stark broadening. To get the most accurate results, a system with (comparatively) less complexity like the $H_\alpha$ spectral line of the Balmer series can be analyzed. The Stark width can be estimated by fitting a (pseudo-) Voigt line profile onto the spectral line:

$$I = I_0 + A \cdot \left[ m_u \frac{2}{\pi} \frac{\Delta\lambda_L}{4(\lambda-\lambda_c)^2 + \Delta\lambda_L^2} + (1-m_u) \frac{\sqrt{4 \ln 2}}{\sqrt{\pi}\Delta\lambda_G} e^{-\frac{4 \ln 2}{\Delta\lambda_G^2}(\lambda-\lambda_c)^2} \right] \quad (9)$$

Here, $I_0$ is the intensity baseline correction, $A$ is the peak area, and $\Delta\lambda_G$ and $\Delta\lambda_L$ are the Gaussian and Lorentzian FWHM, respectively. Additionally, $\lambda_c$ is the center wavelength and the factor $m_u$ describes the fraction of Gaussian to Lorentzian peak shape. The fit results can then be used to estimate the electron density and the electron temperature according to Gigosos et al.[6] Here, it will be assumed that the Lorentzian width of the fit consists mostly of the Stark broadening ($\Delta\lambda_s$).

$$\Delta\lambda_L \approx \Delta\lambda_s \quad (10)$$

Based on the Lorentzian width of the Voigt fitted line profile, the electron density can be calculated according to Konjević et al. [7]

$$For\ H_a: n_e = 10^{17} \cdot \left(\frac{\Delta\lambda_s}{1.098}\right)^{1.47135} \tag{11}$$

$$For\ H_\beta: n_e = 10^{17} \cdot \left(\frac{\Delta\lambda_s}{0.94666}\right)^{1.49} \tag{12}$$

The results of the $H_\alpha$ and $H_\beta$ Voigt fit are shown in **Supplementary Figure 16-Supplementary Figure *19*** and the electron densities are summarized in **Supplementary Figure 20**.

Lastly, we analyzed the OH(A-X) transition by fitting the experimental spectrum with a Voigt line shape (80% Lorentzian) simulated spectrum containing rotational and vibrational transitions, that were generated using the LIFBASE software[3]. With this software, the temperature and broadening can be used as fit parameters. The resulting simulated spectra, the fitted OH(A-X) rotational temperatures, as well as the fit parameter are shown in

**Supplementary Figure *21*-Supplementary Figure 23** and Supplementary Table 2, respectively. The rotational temperature of the OH(A-X) transition is generally a less suitable measure of the neutral gas temperature compared with the $N_2$(C-B) transition as non-thermal distributions of the upper states can be produced during electron impact dissociation of water vapor. Bruggeman et al.[8] gives an overview of some of the relevant considerations for determining neutral gas temperatures from rotational spectra. The resulting neutral gas temperatures inferred from the rotational temperature of plasma-produced excited states i.e. the $N_2$(C-B) transition are summarized in Supplementary Table 3 However, the rotational and vibrational temperatures of the OH(A) state can still be useful for comparing the general plasma properties under variation of operating conditions. From the material dependent experiments (

**Supplementary *Figure 22*** and **Supplementary Figure 23 (a)**) it can be derived, that $T_{rot}$ of the OH(A-X) transition is stable across different materials at the same voltage, thus it can be used to normalize the spectra for the $N_2$(C-B) comparison. Opposed to that, for the voltage dependent experiments (

**Supplementary Figure *21*** and **Supplementary Figure 23 (b)**), a clear increase of the OH(A-X) rotational temperature with increasing voltage can be observed. The large discrepancies in peak correlation and increase in Chi-square for the fit results due to other transitions that are superimposed with the OH(A-X) signal, as for example atomic metal lines and $N_2$(C-B, $\Delta$v=-2).

*Supplementary Table 2:* LIFBASE fit parameter used for the OH(A-X) fit.

| Sample | Baseline corr. (%) | Wavelength shift (Å) | Temperature | | Peak Shape | Pressure (Peak-Broadening) | Peak correlation (%) | Chi-square | Comment |
|---|---|---|---|---|---|---|---|---|---|
| | | | $T_{vib}$ | $T_{rot}$ | | | | | |
| *180 V cold | 1.3 | -0.085 | 6400 | 4100 | L-G(80:20) | 6 | 96.5 | 232.0 | |
| *200 V cold | 1.3 | -0.088 | 11000 | 4900 | L-G(80:20) | 12 | 95.7 | 441.0 | |
| *200 V hot | 0.5 | -0.086 | 9900 | 4700 | L-G(80:20) | 12 | 93.4 | 676.5 | $N_2$(C-B) line |
| *230 V hot | 0.5 | -0.085 | 10000 | 6500 | L-G(80:20) | 10 | 60.4 | 6608.5 | Pt lines |
| +W cold | 0.37 | 0.028 | 5200 | 3900 | L-G(80:20) | 11 | 97.2 | 540.6 | |
| +W hot | 0.6 | 0.023 | 5900 | 4000 | L-G(80:20) | 12 | 96.3 | 742.9 | $N_2$(C-B) line |
| +Pt cold | 0.5 | 0.052 | 4900 | 3800 | L-G(80:20) | 12 | 97.1 | 558.7 | |
| +Pt hot | 0.5 | 0.037 | 4900 | 3600 | L-G(80:20) | 12 | 95.7 | 849.5 | Large BG |

* measured with the ESA 4000 (LLA Instruments, $\Delta\lambda_{instrument}$= 0.11 nm),

+measured with the Aryelle Butterfly (LTB, $\Delta\lambda_{instrument}$= 0.15 nm)

*Supplementary Table 3:* Neutral gas temperature inferred from the rotational temperature of plasma-produced excited states ($N_2$(C-B) transition).

| State | Temperature |
|---|---|
| Pt 180°C “cold” | 1000 K -1400 K |
| Pt 200°C “cold” | 1200 K -1400 K |
| Pt 200°C “hot” | 1400 K -1600 K |
| Pt 230°C “hot” | 3300 K |

Electrode Temperature:

The temperature of the cathode (active electrode) can be estimated by fitting the OES spectrum background continuum with the blackbody (Planck) radiation law (equation 13).

$$I(\lambda, T) = scale \cdot \frac{2hc^2}{\lambda^5} \frac{1}{exp\left(\frac{hc}{\lambda k_B T}\right) - 1} \quad (13)$$

Where h is Planck’s constant, c is the speed of light, $k_B$ is the Boltzmann constant and T is the absolute temperature in Kelvin. A scaling factor is introduced to accommodate for changing intensities when individual spectra with different acquisition parameters (or from the two different spectrometer) are compared. The results for Platinum at different potentials is shown in **Supplementary Figure 24.**

**Supplementary Note 3:**

Enthalpy of Formation at 25°C and 1 atm.

## Ammonia Formation (AF)

| | | |
|---|---|---|
| Nitrogen Reduction: | $N_2 + 6\ H^+ + 6\ e^- \rightarrow 2\ NH_3$ | Cathode |
| Hydrogen Oxidation: | $3\ H_2 \rightarrow 6\ H^+ + 6\ e^-$ | Anode |
| | $1.5\ H_{2(g)} + 0.5\ N_{2(g)} \rightarrow NH_{3(g)}$ | $\Delta_f H^0_{25°C}$= -45.9 kJ mol$^{-1}$ |

**Water electrolysis (WE)**

| | | |
|---|---|---|
| Hydrogen Evolution | $H_2 + 2\ OH^- \rightarrow 2\ H_2O + 2\ e^-$ | Cathode |
| Oxygen Evolution | $0.5\ O_2 + H_2O + 2\ e^- \rightarrow 2\ OH^-$ | Anode |
| | $H_{2(g)} + 0.5\ O_{2(g)} \rightarrow H_2O_{(l)}$ | $\Delta_f H^0_{25°C}$= -285.8 kJ mol$^{-1}$ |
| | $H_{2(g)} + 0.5\ O_{2(g)} \rightarrow H_2O_{(g)}$ | $\Delta_f H^0_{100°C}$=-242.6 kJ mol$^{-1}$ |
| **Combined (Total):** | $2\ N_2 + 12\ H_2O + 12\ e^- \rightarrow 4\ NH_3 + 12\ OH^-$ | Cathode |
| | $6\ H_2O + 12\ OH^- \rightarrow 3\ O_2 + 12\ H_2O + 12\ e^-$ | Anode |
| | $0.5\ N_{2(g)} + 1.5\ H_2O_{(l)} \rightarrow NH_{3(g)} + 0.75\ O_{2(g)}$ | $\Delta_f H^0_{25°C}$= 382.8 kJ mol$^{-1}$ |
| | $0.5\ N_{2(g)} + 1.5\ H_2O_{(v)} \rightarrow NH_{3(g)} + 0.75\ O_{2(g)}$ | $\Delta_f H^0_{100°C}$=316.3 kJ mol$^{-1}$ |

## Standard Entropy and Enthalpy:

*Supplementary Table 4:* Thermochemistry parameters at 25°C and 100°C (calculated according to NIST[9]).

| | $H_{2,(g)}$ | $O_{2,(g)}$ | $N_{2,(g)}$ | $NH_{3,(g)}$ | $H_2O_{,(g)}$ | $H_2O_{,(l)}$ |
|---|---|---|---|---|---|---|
| $H^0_{(25°C)}$ in kJ mol$^{-1}$ | 0 | 0 | 0 | -45.94 | -241.83 | -285.83 |
| $H^0_{(100°C)}$ in kJ mol$^{-1}$ | 2.15 | 2.19 | 2.16 | -43.22 | -239.33 | -288.33 |
| $S^0_{(25°C)}$ in JK$^{-1}$mol$^{-1}$ | 130.68 | 205.15 | 191.61 | 192.77 | 188.84 | 69.95 |
| $S^0_{(100°C)}$ in JK$^{-1}$mol$^{-1}$ | 137.11 | 211.71 | 198.07 | 200.90 | 196.33 | 196.33 |
| $C_{p,(25°C)}$ in J Kg$^{-1}$K$^{-1}$ | 28.84 | 29.38 | 29.12 | 35.65 | 33.59 | 75.37 |
| $C_{p,(100°C)}$ in J $Kg^{-1}K^{-1}$ | 29.134 | 29.87 | 29.20 | 37.85 | 34.04 | 34.03 |

With this, the change in entropy and enthalpy can be calculated according to:

$$\Delta S^0_{XY} = \sum n\, S^0_{XY} - \sum n\, S^0_{(X,Y)} \quad (14)$$

$$\Delta H^0_{XY} = \sum n\, H^0_{XY} - \sum n\, H^0_{(X,Y)} \quad (15)$$

$\Delta S^0_{AF} = \frac{1}{2}[\,2\,S^0_{NH_3} - (S^0_{N_2} + 3\,S^0_{H_2})] =$ -99.05 $JK^{-1}mol^{-1}$ (g)

$\Delta S^0_{WE} = \left(S^0_{H_2} + \frac{1}{2}S^0_{O_2}\right) - S^0_{H_2O(l)} =$ 163.3 $JK^{-1}mol^{-1}$ (l)

$\Delta S^0_{WE} = \left(S^0_{H_2} + \frac{1}{2}S^0_{O_2}\right) - S^0_{H_2O(g)} =$ 46.64 $JK^{-1}mol^{-1}$ (g)

$\Delta S^0_{combined} = \frac{1}{2}[(2\,S^0_{NH_3} + 1.5\,S^0_{O_2}) - (S^0_{N_2} + 3S^0_{H_2O(l)})] =$ 145.90 $JK^{-1}mol^{-1}$ (l)

$\Delta S^0_{combined} = \frac{1}{2}[(2\,S^0_{NH_3} + 1.5\,S^0_{O_2}) - (S^0_{N_2} + 3S^0_{H_2O(g)})] =$ -33.84 $JK^{-1}mol^{-1}$ (g)

**Gibbs Free Energy:**

$$\Delta G^0 = \Delta H^0 - T\Delta S^0 \quad (16)$$

Assuming 298.15 K = RT and for water vapor formation 373.15 K:

$\Delta G^0_{RT,AF}$ = -16.41 kJ $mol^{-1}$

$\Delta G^0_{RT,WE}$ = 237.14 kJ $mol^{-1}$ (l), $\Delta G^0_{100°C,WE}$ = 225.21 kJ $mol^{-1}$ (v)

$\Delta G^0_{RT,combined}$ = 339.31 kJ $mol^{-1}$ (l), $\Delta G^0_{100°C,combined}$ = 328.92 kJ $mol^{-1}$ (v)

In the beginning of the reaction, the change in standard Gibbs free energy is positive, thus it is an endergonic reaction ($\Delta G^0 > 0$). After sufficient energy is put into the system, the water transitions into water vapor, at which point the standard Gibbs free energy change is increasing. The entropy drops at the transition from liquid water to vapor. Due to the inhomogeneous interface and localized heating, we assume the presence of liquid water and vapor during the plasma. A schematic representation of this with a temperature range above the 1600 K we have observed in OES is shown in **Supplementary Figure 25-26.**

Equilibrium Constants:

*Thermodynamic equilibrium constants*

The respective thermodynamic equilibrium constants are calculated according to:

$$\Delta G_T^o = -RT\ ln(K_T^\theta) \quad (17)$$

$$K_T^\theta = e^{-\frac{\Delta G_T^o}{RT}} \quad (18)$$

$K_{298K}^\theta$ (AF) = 782.93

$K_{298K}^\theta$ (WE, liquid) = 2.85 x $10^{-42}$

$K_{373K}^\theta$ (WE, gas) = 2.46 x $10^{-32}$

$K_{298K}^\theta$ (Comb., liquid) = 3.60 x $10^{-60}$

$K_{373K}^\theta$ (Comb., gas) = 6.82 x $10^{-47}$

Equilibrium constants:
The equilibrium constant $K_c$ can be calculated via:

$$K_{c,T} = \frac{K_T^\theta}{\Gamma_T} \quad (19)$$

$$K_{c,298K}(AF) = \frac{K_{298K,AF}^\theta}{\Gamma_{298K,AF}} = 6.80 \text{ x } 10^5$$

$$K_{c,298K}(WE) = \frac{K_{298K,WE}^\theta}{\Gamma_{298K,WE}} = 1.01 \text{ x } 10^{-14}$$

$$K_{c,373K}(WE) = \frac{K_{373K,WE}^\theta}{\Gamma_{373K,WE}} = 1.01 \text{ x } 10^{-14}$$

$$K_{c,298K}(Comb.) = \frac{K_{298K,Comb.}^\theta}{\Gamma_{298K,Comb.}} = 1.76 \text{ x } 10^{-5}$$

$$K_{c,373K}(Comb.) = \frac{K_{373K,Comb.}^\theta}{\Gamma_{373K,Comb.}} = 1.76 \text{ x } 10^{-5}$$

**Supplementary Note 4:**

Electrochemical Literature Comparison:

*Supplementary Table 5:* List of Publications for figure 3b, adapted from [10]:

| Catalyst | Reaction | Partial current densities (mA $cm^{-2}_{geo}$) | Ammonia Production rate (µmol $h^{-1}$ $cm^{-2}$) | $N_2$ Pressure (bar) | Reference |
|---|---|---|---|---|---|
| Autoclave\|$LiNTf_2$ | $N_2/H_2O$ | 429.17 | 5337.69 | 15 | [11] |
| Autoclave\|$LiNTf_2$ | $N_2/H_2O$ | 29.73 | 369.721 | 20 | [11] |
| Autoclave\|$LiBF_4$ | $N_2/H_2O$ | 95 | 1181.55 | 20 | [12] |
| Autoclave\|$LiClO_4$ | $N_2/H_2O$ | 3.19 | 39.7 | 20 | [13] |
| Autoclave\|$LiNTf_2$ | $N_2/H_2O$ | 34.23 | 425.772 | 10 | [11] |
| Autoclave\|$LiBF_4$ | $N_2/H_2O$ | 50 | 621.867 | 10 | [14] |
| GDE cell\|$LiBF_4$ | $N_2/H_2O$ | 4.13 | 51.341 | 1 | [15] |
| Autoclave\|$LiNTf_2$ | $N_2/H_2O$ | 22.56 | 280.545 | 5 | [11] |
| Autoclave\|$LiBF_4$ | $N_2/H_2O$ | 60.75 | 755.569 | 5 | [14] |
| Autoclave\|$LiClO_4$ | $N_2/H_2O$ | 1.15 | 14.353 | 50 | [16] |
| Autoclave\|$LiClO_4$ | $N_2/H_2O$ | 2.19 | 27.263 | 10 | [13] |
| Autoclave\|$LiBF_4$ | $N_2/H_2O$ | 0.54 | 6.741 | 50 | [16] |
| GDE cell\|$LiBF_4$ | $N_2/H_2O$ | 10.04 | 124.871 | 2 | [17] |
| Autoclave\|$LiClO_4$ | $N_2/H_2O$ | 0.75 | 9.368 | 30 | [18] |
| Autoclave\|$LiBF_4$ | $N_2/H_2O$ | 23.92 | 297.564 | 2 | [14] |
| Autoclave\|$LiClO_4$ | $N_2/H_2O$ | 0.56 | 7.012 | 5 | [18] |
| Glass cell\|$LiClO_4$ | $N_2/H_2O$ | 0.56 | 6.94 | 1 | [19] |
| Autoclave\|$LiBF_4$ | $N_2/H_2O$ | 25.43 | 316.323 | 15 | [11] |
| Compartment cell\|$LiBF_4$ | $N_2/H_2O$ | 1.48 | 18.407 | 1 | [20] |
| Autoclave\|$LiNTf_2$ | $N_2/H_2O$ | 3.15 | 39.136 | 1 | [11] |
| Autoclave\|$LiClO_4$ | $N_2/H_2O$ | 0.28 | 3.458 | 1 | [21] |
| Autoclave\|$LiClO_4$ | $N_2/H_2O$ | 20.53 | 255.38 | 15 | [11] |
| Compartment cell\|$LiNTf_2$ | $N_2/H_2O$ | 0.4 | 4.933 | 1 | [22] |
| **Plasmareactor\| Pt\|Plasma** | **$N_2/H_2O$** | **192.57** | **2870.02** | **1** | **This Work** |
| Compartment cell\|$LiBF_4$ | $N_2/H_2O$ | 0.09 | 1.131 | 0.75 | [20] |
| Compartment cell\|$LiBF_4$ | $N_2/H_2O$ | 0.09 | 1.075 | 0.5 | [20] |
| Compartment cell\|$LiBF_4$ | $N_2/H_2O$ | 0.04 | 0.451 | 0.25 | [20] |
| Compartment cell\|$LiBF_4$ | $N_2/H_2O$ | 0.01 | 0.142 | 0.125 | [20] |
| Compartment cell\|$LiBF_4$ | $N_2/H_2O$ | 0.01 | 0.067 | 0.05 | [20] |
| Compartment cell\|$LiBF_4$ | $N_2/H_2O$ | 0 | 0 | 0 | [20] |

**Supplementary Note 5:**

**Classic NRR mechanism.**

The catalytic process exhibits both associative and dissociative mechanisms, as illustrated in Table 6. The associative mechanism includes distal and alternating pathways, akin to enzymatic mechanisms. Within the distal pathway, hydrogenation primarily occurs on the N atom farthest from the surface, resulting in the release of the initial ammonia molecule. The remaining N persists as M≡N, and further hydrogenation yields the second ammonia. In the alternating pathway, hydrogenation alternates between two N atoms, leading to the sequential release of $NH_3$ molecules.[23, 24]

*Supplementary Table 6*: Reaction mechanism steps of NRR

| | | |
|---|---|---|
| **Associative** | $* + N_2 \rightarrow *N_2$ | 1 |
| | $*N_2 + 6\ (H^+ + e^-) \rightarrow *N_2H + 5\ (H^+ + e^-)$ | 2 |
| | $*N_2H + 5\ (H^+ + e^-) \rightarrow *NHNH + 4\ (H^+ + e^-)$ | $3^a$ |
| | $*N_2H + 5\ (H^+ + e^-) \rightarrow *NNH_2 + 4\ (H^+ + e^-)$ | $3^b$ |
| | $*NHNH + 4\ (H^+ + e^-) \rightarrow *NHNH_2 + 3\ (H^+ + e^-)$ | $4^a$ |
| | $*NNH_2 + 4\ (H^+ + e^-) \rightarrow *N + NH_3 + 3\ (H^+ + e^-)$ | $4^b$ |
| | $*N + 3\ (H^+ + e^-) \rightarrow *NH + 2\ (H^+ + e^-)$ | 5 |
| | $*NH + 2\ (H^+ + e^-) \rightarrow *NH_2 + (H^+ + e^-)$ | 6 |
| | $*NH_2 + (H^+ + e^-) \rightarrow *NH_3$ | 7 |
| | $*NH_3 \rightarrow NH_3 + *$ | 8 |
| **Dissociative** | $2* + N_2 \rightarrow 2*N$ | 1 |
| | $2*N + 6\ (H^+ + e^-) \rightarrow *N + *NH + 5\ (H^+ + e^-)$ | 2 |

$$*N + *NH + 5 (H^+ + e^-) \rightarrow 2*NH + 4 (H^+ + e^-) \quad 3$$

$$2*NH + 4 (H^+ + e^-) \rightarrow *NH + *NH_2 + 3 (H^+ + e^-) \quad 4$$

$$*NH + *NH_2 + 3 (H^+ + e^-) \rightarrow 2*NH_2 + 2 (H^+ + e^-) \quad 5$$

$$2*NH_2 +2 (H^+ + e^-) \rightarrow *NH_2 + *NH_3 + (H^+ + e^-) \quad 6$$

$$*NH_2 + *NH_3 + H^+ + e^- \rightarrow 2*NH_3 \quad 7$$

$$2*NH_3 \rightarrow *NH_3 + NH_3 + * \quad 8$$

$$*NH_3 + NH_3 + * \rightarrow 2NH_3 + 2* \quad 9$$

---

(* equals surface site)

In- Liquid Plasma NRR:

Inside a plasma, reactions of excited gas molecules as well as on reactor walls and catalysts can occur. In contrast to pure thermal catalysis, in plasma "non-thermal" Eley-Rideal (ER) pathways are possible in addition to the Langmuir-Hinshelwood (LH) mechanism. For this to be possible in NRR, the nitrogen molecules need to be activated by the plasma. A summary of the activation processes, the required minimum energy are summarized below. # marks excitation methods observed by OES in this work. It should be noted that dissociation and electronic excitation require less energy than ionization and are thus also likely to occur, but not detectable by OES. The lifetimes of the excited states depend strongly on the quenching (pressure) and is estimated to be around a few ns to µs (in high vacuum up to seconds) at most. For dissociation and Ionization, a lifetime is not applicable, the electron is continuum-bound. The resulting states ($N_2^+$ and N) have their own respective lifetimes.

<u>Nitrogen Activation:</u>

1. Vibrational Excitation#:
   $N_2(B, v=0) \rightarrow N_2(B, v=1)$ Req. energy ($\omega_e$): 0.22 eV [25]

2. Electronic Excitation#:
   $N_2^+(X) \rightarrow N_2^+(B)$ Req. energy: 3.17 eV [25]
   (found at 391.4 nm)
3. Dissociation:
   $N_2 + e^- \rightarrow 2 N + e^-$ Req. energy: 9.79 eV [1]
   (Bond dissociation energy $D_0$ for forming two ground-state N-atoms)
4. Ionization#:

$N_2 + e^- \rightarrow N_2^+ + 2\ e^-$ Req. energy: 15.6 eV [1]

Similar to thermal catalysis, in-liquid plasma catalysis formation of ammonia correlates with the presence of NH radicals. NH is a crucial intermediate, however the following pathways heavily depend on the plasma parameters, the catalyst and the pressure. In the following, the most likely mechanistic steps are summarized.[26-30] If the electrode temperature exceeds 1000 K as e.g. the Pt "hot" state at 200 V, thermolysis of water into hydrogen radicals becomes possible as well, adding further to the formation of $NH_3$.

<u>Mechanism:</u>

1. Water decomposition:

$$H_2O + e^- \rightarrow OH + H + e^-$$
$$H_2O + UV \rightarrow OH + H$$
$$H_2O + Heat \rightarrow OH + H$$

2. NH radical formation:

$$2\ H + 2\ OH + {}^*N_2 \rightarrow 2\ OH + 2\ NH$$
$$H_2O + 0.5\ N_2 \rightarrow OH + NH$$

3. $NH_2$ radical formation:

$$NH + {}^*H \rightarrow {}^*NH_2$$
$$H + {}^*NH \rightarrow {}^*NH_2$$
$${}^*H + {}^*NH \rightarrow {}^*NH_2$$

4. $NH_3$ formation:

$${}^*NH_2 + {}^*H \rightarrow {}^*NH_3 \quad \text{(E-R at high pressure)}$$
$${}^*NH_2 + H \rightarrow {}^*NH_3 \quad \text{(L-H at low pressure)}$$

Direct pathway: $^*NH + H_2 \rightarrow NH_3$

**Supplementary Note 6:**

Energy Efficiency:

$$EE_{NH_3} = \frac{E_{out}}{E_{in}} = \frac{\Delta G_R^0 \cdot m_{NH_3}}{\int V_{cell}(t) \cdot I(t)dt} \quad (20)$$

Calculation of ammonia concentration from the calibration curve:

$$c(NH_3) = 69.05508 * A_{Peak} - 1.91586 \quad (21)$$

Calculation of synthesis rate:

Production rate: $$r_S = \frac{n(NH_3)}{t * m_{catalyst}} \quad (22)$$

Calculation of heating power & efficiency:

In order to assess the impact of joule heating and temperature increases resulting from the reaction, measurements of the electrolyte temperature were conducted. An infrared (IR) thermometer was employed to gauge the temperature at the apex of the reactor, providing a representative measurement of the bulk electrolyte temperature. Additionally, a thermocouple was strategically positioned near the plasma source at the bottom of the reactor to capture localized temperature variations.

The temperature difference between these two measurements was minimal, ranging only between 1-2°C. This slight variance was attributed to effective convection induced by the emergence of bubbles within the electrolyte. The known mass of the electrolyte served as the basis for calculating the heating efficiency. This involved an initial determination of the power required for achieving the observed temperature increase, followed by the division of this calculated power by the total power input into the system. This approach ensured a comprehensive evaluation of the efficiency of joule heating and temperature rise within the experimental setup. The equations for calculating the joule heating power and the respective heating efficiency are shown below.

Joule heating power: $$P_{JH} = \frac{g_{H2O} \cdot \Delta T \cdot c_{H2O}}{t} \quad (23)$$

Heating efficiency: $$\eta_H = [\%] = X\% \frac{P_{JH}}{P_{Total}} \quad (24)$$

**Supplementary Figure 27** shows the increase over time averaged over five measurements for a Pt cathode at 200 V with a $N_2$ flow of 0.07 L min$^{-1}$. The rise in temperature was measured simultaneous via a thermocouple and a laser thermometer.

## Supplementary Note 7:

*Supplementary Table 7:* XPS Fits

| Element | Before | | | After | | |
|---|---|---|---|---|---|---|
| | Peak (eV) | Shape | Composition | Peak | Shape | Composition |
| Cu 2p | 931.92 (BE) | GL(80) | 70% $Cu^{0/+}$ | 932.02 (BE) | GL(80) | 68% $Cu^{0/+}$ |
| | 933.92 (BE) | GL(80) | 30% $Cu^{2+}$ | 934.13 (BE) | GL(80) | 32% $Cu^{2+}$ |
| Cu LMM Auger | 568-570 (BE) | Reference spectrum | 0% $Cu^{0}$<br>56% $Cu^{+}$<br>44% $Cu^{2+}$ | 568-570 (BE) | Reference spectrum | 27% $Cu^{0}$<br>40% $Cu^{+}$<br>33% $Cu^{2+}$ |
| Pt 4f | 71.03 (BE) | LA(1.2,85,70) | 100% $Pt^{0}$ | 71.02 (BE) | LA(1.2,85,70) | 100% $Pt^{0}$ |
| W 4f | 36.17 (BE) | GL(80) | 100% $W^{6+}$ | 36.10 (BE) | GL(80) | 100% $W^{6+}$ |

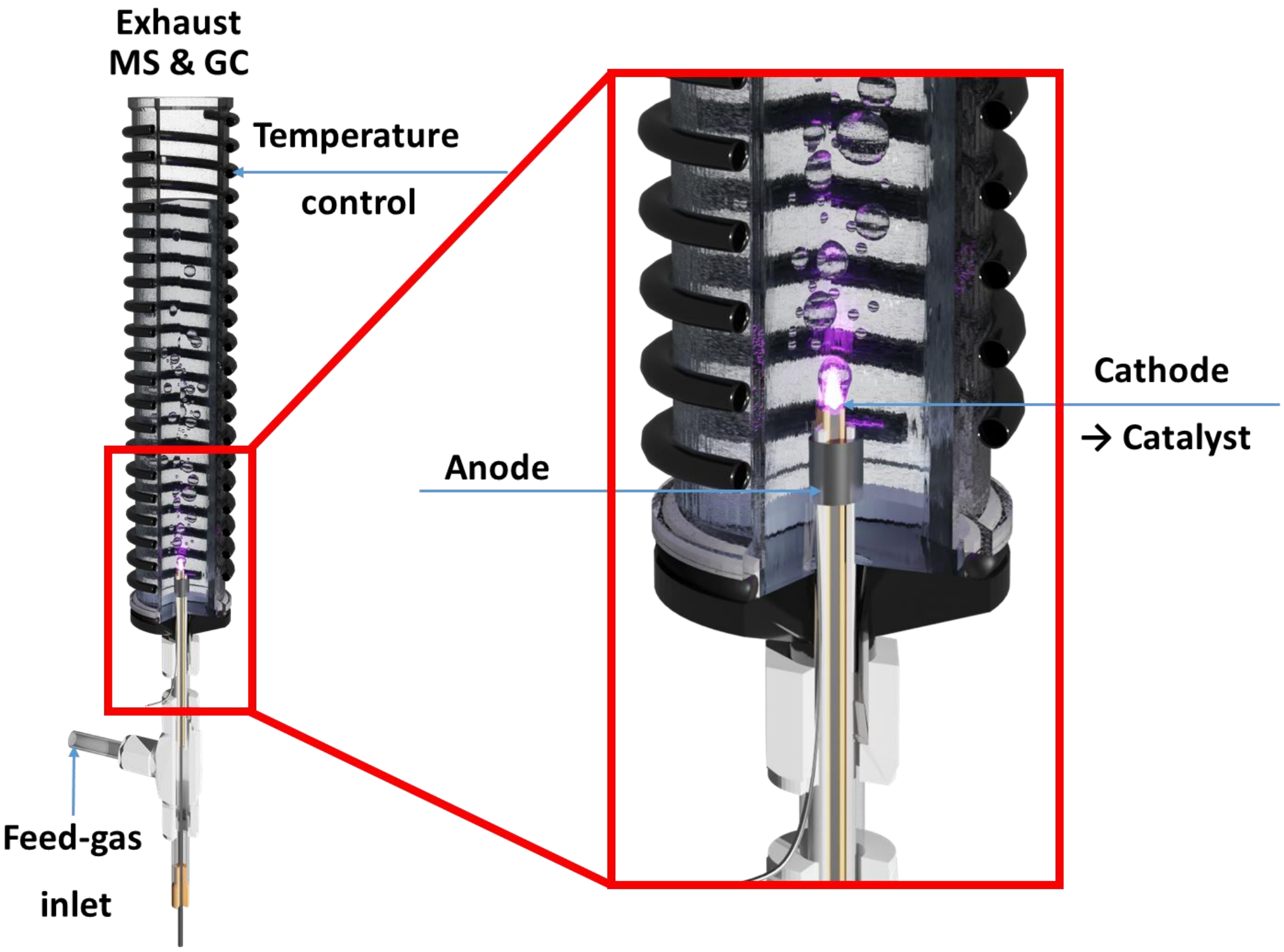

**Supplementary Figure 1:** Experimental setup of the plasma reactor with magnified cathode (Pt-wire) and anode (Pt-foil). The 0.5-4 mm long cathode is encased in a ceramic tube with a 0.5 mm larger inner diameter, so that the $N_2$ gas can pass through it as well. Only the tip of the cathode wire (0.5-4.0 mm) is exposed inside of the electrolyte. Outside of the ceramic tube, a Pt foil is wrapped around the outside of the ceramic and acts as the anode. The shortest distance between the cathode and anode is 5 mm, measured from the upper rim of the Pt-foil and the lowest part of the Pt-wire that sticks out from the ceramic. A shorter distance results in arc formation.

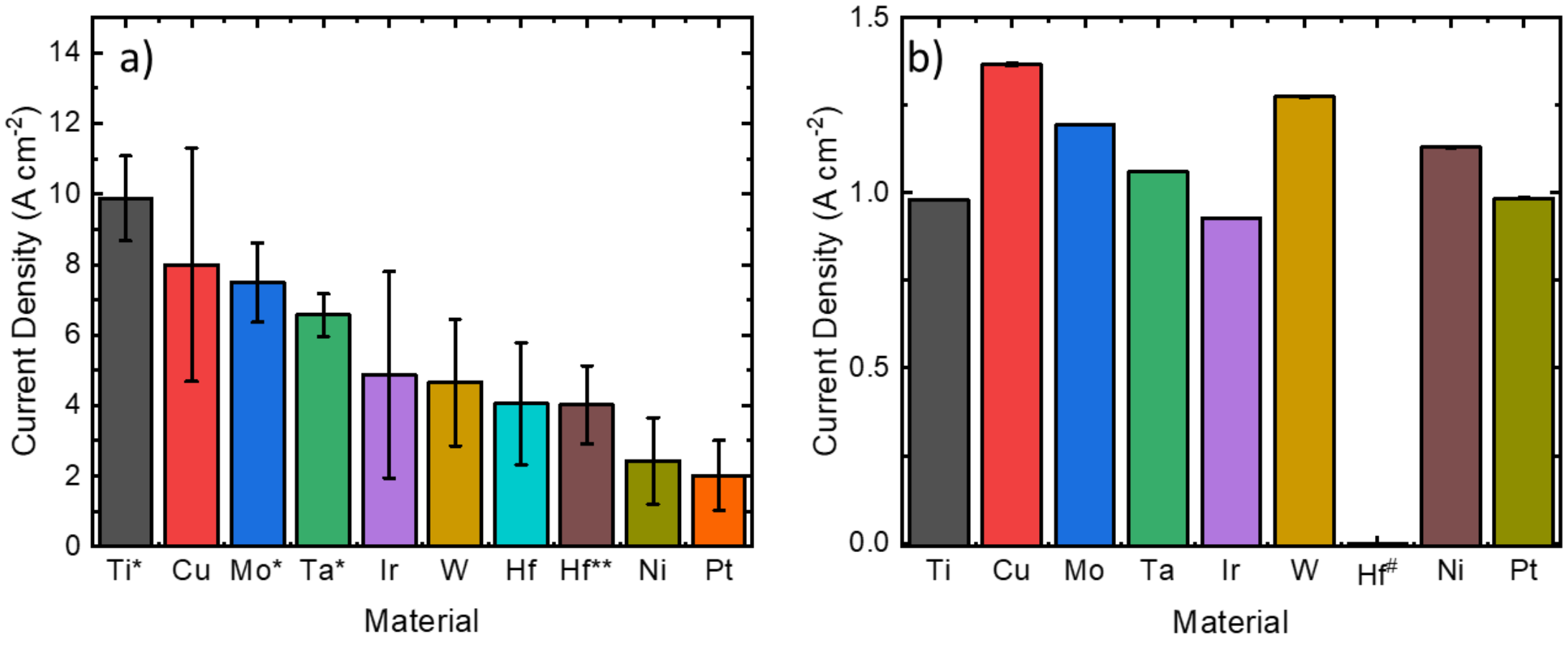

**Supplementary Figure 2:** Current densities at 200 V **(a)** and 10 V **(b)**, 0.07 L min$^{-1}$ of $N_2$ Flow, averaged over three consecutive measurements. In a), * Indicates unstable materials, **indicates materials, that only become unstable after several minutes (15 min+) at 200 V. In b), $^{\#}$ indicates that at 10 V, the natural passivation layer of Hf prevents current from flowing.

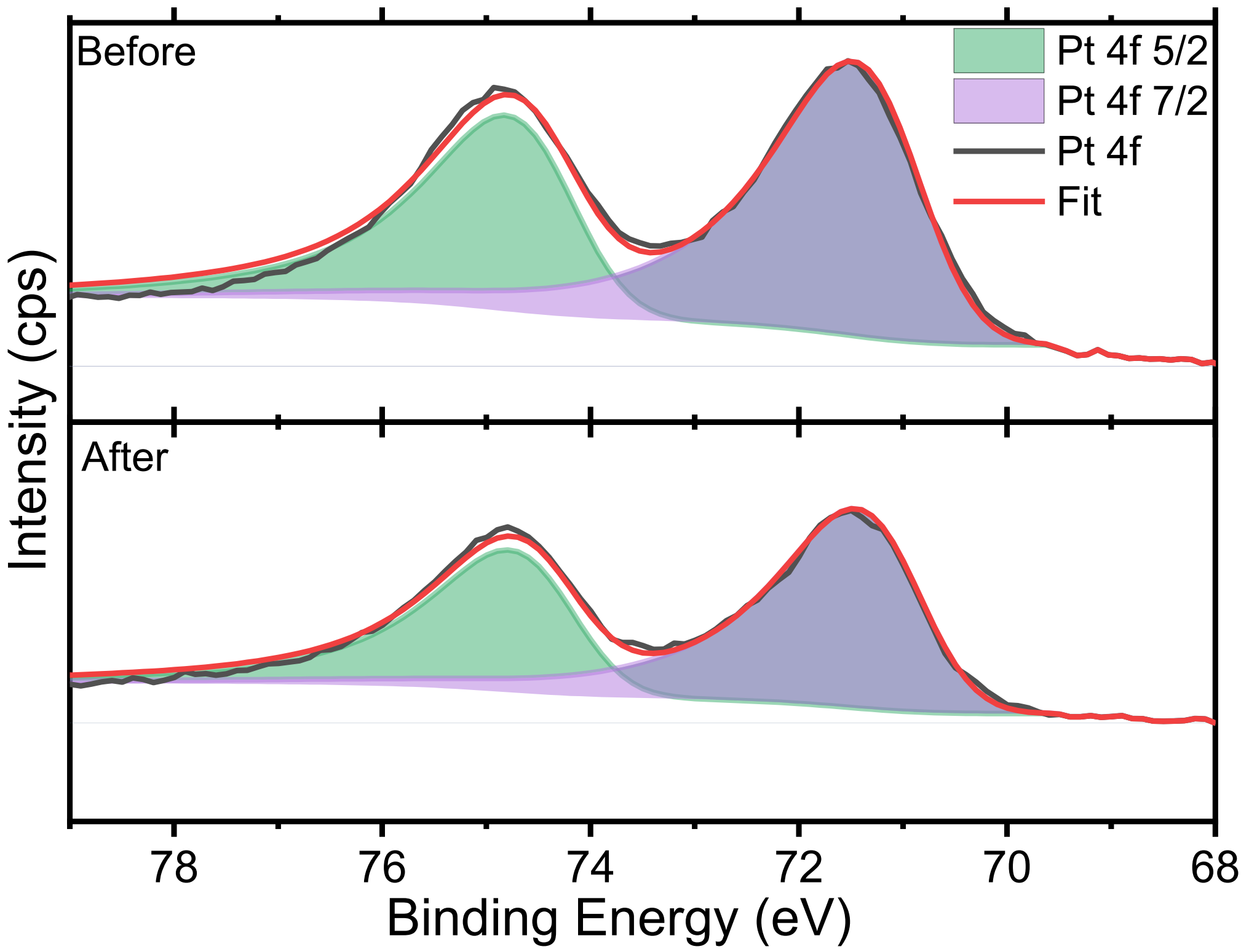

**Supplementary Figure 3:** XPS data from the Pt 4f region measured ex situ on the Pt cathode before and after 5 min of plasma. No platinum oxide species could be reasonably fitted. The metallic species were fitted with an asymmetric line shape in CasaXPS. For metallic Pt, the position of the Pt 4f 5/2 peak is at 74.33 eV binding energy and the 7/2 peak at 69.50 eV binding energy. The ratio of the area is 42:58.

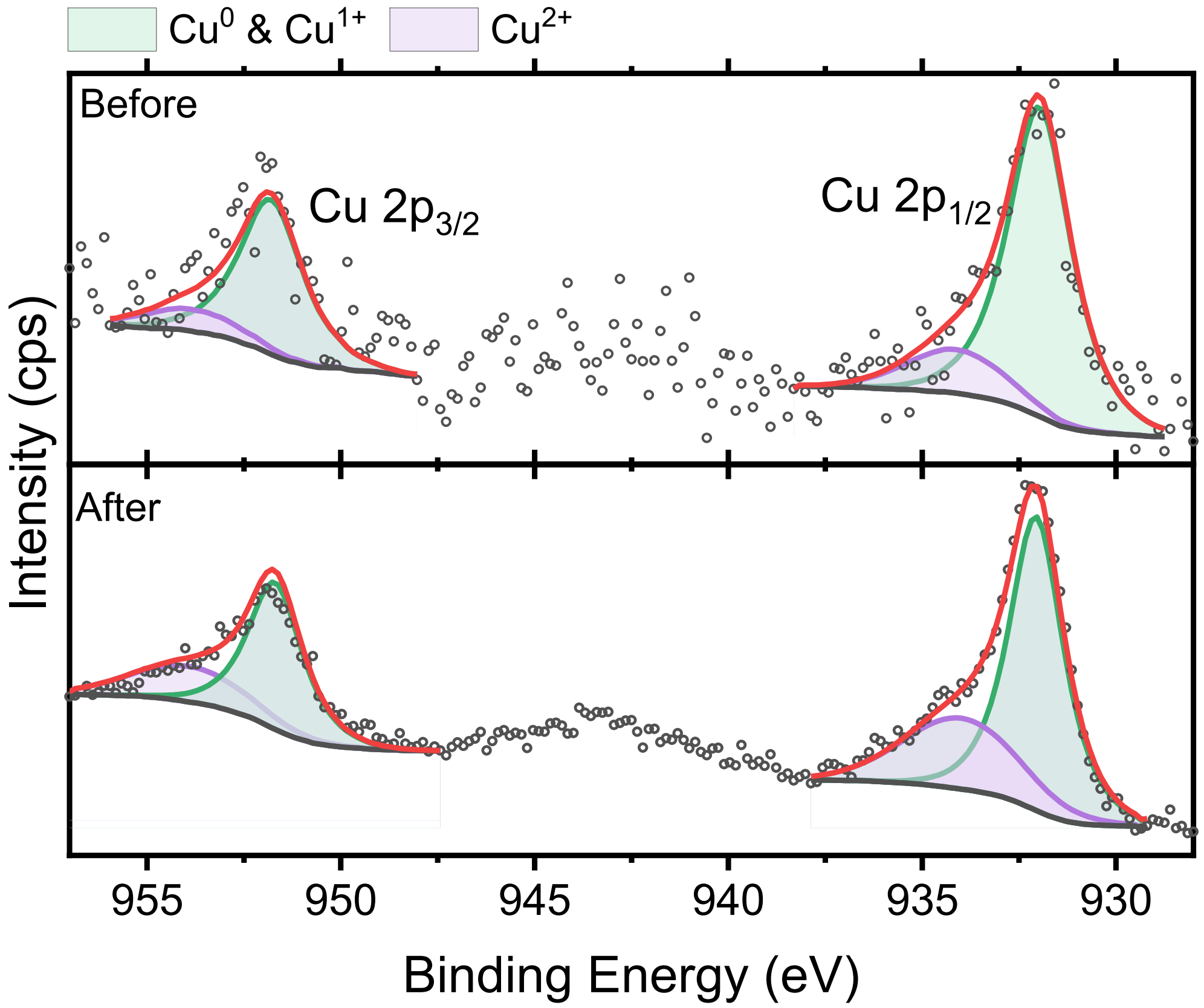

**Supplementary Figure 4:** Ex situ acquired XPS data of the Cu 2p region measured on a Cu-wire that was used as the cathode at 200 V plasma voltage and 0.07 L min$^{-1}$ of $N_2$ flow for 5 min. The peaks were fitted in CasaXPS with a Gaussian-Lorentzian-peak shape.

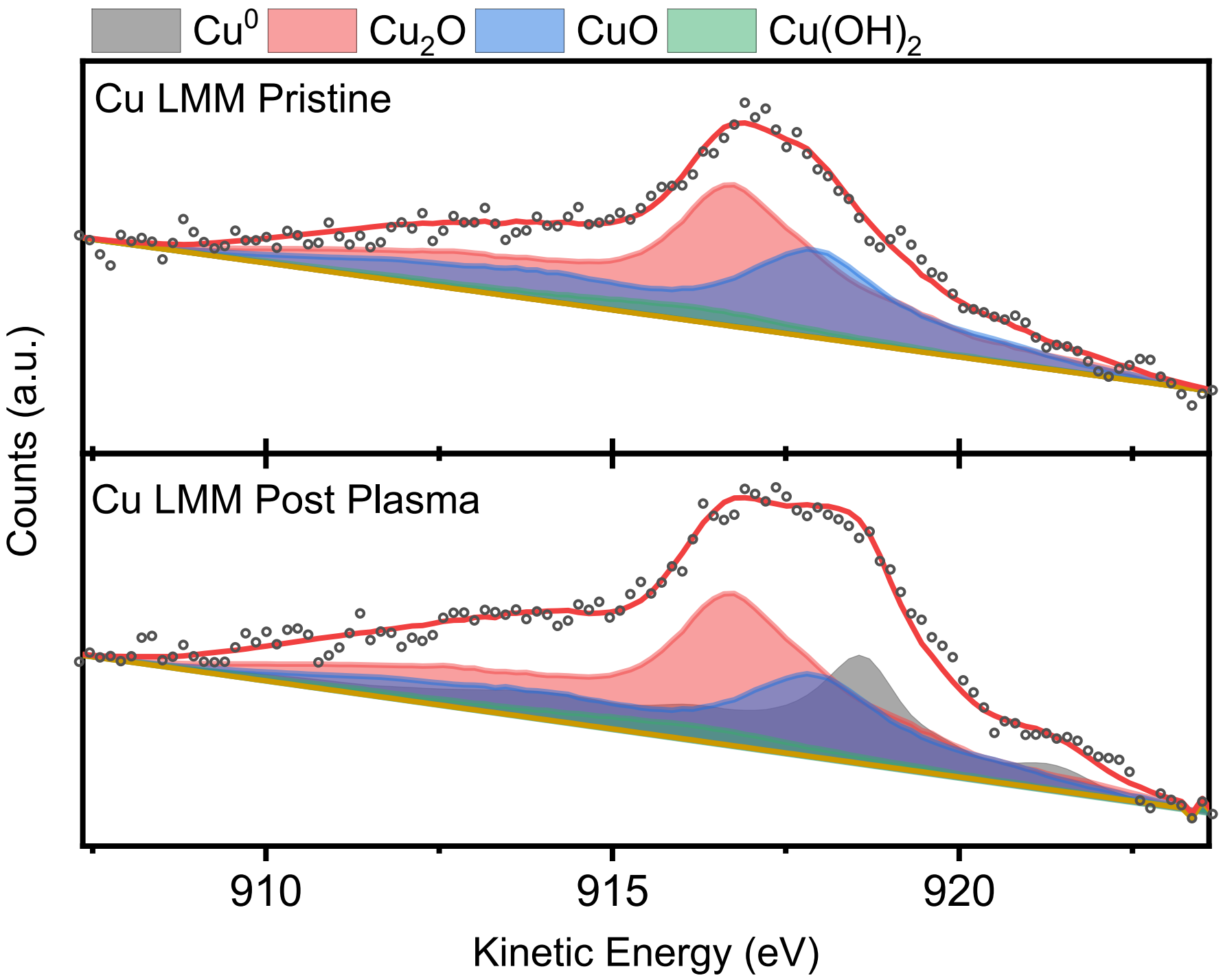

**Supplementary Figure 5:** Ex situ acquired Cu LMM Auger peak with linear deconvolution of $Cu^0$, $Cu_2O$, $Cu(OH)_2$, and CuO reference spectra. These spectra were collected from the same samples as the Cu 2p from Supplementary Figure 4 to distinguish the $Cu^0$ and $Cu^{(I)}$ ($Cu_2O$) contributions.

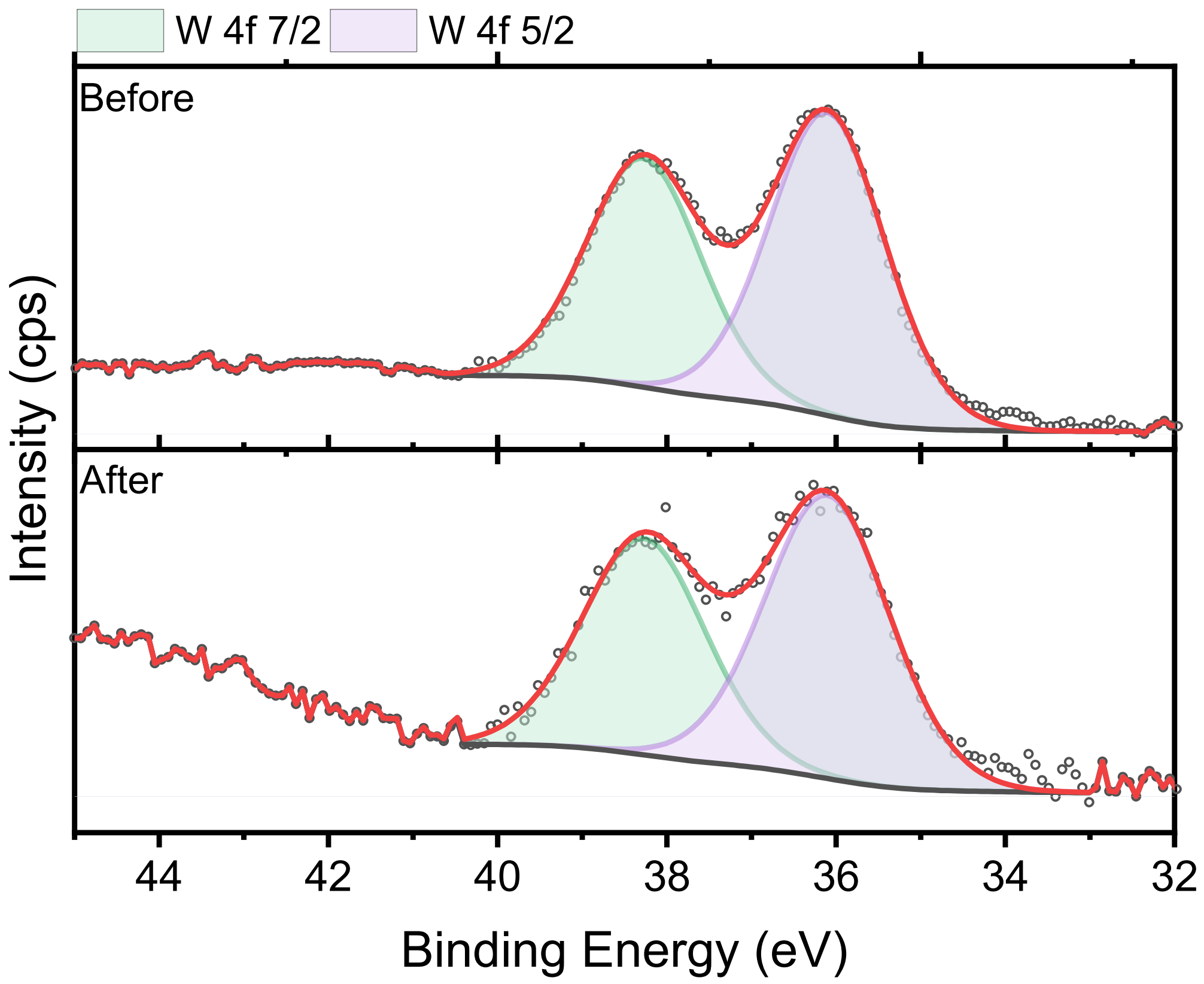

**Supplementary Figure 6**: Ex situ acquired XPS data of the W 4f region measured on a W-wire that was used as the cathode at 200 V plasma voltage and 0.07 L min$^{-1}$ of $N_2$ flow for 5 min. Only $W^{6+}$ ($WO_3$) was identified with no metallic species observed. The peaks were fitted in CasaXPS with a Gaussian-Lorentzian-peak shape.

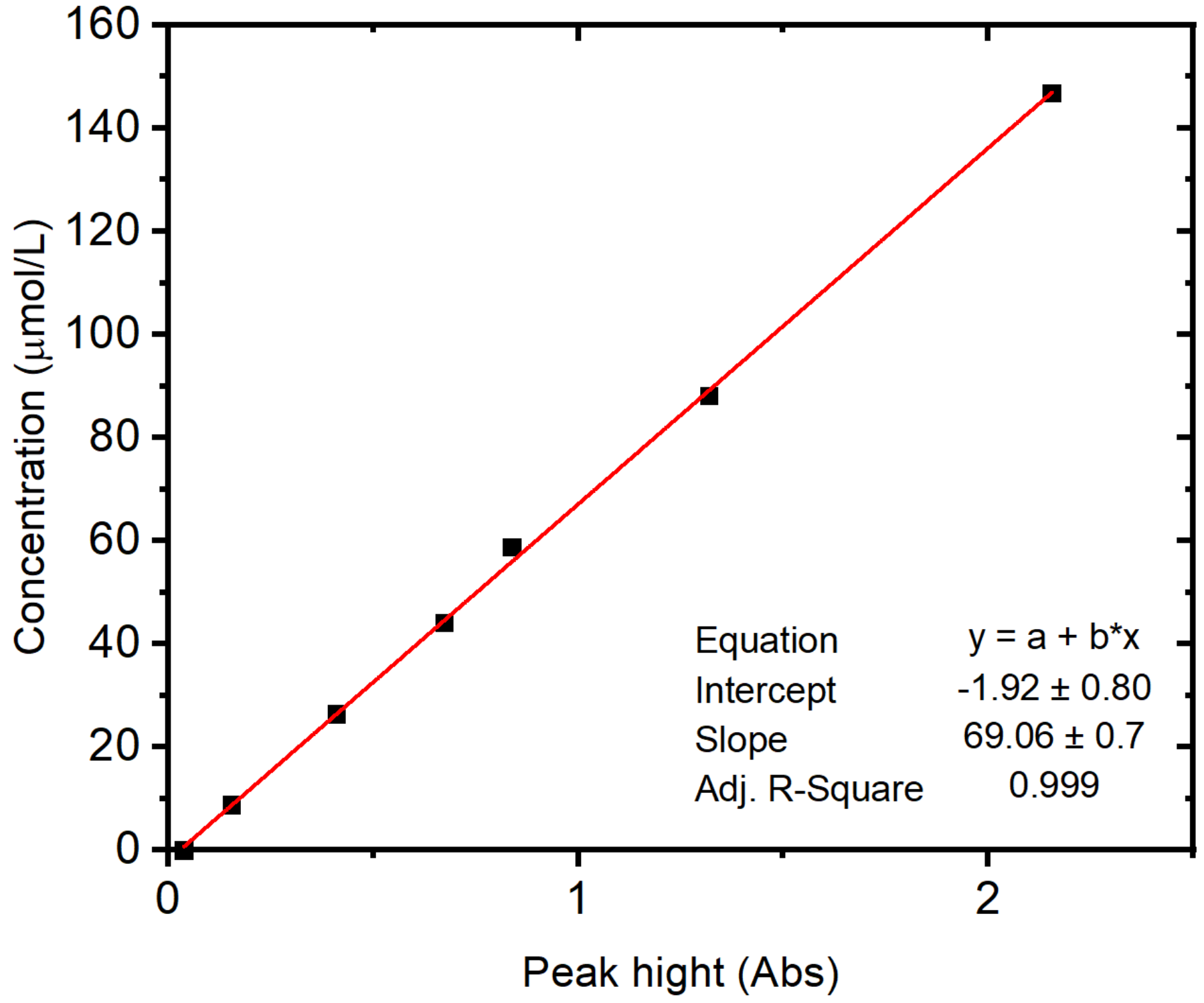

**Supplementary Figure 7:** Calibration curve for the detection of ammonia via a known $NH_4Cl$ standard solution (0, 10, 25, 40, 60, 90 and 150 μmol/L) with UV-Vis spectroscopy.

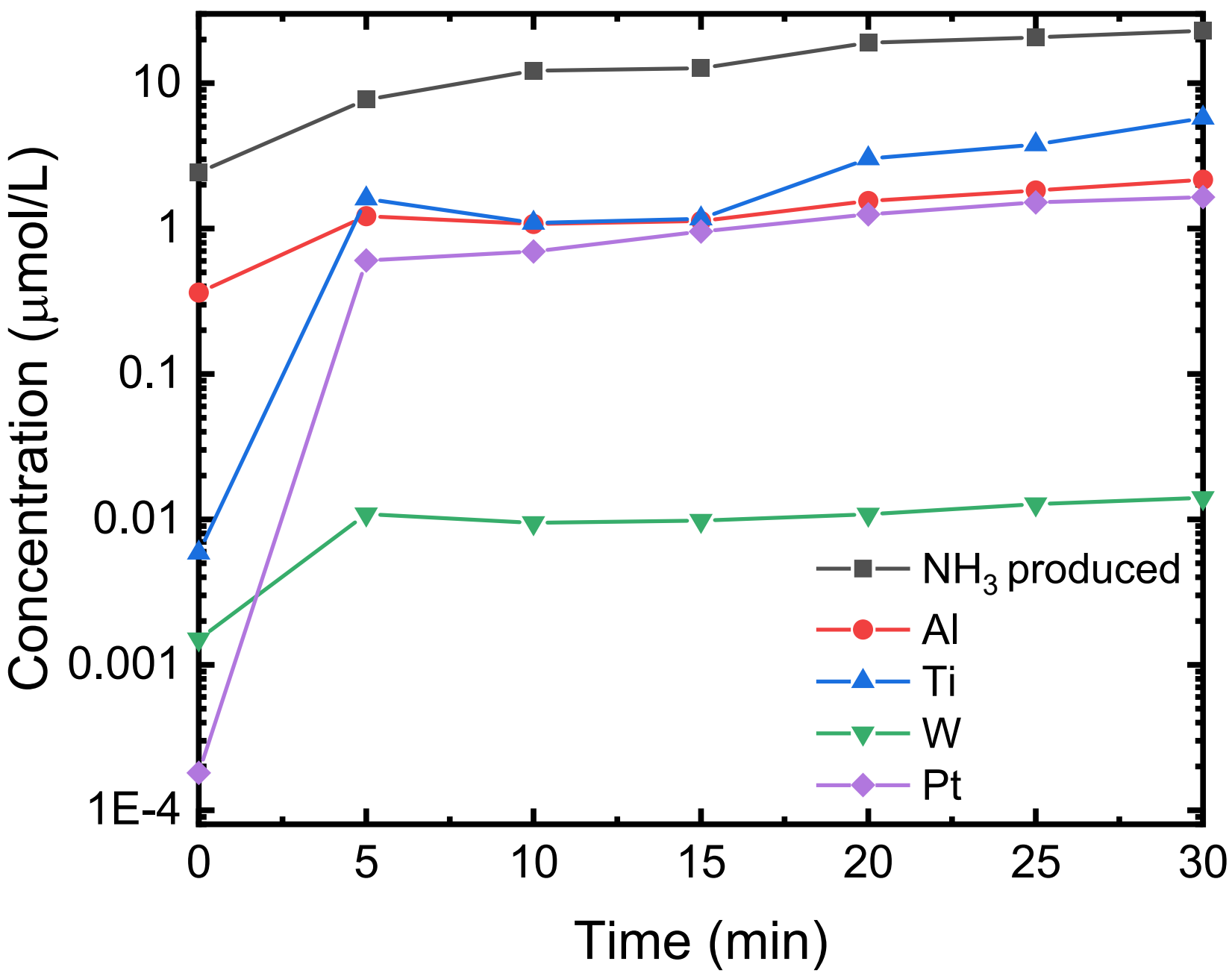

**Supplementary Figure 8.** Results of ICP-MS metal concentrations of Al, Ti, W, and Pt plotted together with the concentration increase of produced ammonia from the reaction in the solution. Each timestep consists of 5 min of plasma on, followed by taking a sample and the subsequent continuation of the plasma for the next timestep. These electrolyte samples were taken from the same experiment on a Pt electrode at 200 V as the one presented in **Supplementray Figure 9**. The elements found here are coming from the an experiement with a Pt electrode. Al, Ti, and W are possible contaminents from different parts of the reactor that could be in contact with the electrolyte (e.g. the ceramic).

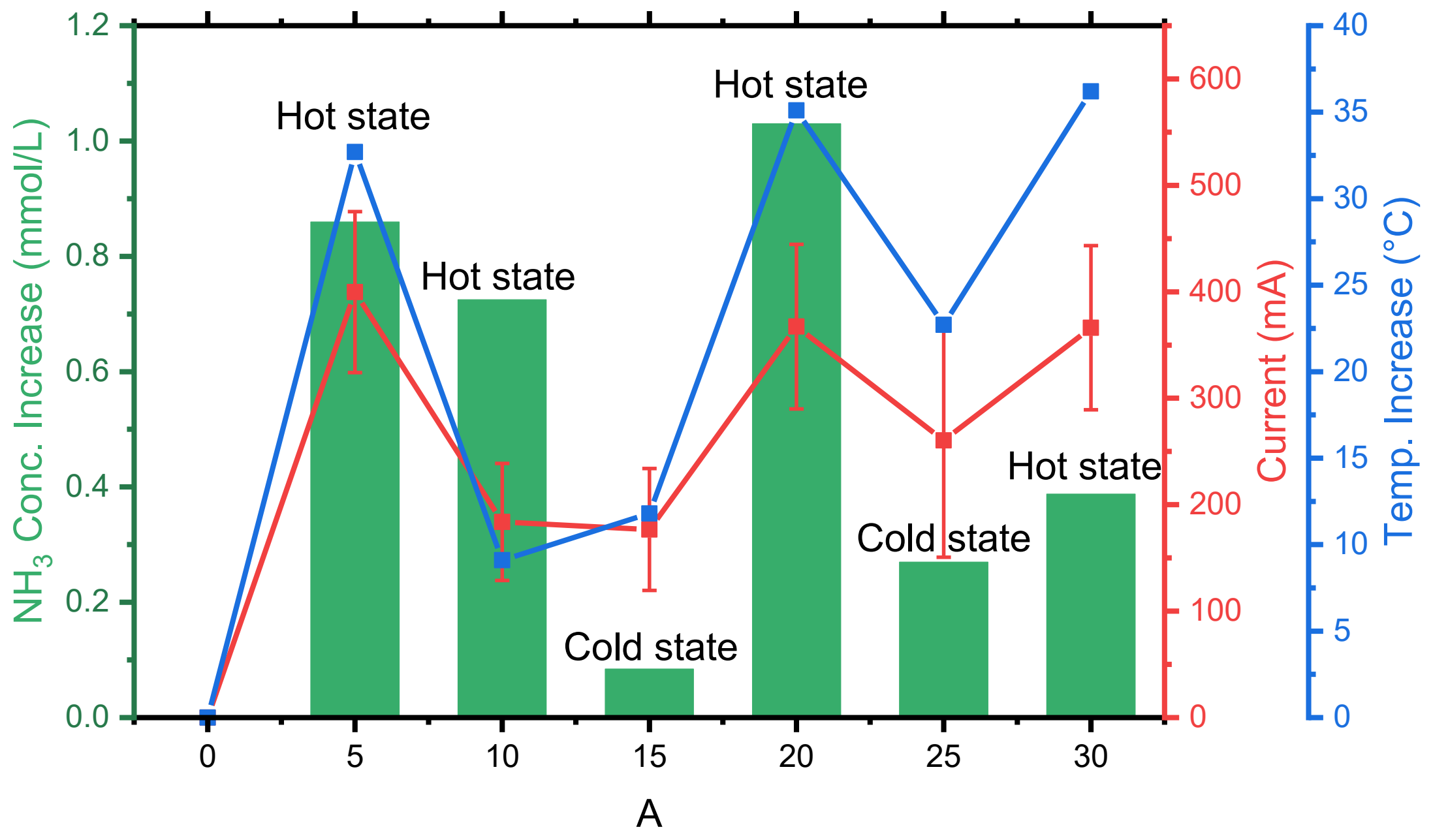

**Supplementary Figure 9.** Time dependent measurement of plasma at 200 V on a 2.0 mm Pt cathode and 0.07 L $min^{-1}$ of nitrogen flow. After every 5 min of plasma on section, the reactor was cooled back down to room temperature. The resulting ammonia concentration in the electrolyte is plotted with the respective increase of the temperature and the current. At 10 and 30 min, a transition from the cold to the hot state was observed after ~2 and 4 min of plasma, respectively, and therefore, these have to be considered mixed states.

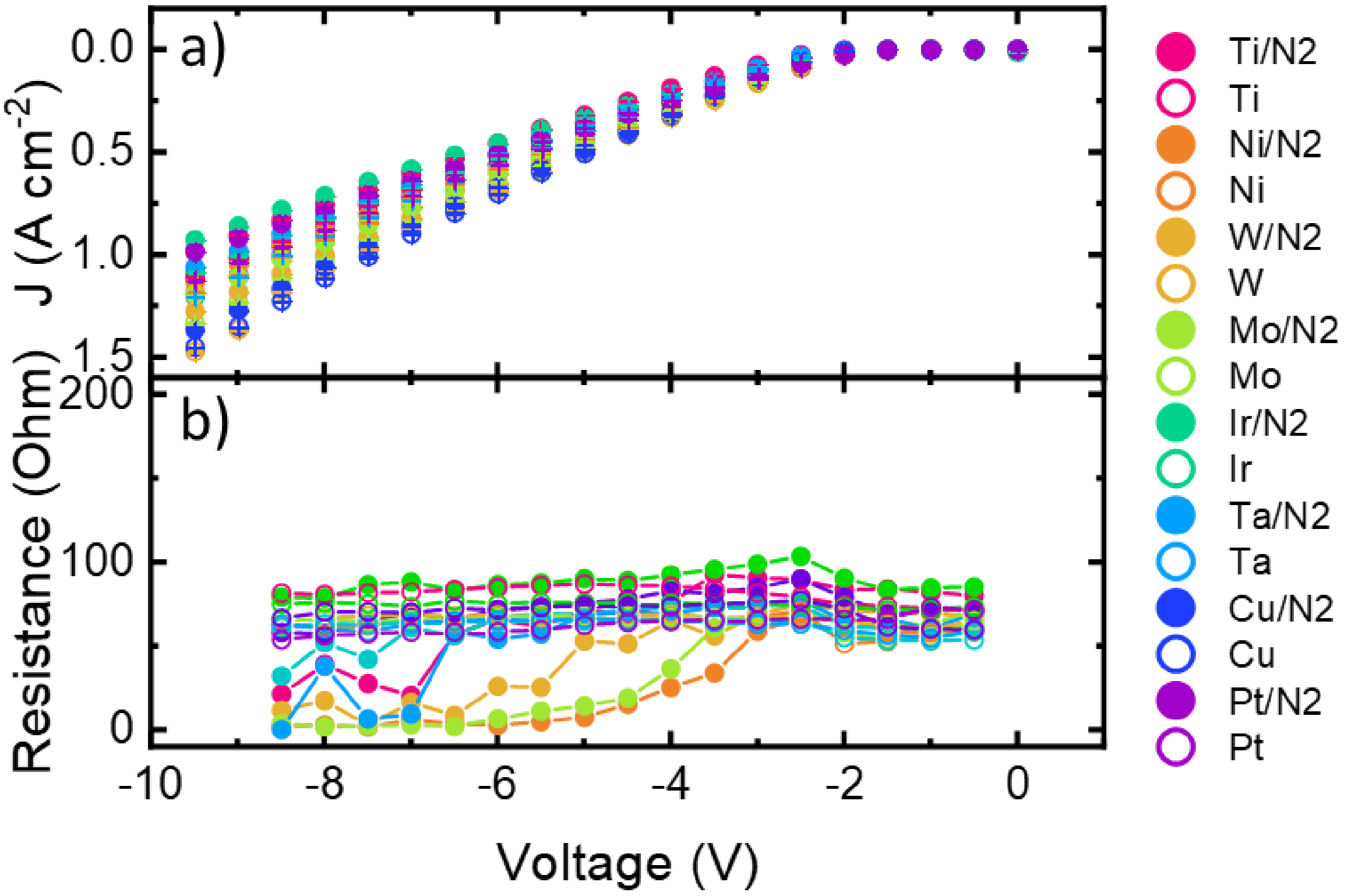

**Supplementary Figure 10:** SPEIS measurements of all tested metal cathode materials with and without 0.07 L min$^{-1}$ of $N_2$ flow. In **(a)** the polarization curves are shown. Hf was omitted because it is not conductive at these potentials due to its natural oxide passivation layer. The respective solution resistances are shown in **(b)**. The bottom figure shows the respective electrolyte resistance, calculated from the fitted Nyquist plot.

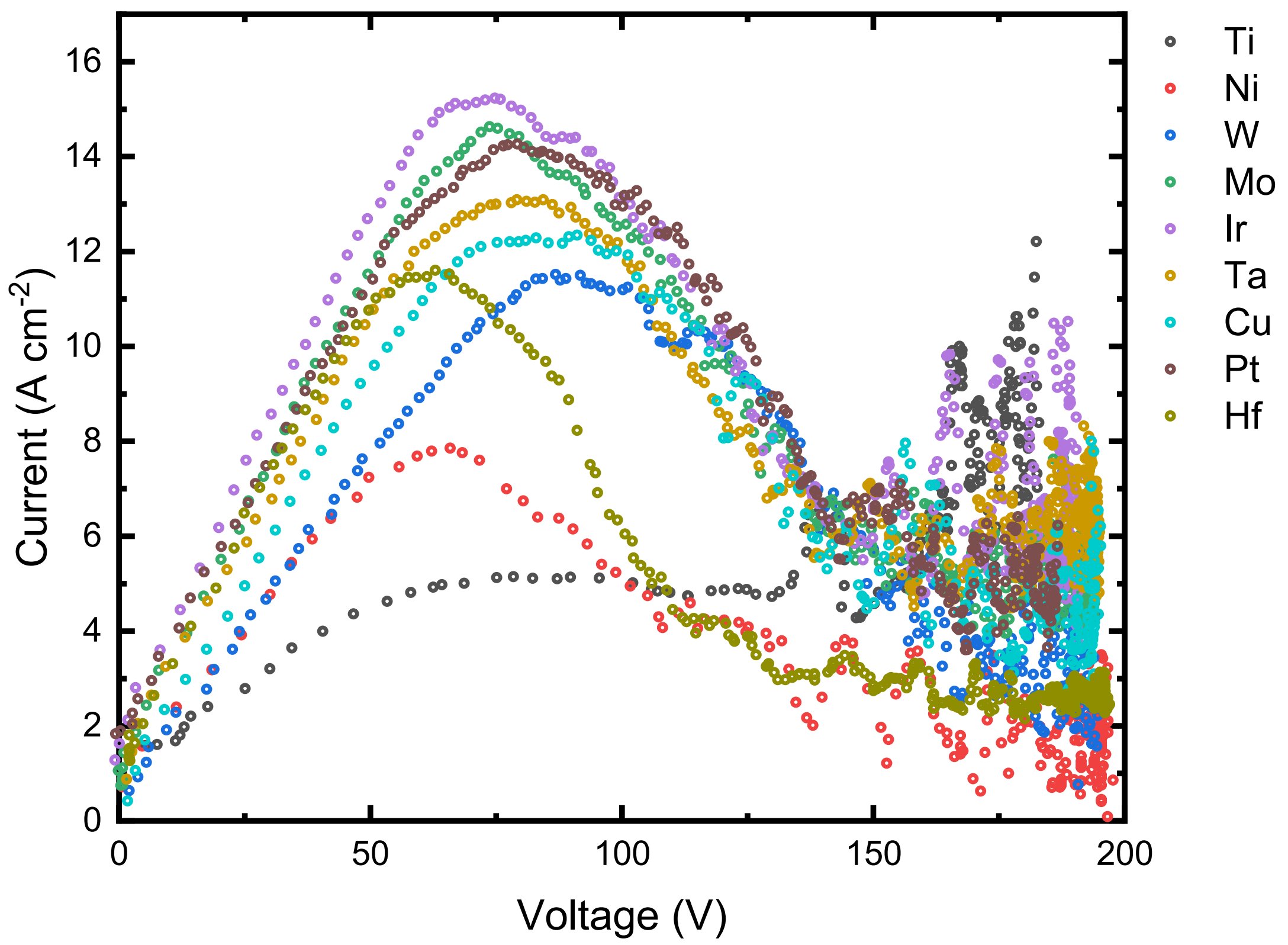

**Supplementary Figure 11:** Current versus voltage data obtained for all cathode materials employed using a high voltage probe during plasma ignition with a $N_2$ flow of 0.07 L min$^{-1}$ .

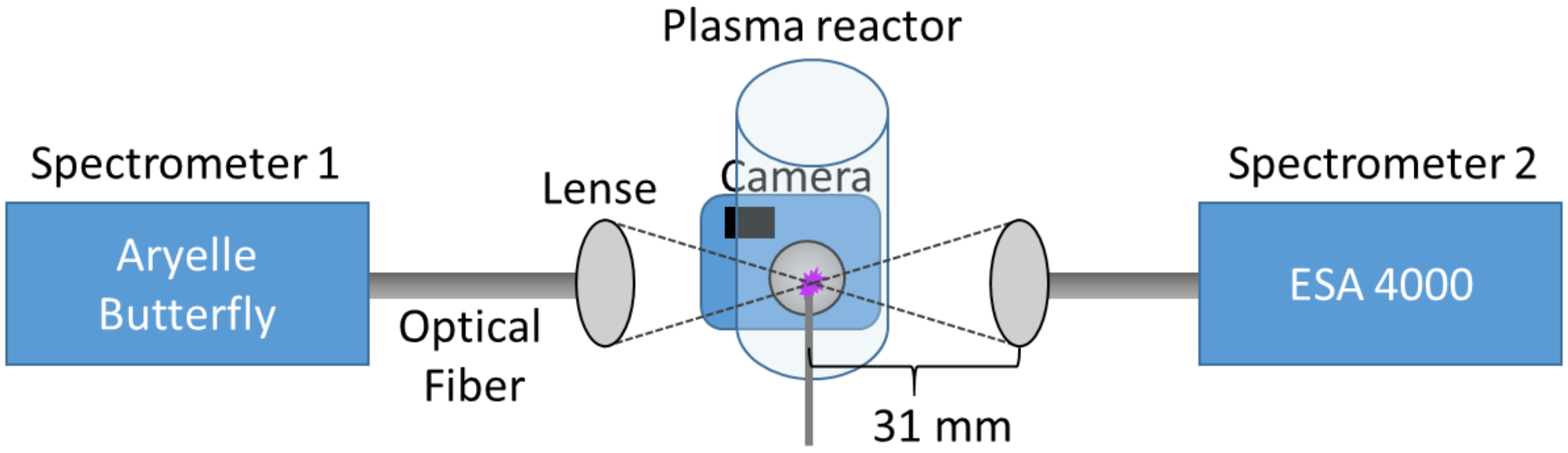

**Supplementary Figure 12:** OES measurement setup configuration. Two OES spectrometer with attached optical fibers and focusing lenses were placed on opposite sides of the cathode wire. A high-speed camera was placed orthogonally to capture the bubble formation and stochastic discharge distribution.

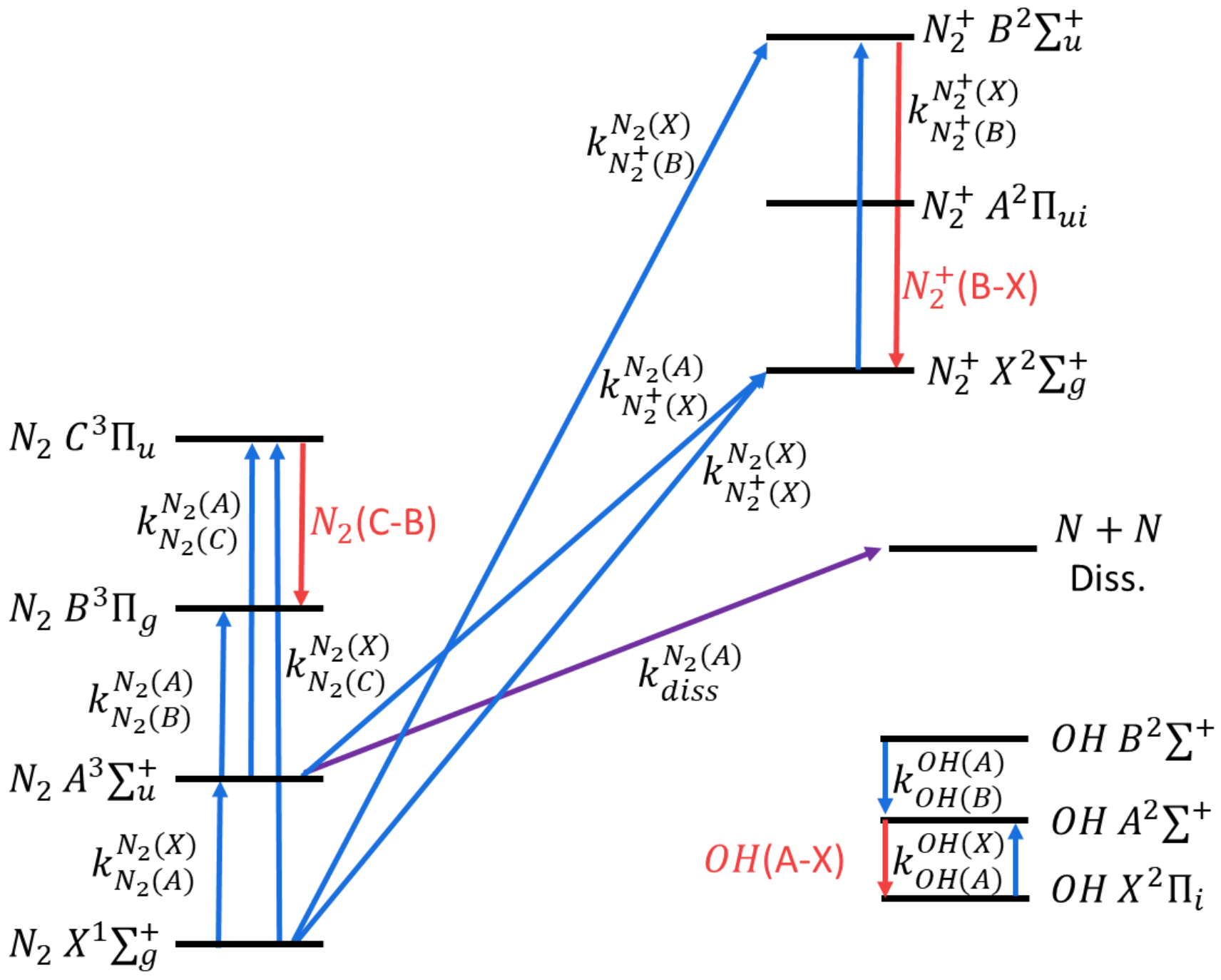

**Supplementary Figure 13:** Collisional-radiative model of OH and $N_2$ molecules. The red arrows indicate the (spontaneous) emission observable in the spectral range, the blue arrows indicate the electron impact excitation. Adapted from Ref. [2].

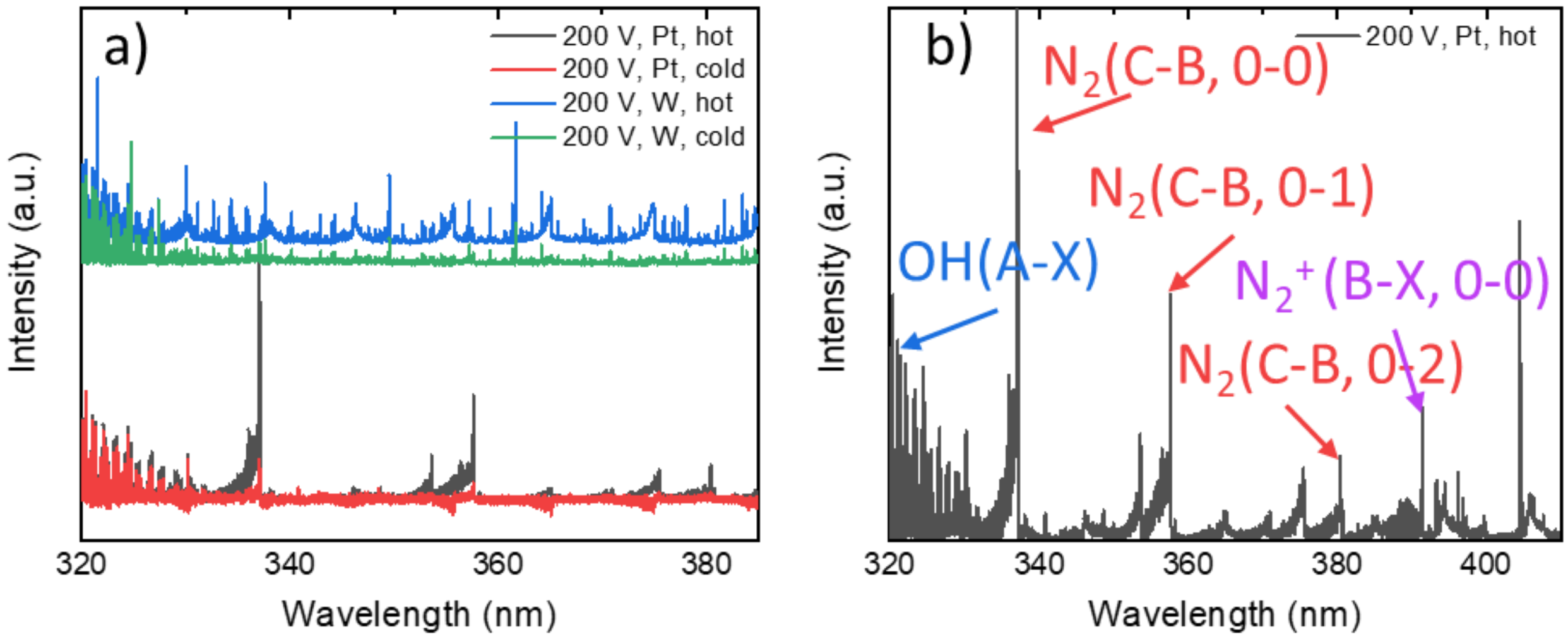

***Supplementary Figure 14.*** *Full range of observed molecular nitrogen transitions on Pt and W electrodes (a). The hot and cold state are superimposed for better comparison. In b) rovibrational transitions within the $N_2$(C-B), $N_2^+$(B-X), and OH(A-X) spectra can be observed.*

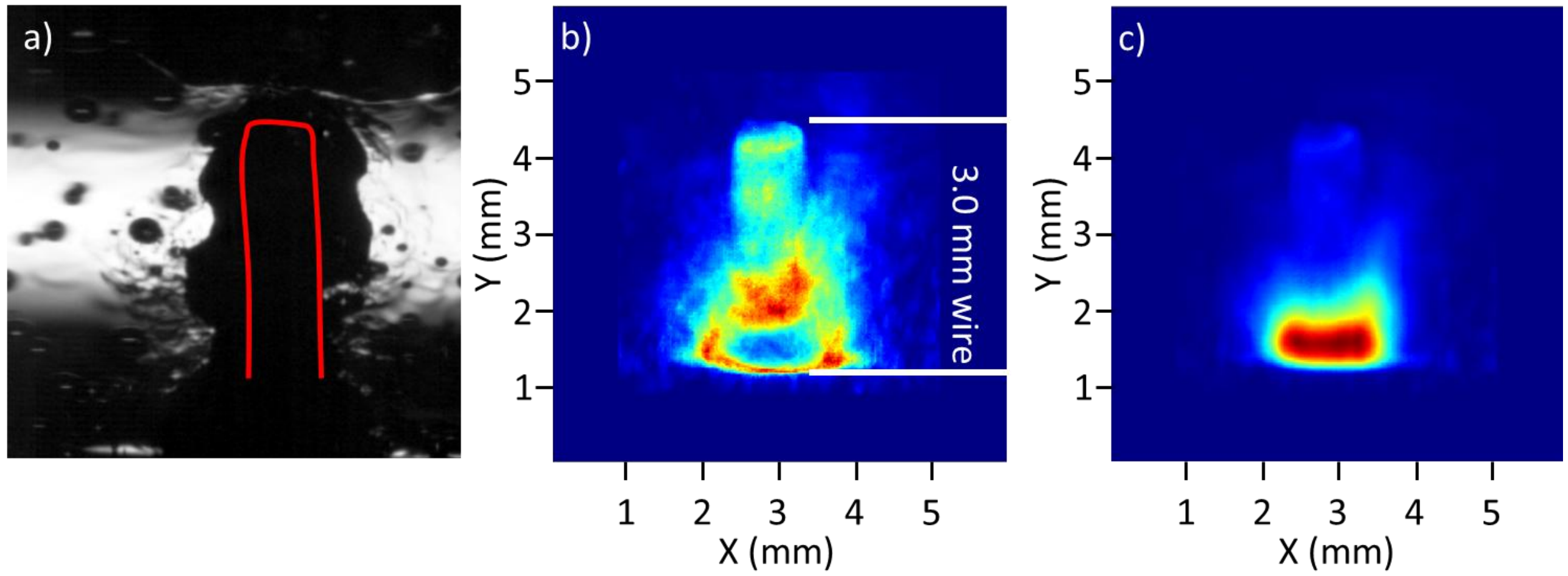

**Supplementary Figure 15:** Illustration of camera images of the cathode captured in back-illuminated shadowgraphy mode **(a)**, showcasing solely the low-intensity plasma emission **(b)**, and the entire plasma emission intensity **(c)**. The depicted cathode is a 3.0 mm Pt wire subjected to a voltage of 200 V with a $N_2$ flow rate of 30 L min$^{-1}$.

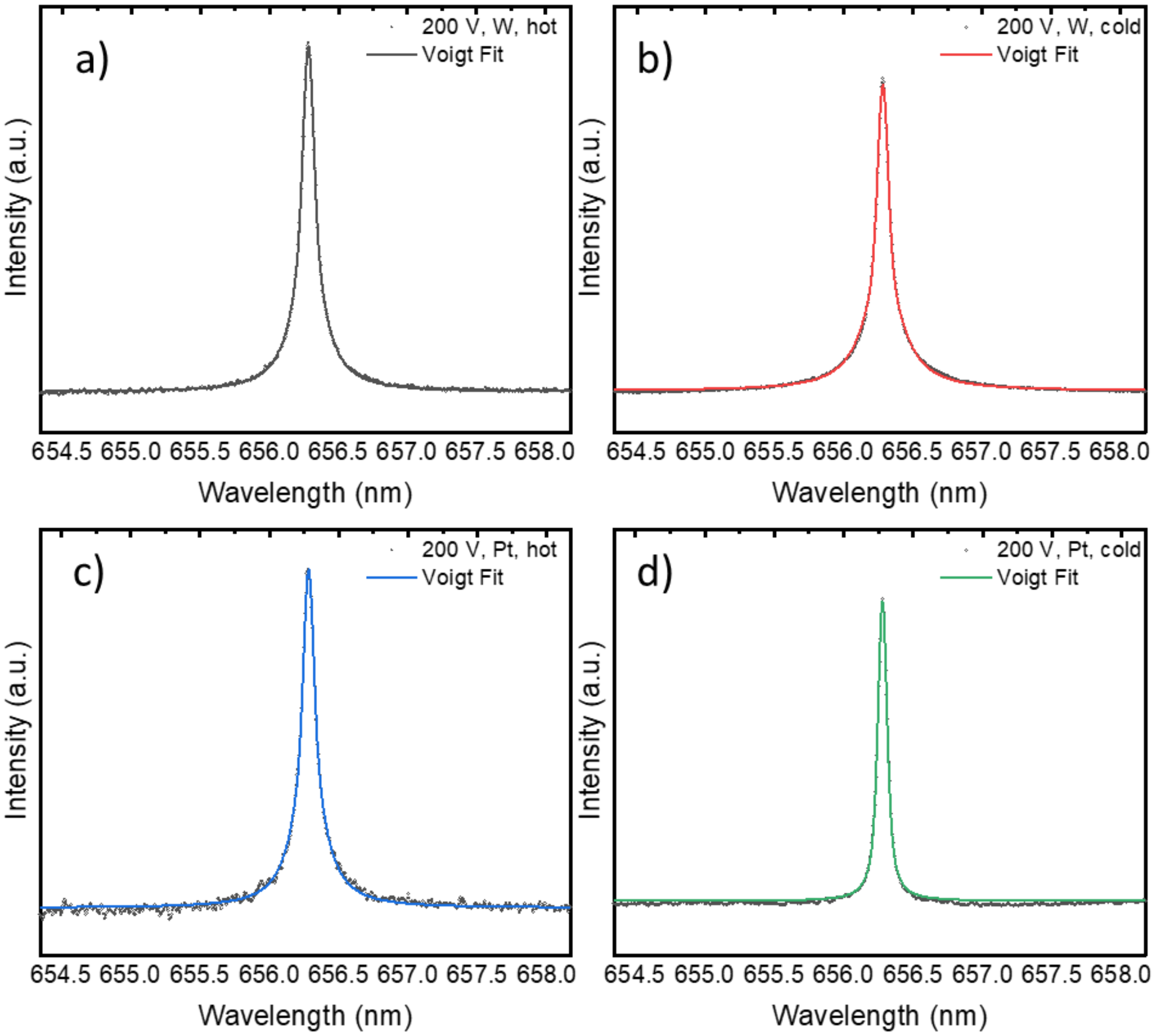

**Supplementary Figure 16:** Voigt fits of the $H_\alpha$ line for material-specific measurements. Panels **(a)** and **(b)** depict the fitted spectra of W in the hot and cold states, respectively. Panels **(c)** and **(d)** show the corresponding spectra for Pt. All measurements were performed at a $N_2$ flow of 0.07 L $min^{-1}$ in 0.1 M KOH electrolyte.

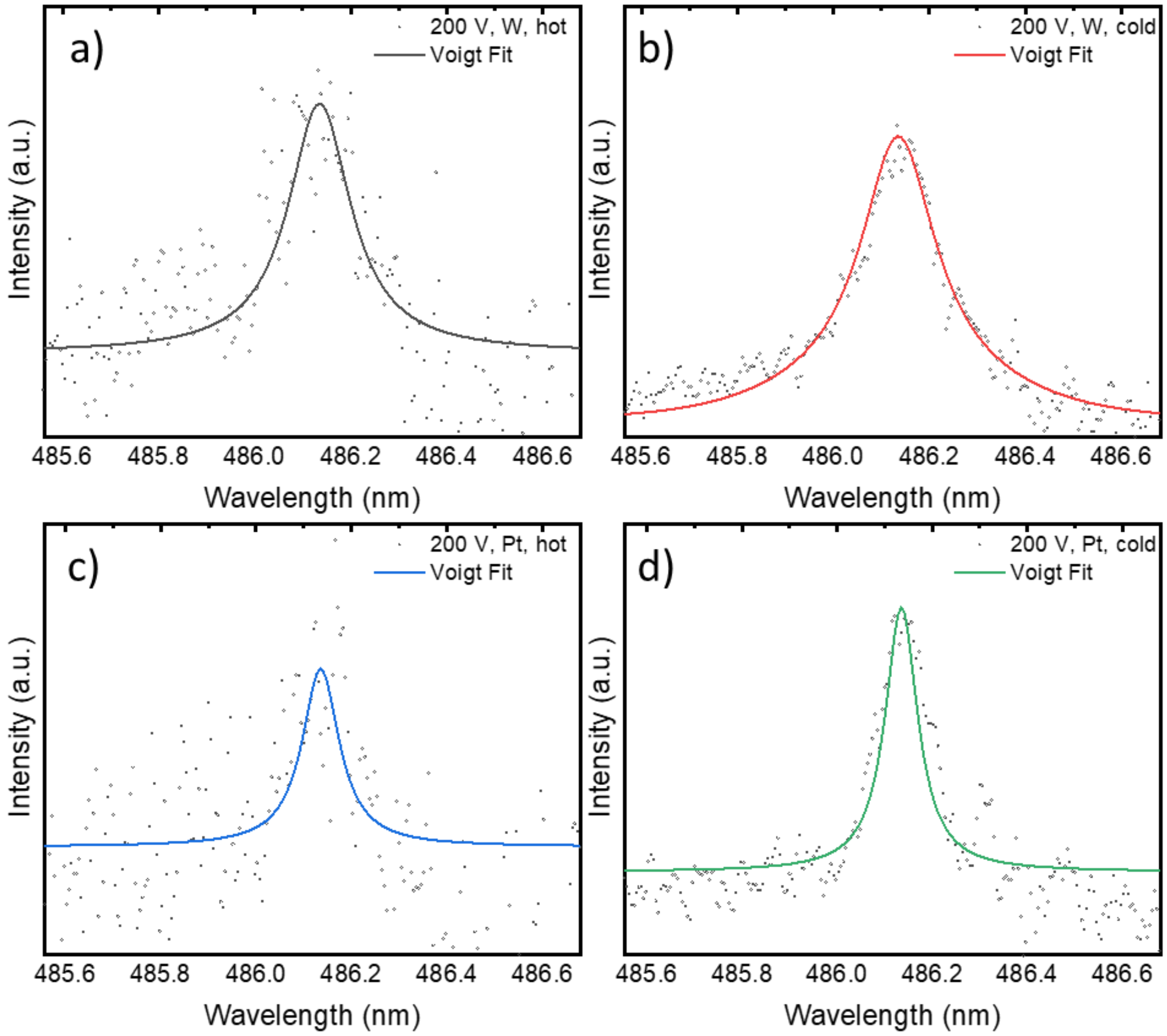

**Supplementary Figure 17:** Voigt fits of the $H_\beta$ line for material-specific measurements. Panels **(a)** and **(b)** depict the fitted spectra of W in the hot and cold states, respectively. Panels **(c)** and **(d)** show the corresponding spectra for Pt. All measurements were performed at a $N_2$ flow of 0.07 L $min^{-1}$ in 0.1 M KOH electrolyte.

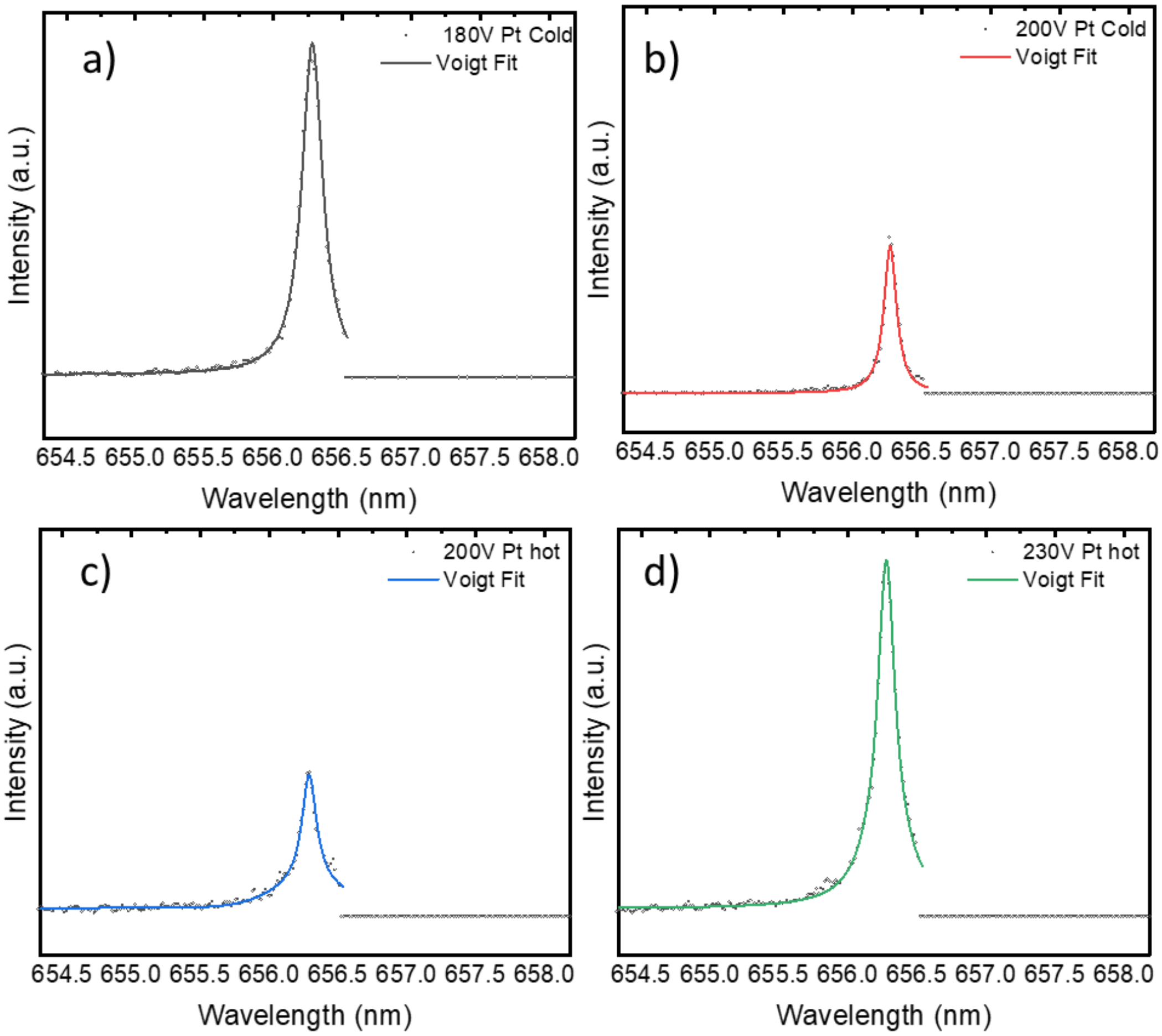

**Supplementary Figure 18:** Voigt fits of the $H_\alpha$ line for voltage-dependent measurements. Panels **(a)** and **(b)** depict the fitted spectra of the cold states at 180 V and 200 V, respectively. Panels **(c)** and **(d)** show the corresponding spectra for the hot states at 200 V and 230 V. All measurements were performed at a $N_2$ flow of 0.07 L $min^{-1}$ in 0.1 M KOH electrolyte. The right side of the $H_\alpha$ line is cut off by a sensitivity gap in the spectrometer.

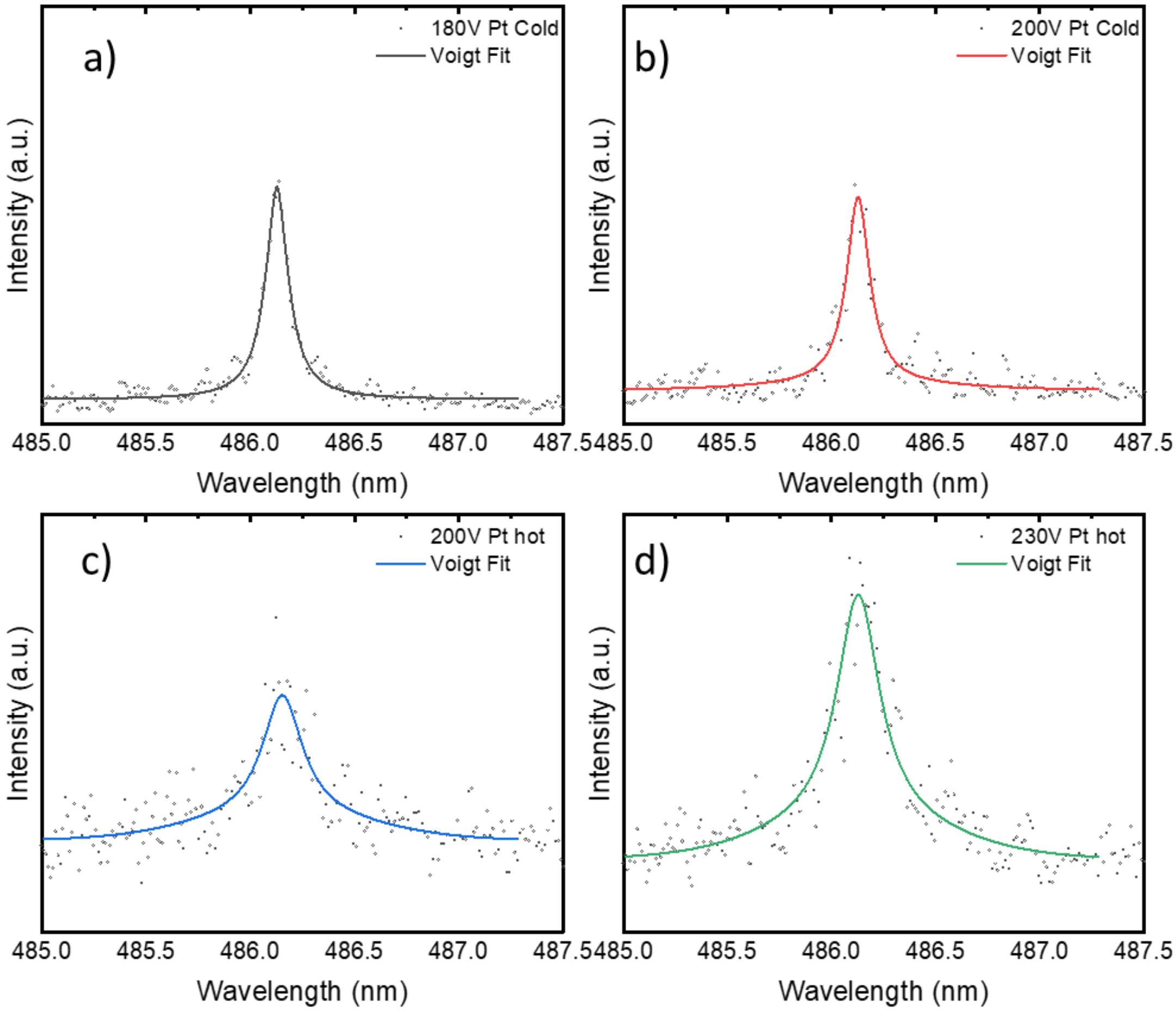

**Supplementary Figure 19:** Voigt fits of the $H_\beta$ line for voltage-dependent measurements. Panels **(a)** and **(b)** depict the fitted spectra of the cold states at 180 V and 200 V, respectively. Panels **(c)** and **(d)** show the corresponding spectra for the hot states at 200 V and 230 V. All measurements were performed at a $N_2$ flow of 0.07 L $min^{-1}$ in 0.1 M KOH electrolyte.

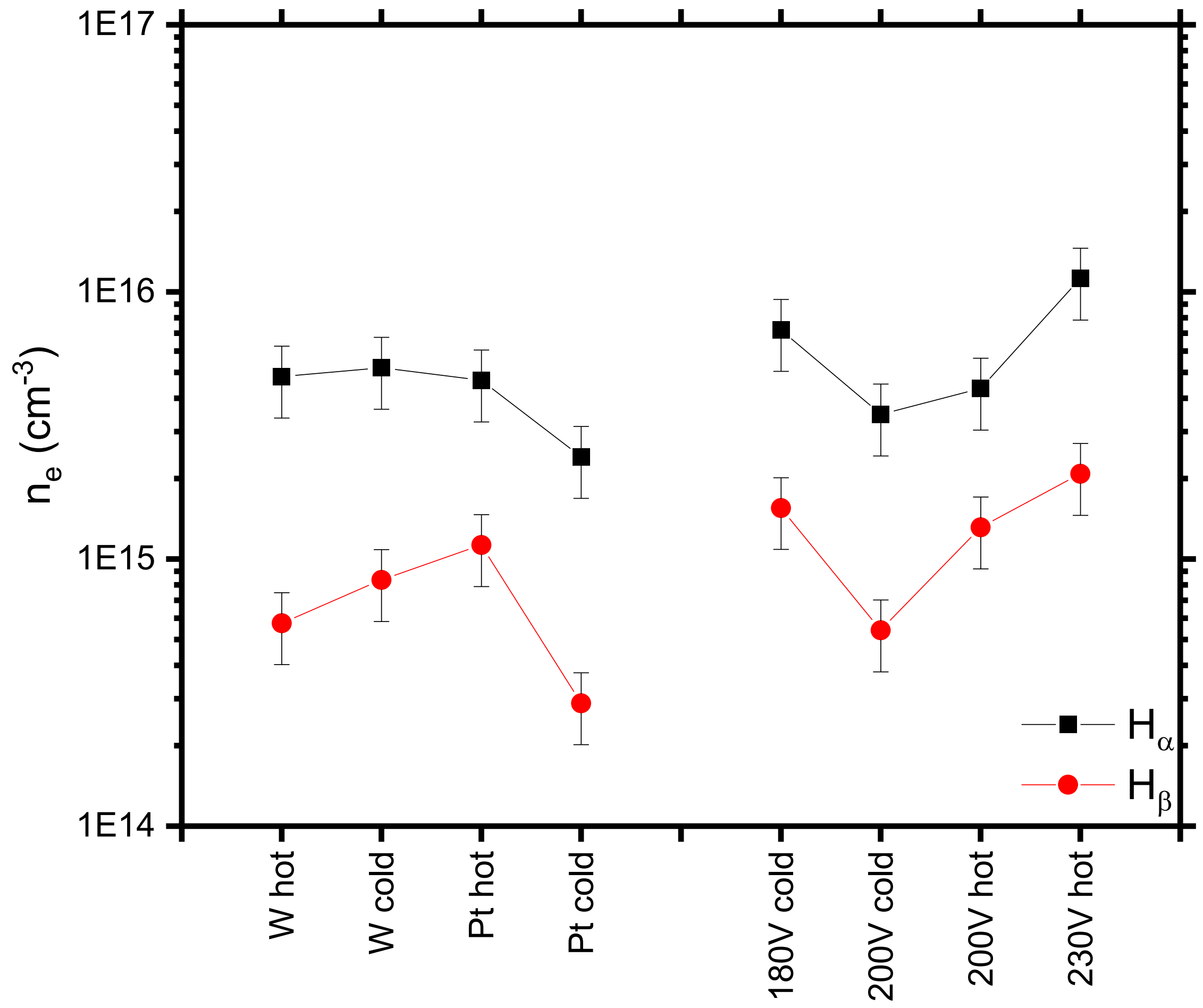

**Supplementary Figure 20:** Electron density calculated from the $H_\alpha$ and $H_\beta$ Lorentzian line width for the material- and voltage-dependent measurements.

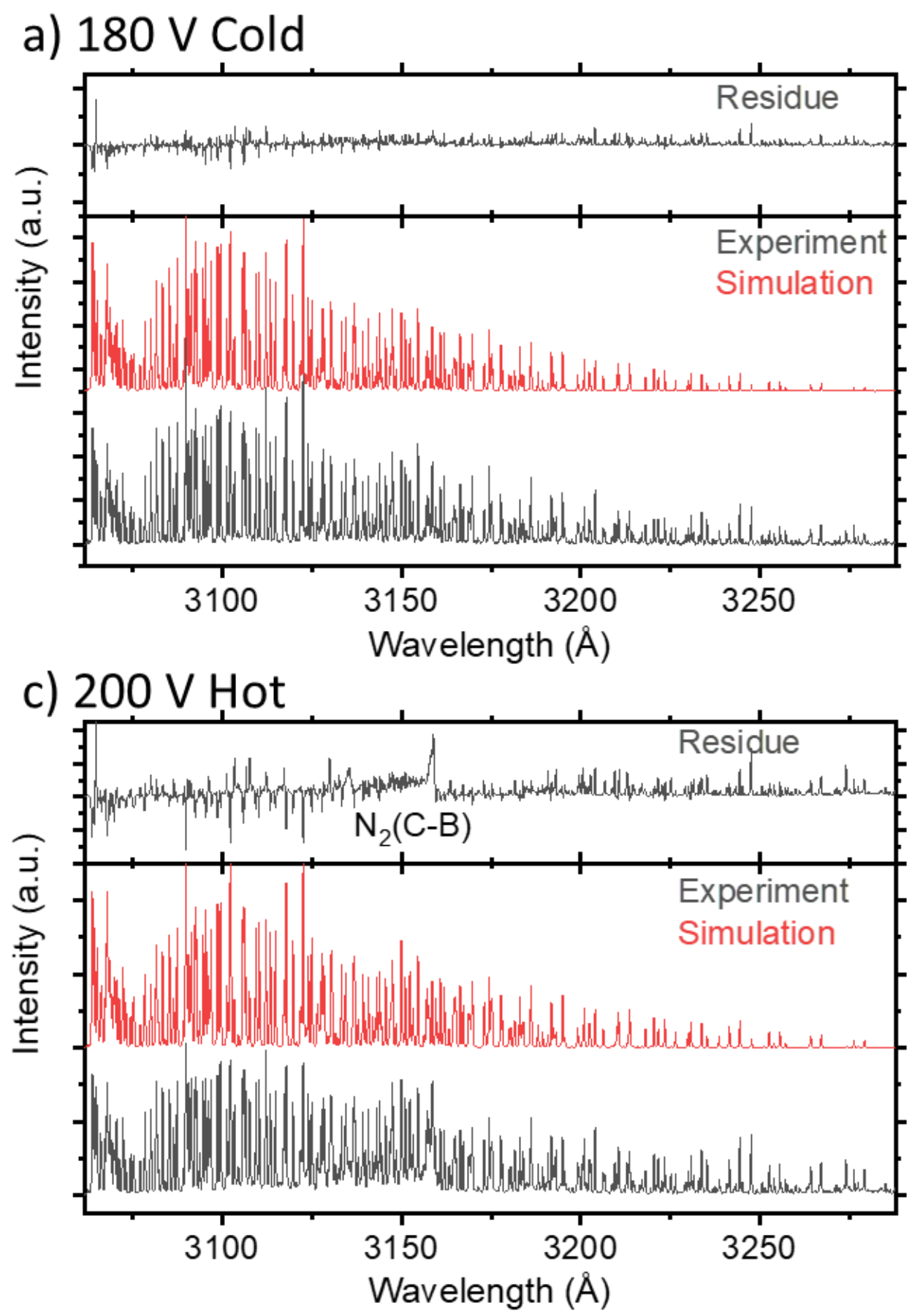

**Supplementary Figure 21:** Voltage dependency on the OH(A-X) transitions at different voltages and 0.07 L min$^{-1}$ of $N_2$ flow in 0.1 M KOH electrolyte on a Pt cathode. At the 200 V hot state **(c)**, an additional peak from the $N_2$(C-V, $\Delta v$ = -2) transition can be seen in the residue. Atomic lines of the Pt cathode can be seen overlaid with the OH(A-X) transitions at 230 V **(d)**. The experimental data was collected with the Aryelle Spectrometer.

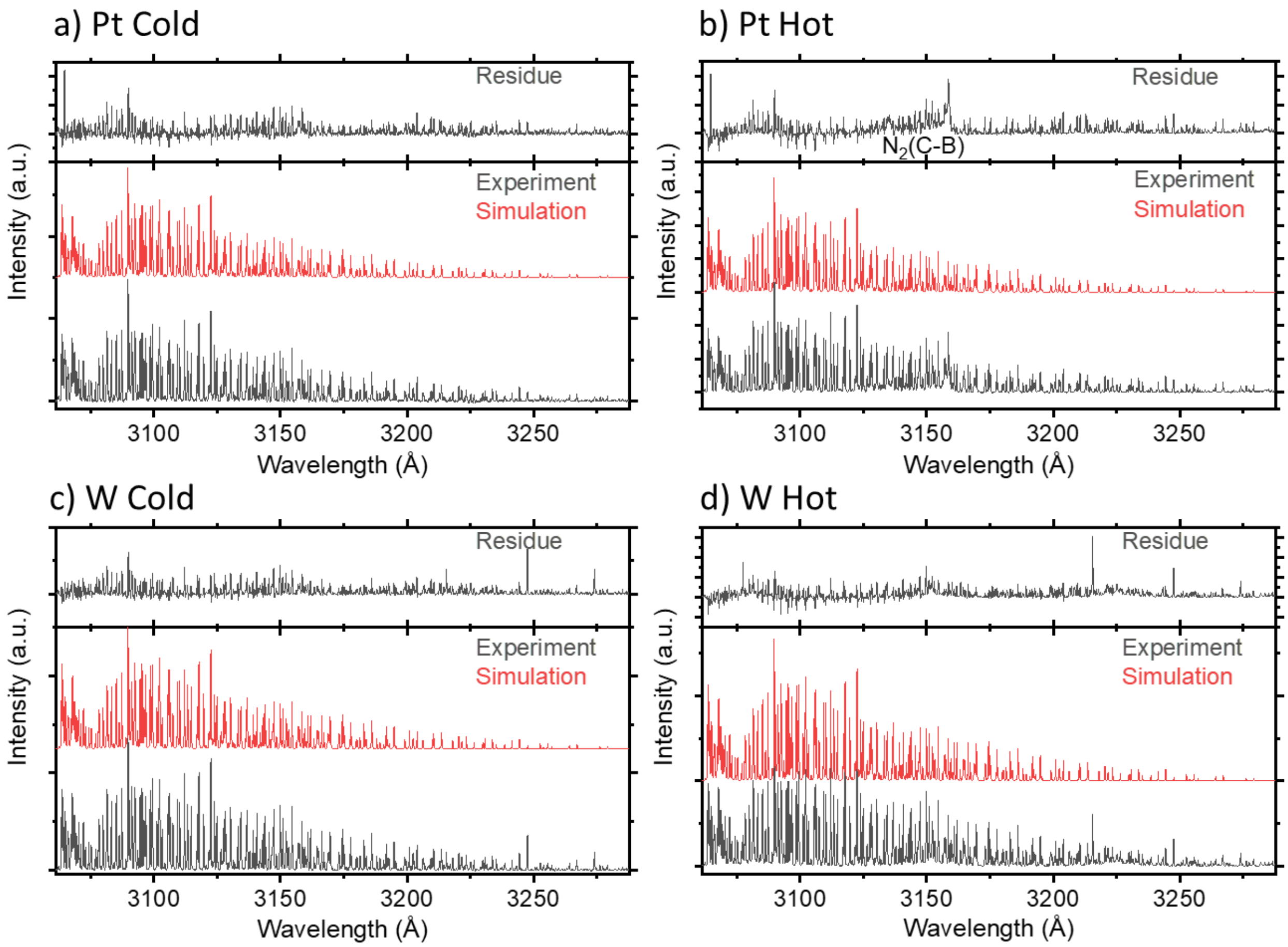

**Supplementary Figure 22:** Material dependency on the OH(A-X) transitions at 200 V and 0.07 L min$^{-1}$ of $N_2$ flow in 0.1 M KOH electrolyte. In **(a)** and **(b)**, Pt was used as the cathode material, in **(c)** and **(d)** W was used. At the 200 V hot state on Pt (b), the same additional peak from the $N_2$(C-V, $\Delta v = -2$) transition as in

**Supplementary Figure 21** can be observed. The atomic lines of W can always be seen superimposed on the OH(A-X) transitions (c-d). The experimental data was collected with the second echelle spectrometer.

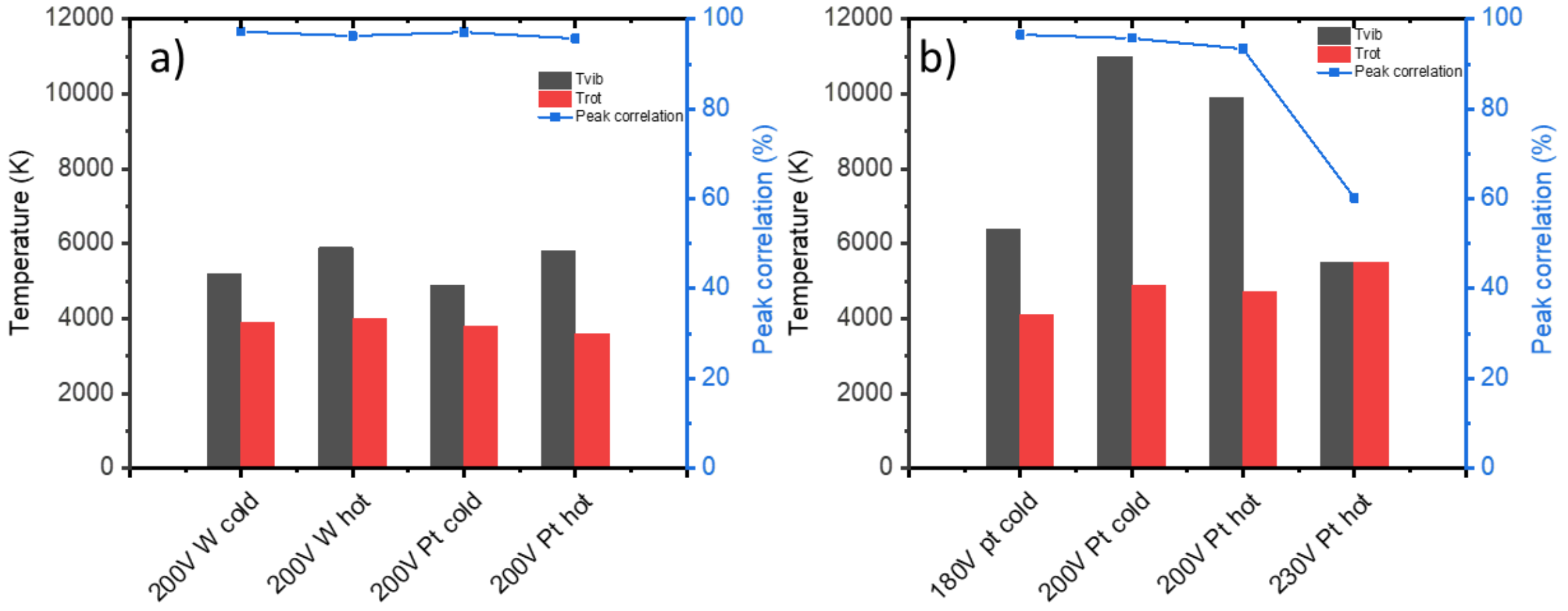

**Supplementary Figure 23:** Fitted rotational and vibrational gas temperatures for the material dependent (a) and the voltage dependent (b) experiments. The blue dots indicate the peak correlation between simulated and experimental spectrum. The overlap with other peaks besides the OH(A-X) peak drastically reduces this value.

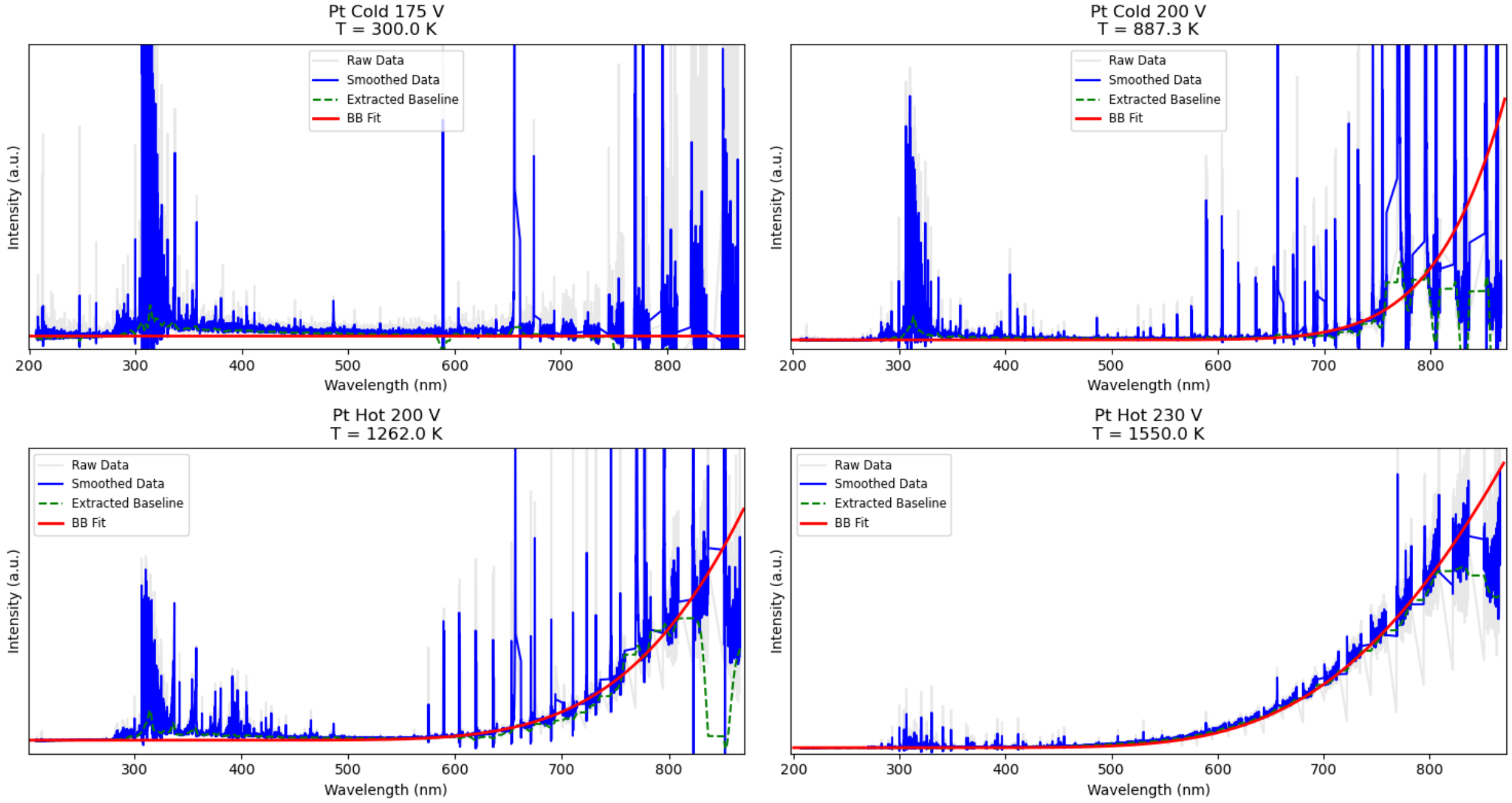

**Supplementary Figure 24:** Background continuum fitting of OES spectra with a Pt wire plasma active electrode (cathode) at varying applied voltages.

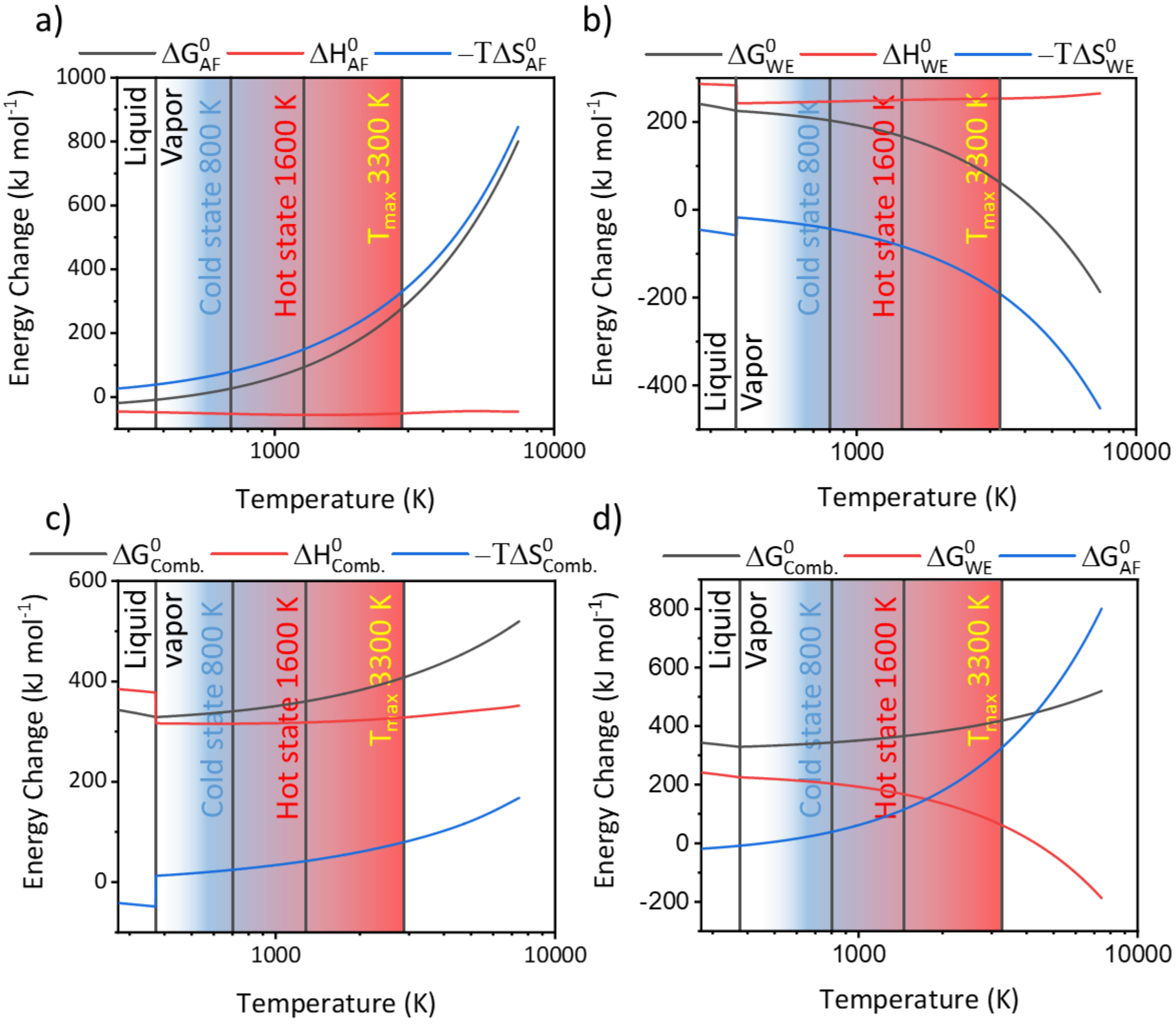

**Supplementary Figure 25:** Calculated energies (ΔG, TΔS, and ΔH) are presented for temperatures spanning from 0°C to 7000°C, corresponding to the formation of ammonia (a), the electrolysis of water (b), and the combination of both reactions (c) under conditions of liquid water and water vapor, respectively. In (d), the Gibbs free energy ($\Delta G^0$) of all three reactions is compared. The ranges for the cold and hot state temperatures of the gas and electrode temperature are indicated as well as the highest measured temperature. The jump in the phase transfer is due to the loss of hydrogen bonds from solvation.

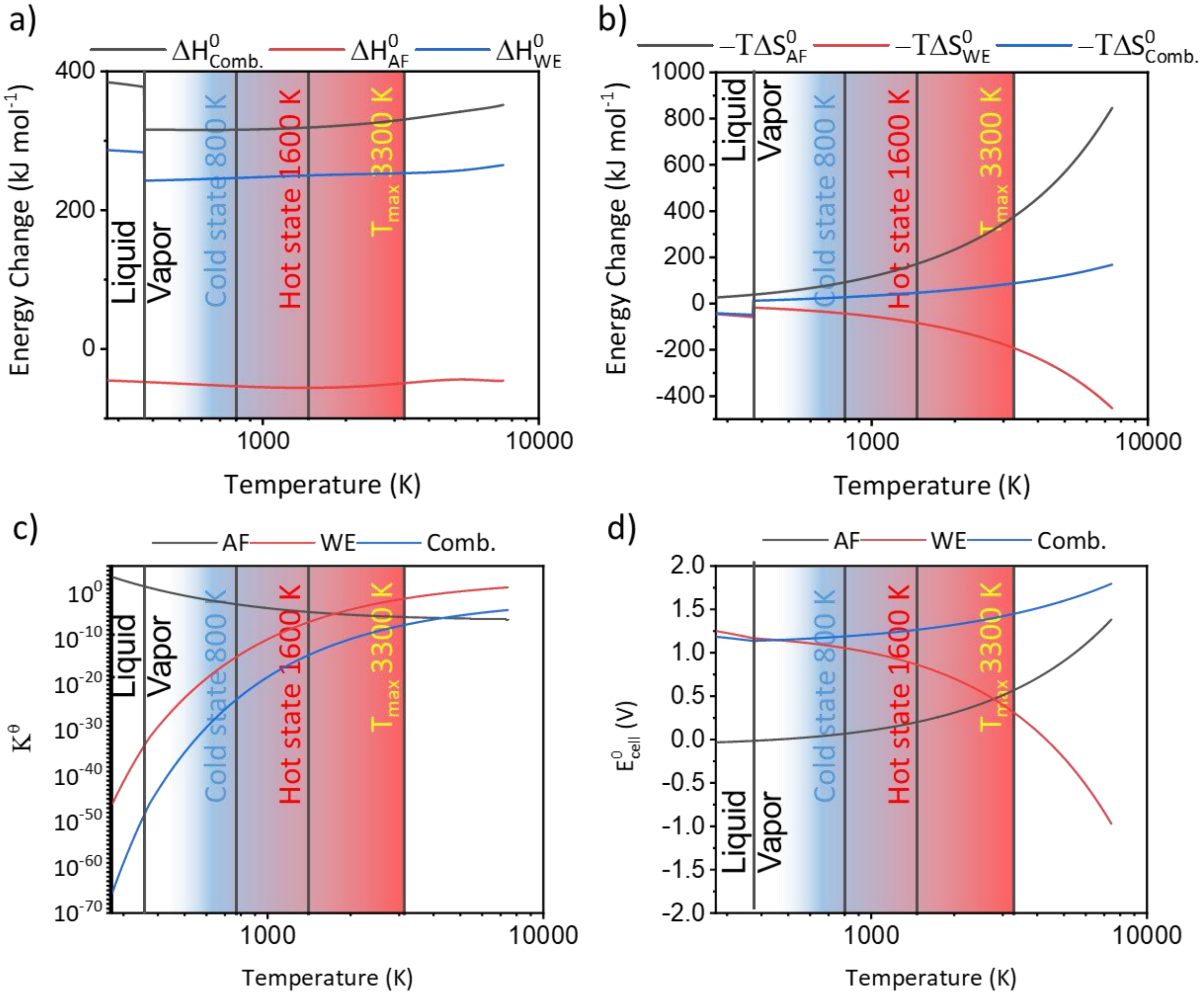

**Supplementary Figure 26:** Illustration of the temperature-dependent variation in enthalpy (a) and change of entropy (b) with temperature for ammonia formation (AF), water electrolysis (WE), and the combined reactions of both, considering liquid water and water vapor phases, respectively. In (c), the rate constants are shown and the corresponding cell potentials are shown in (d). The ranges for the cold and hot state temperatures of the gas and electrode temperature are indicated as well as the highest measured temperature. The jump in the phase transfer is due to the loss of hydrogen bonds from solvation.

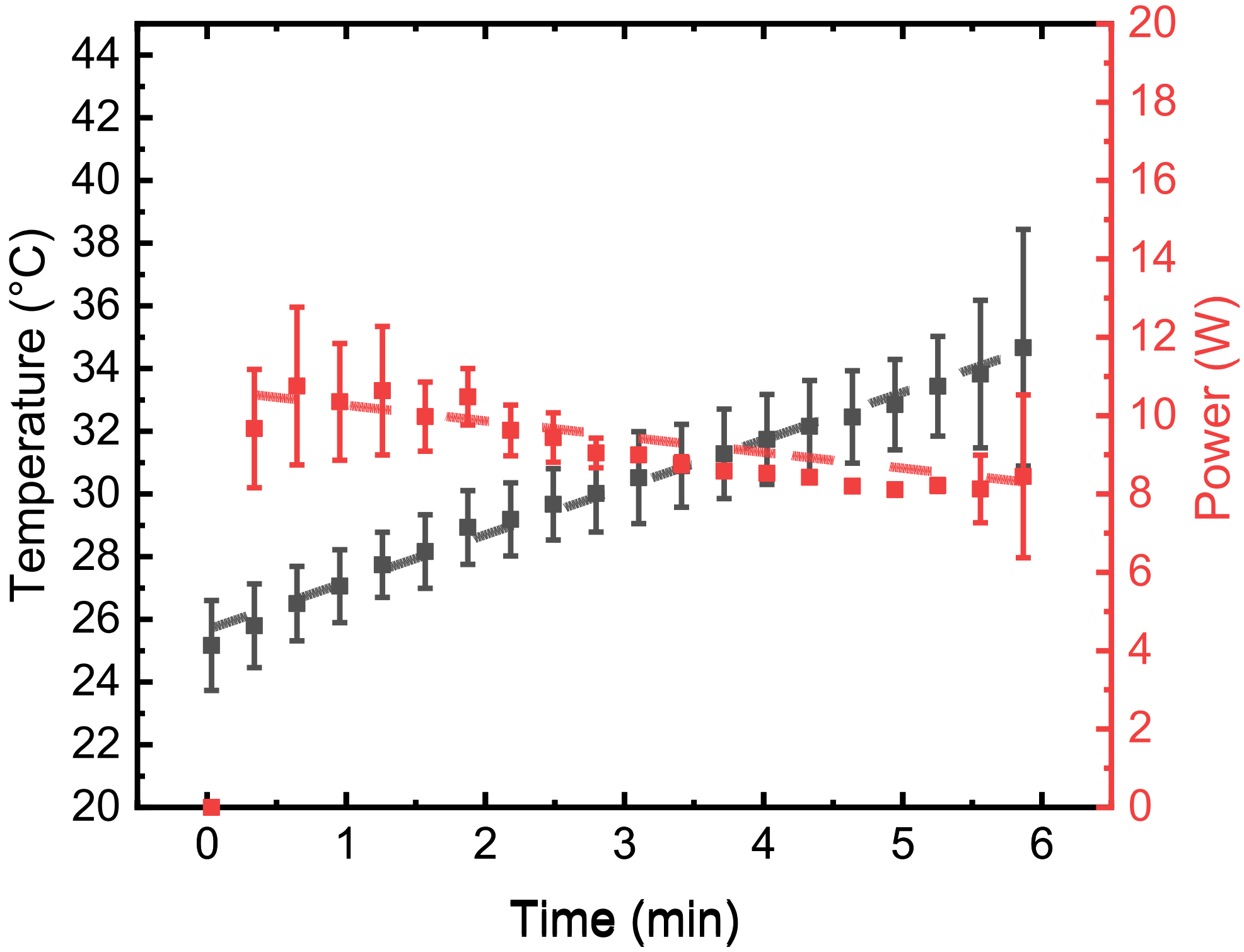

**Supplementary Figure 27:** The power required for heating, calculated based on the observed temperature rise, was plotted against the average current at 200 V (during plasma). The red and black dotted lines represent a linear fit to the data.

## Supplementary References:

1. Klaus P. Huber, G. H. H., Constants of Diatomic Molecules. *NIST Chemistry WebBook* **1977,** *NIST Standard Reference Database Number 69*.
2. Gröger, S.; Fiebrandt, M.; Hamme, M.; Bibinov, N.; Awakowicz, P., Characterization of a transient spark micro-discharge in nitrogen using simultaneous two-wavelength diagnostics. *Measurement Science and Technology* **2020,** *31* (7).
3. Luque, J.; Crosley, D. R. J. S. i. r. M., LIFBASE: Database and spectral simulation program (version 1.5). **1999,** *99* (009).
4. Dyatko, N.; Kochetov, I.; Napartovich, A.; Sukharev, A., *EEDF: The software package for calculations of Electron Energy Distribution Function*. 2015.
5. Behringer, K., Diagnostics and modelling of ECRH microwave discharges. *Plasma Physics and Controlled Fusion* **1991,** *33* (9), 997.
6. Gigosos, M. A.; González, M. A.; Cardeñoso, V., Computer simulated Balmer-alpha, -beta and -gamma Stark line profiles for non-equilibrium plasmas diagnostics. *Spectrochimica Acta Part B-Atomic Spectroscopy* **2003,** *58* (8), 1489-1504.
7. Konjević, N.; Ivković, M.; Sakan, N., Hydrogen Balmer lines for low electron number density plasma diagnostics. *Spectrochimica Acta Part B: Atomic Spectroscopy* **2012,** *76*, 16-26.
8. Bruggeman, P. J.; Sadeghi, N.; Schram, D. C.; Linss, V., Gas temperature determination from rotational lines in non-equilibrium plasmas: a review. *Plasma Sources Science and Technology* **2014,** *23* (2), 023001.
9. Chase, M. W., Jr., NIST-JANAF Themochemical Tables, Fourth Edition. *J. Phys. Chem. Ref. Data* **1998,** *Monograph 9*, 1-1951.
10. Chang, W.; Jain, A.; Rezaie, F.; Manthiram, K., Lithium-mediated nitrogen reduction to ammonia via the catalytic solid–electrolyte interphase. *Nature Catalysis* **2024,** *7* (3), 231-241.
11. Du, H.-L.; Chatti, M.; Hodgetts, R. Y.; Cherepanov, P. V.; Nguyen, C. K.; Matuszek, K.; MacFarlane, D. R.; Simonov, A. N., Electroreduction of nitrogen with almost 100% current-to-ammonia efficiency. *Nature* **2022,** *609* (7928), 722-727.
12. Li, S.; Zhou, Y.; Li, K.; Saccoccio, M.; Sažinas, R.; Andersen, S. Z.; Pedersen, J. B.; Fu, X.; Shadravan, V.; Chakraborty, D.; Kibsgaard, J.; Vesborg, P. C. K.; Nørskov, J. K.; Chorkendorff, I., Electrosynthesis of ammonia with high selectivity and high rates via engineering of the solid-electrolyte interphase. *Joule* **2022,** *6* (9), 2083-2101.
13. Li, K.; Shapel, S. G.; Hochfilzer, D.; Pedersen, J. B.; Krempl, K.; Andersen, S. Z.; Sažinas, R.; Saccoccio, M.; Li, S.; Chakraborty, D.; Kibsgaard, J.; Vesborg, P. C. K.; Nørskov, J. K.; Chorkendorff, I., Increasing Current Density of Li-Mediated Ammonia Synthesis with High Surface Area Copper Electrodes. *ACS Energy Letters* **2022,** *7* (1), 36-41.
14. Suryanto, B. H. R.; Matuszek, K.; Choi, J.; Hodgetts, R. Y.; Du, H.-L.; Bakker, J. M.; Kang, C. S. M.; Cherepanov, P. V.; Simonov, A. N.; MacFarlane, D. R., Nitrogen reduction to ammonia at high efficiency and rates based on a phosphonium proton shuttle. **2021,** *372* (6547), 1187-1191.
15. Fu, X.; Pedersen, J. B.; Zhou, Y.; Saccoccio, M.; Li, S.; Sažinas, R.; Li, K.; Andersen, S. Z.; Xu, A.; Deissler, N. H.; Mygind, J. B. V.; Wei, C.; Kibsgaard, J.; Vesborg, P. C. K.; Nørskov, J. K.; Chorkendorff, I., Continuous-flow electrosynthesis of ammonia by nitrogen reduction and hydrogen oxidation. **2023,** *379* (6633), 707-712.
16. Tsuneto, A.; Kudo, A.; Sakata, T., Lithium-mediated electrochemical reduction of high pressure N2 to NH3. *Journal of Electroanalytical Chemistry* **1994,** *367* (1), 183-188.
17. Lazouski, N.; Chung, M.; Williams, K.; Gala, M. L.; Manthiram, K., Non-aqueous gas diffusion electrodes for rapid ammonia synthesis from nitrogen and water-splitting-derived hydrogen. *Nature Catalysis* **2020,** *3* (5), 463-469.

18. Tsuneto, A.; Kudo, A.; Sakata, T., Efficient Electrochemical Reduction of N2 to NH3 Catalyzed by Lithium. *Chemistry Letters* **2006,** *22* (5), 851-854.
19. Spry, M.; Westhead, O.; Tort, R.; Moss, B.; Katayama, Y.; Titirici, M.-M.; Stephens, I. E. L.; Bagger, A., Water Increases the Faradaic Selectivity of Li-Mediated Nitrogen Reduction. *ACS Energy Letters* **2023,** *8* (2), 1230-1235.
20. Lazouski, N.; Schiffer, Z. J.; Williams, K.; Manthiram, K., Understanding Continuous Lithium-Mediated Electrochemical Nitrogen Reduction. *Joule* **2019,** *3* (4), 1127-1139.
21. Andersen, S. Z.; Statt, M. J.; Bukas, V. J.; Shapel, S. G.; Pedersen, J. B.; Krempl, K.; Saccoccio, M.; Chakraborty, D.; Kibsgaard, J.; Vesborg, P. C. K.; Nørskov, J.; Chorkendorff, I., Increasing stability, efficiency, and fundamental understanding of lithium-mediated electrochemical nitrogen reduction. *Energy & Environmental Science* **2020,** *13* (11), 4291-4300.
22. Tort, R.; Westhead, O.; Spry, M.; Davies, B. J. V.; Ryan, M. P.; Titirici, M.-M.; Stephens, I. E. L., Nonaqueous Li-Mediated Nitrogen Reduction: Taking Control of Potentials. *ACS Energy Letters* **2023,** *8* (2), 1003-1009.
23. Ertl, G., Surface Science and Catalysis—Studies on the Mechanism of Ammonia Synthesis: The P. H. Emmett Award Address. *Catalysis Reviews* **1980,** *21* (2), 201-223.
24. Zhao, X.; Hu, G.; Chen, G. F.; Zhang, H.; Zhang, S.; Wang, H. J. A. M., Comprehensive understanding of the thriving ambient electrochemical nitrogen reduction reaction. **2021,** *33* (33), 2007650.
25. Lofthus, A.; Krupenie, P. H., The spectrum of molecular nitrogen. *Journal of Physical and Chemical Reference Data* **1977,** *6* (1), 113-307.
26. Uyama, H.; Matsumoto, O., Synthesis of ammonia in high-frequency discharges. II. Synthesis of ammonia in a microwave discharge under various conditions. *Plasma Chemistry and Plasma Processing* **1989,** *9* (3), 421-432.
27. Uyama, H.; Matsumoto, O., Synthesis of ammonia in high-frequency discharges. *Plasma Chemistry and Plasma Processing* **1989,** *9* (1), 13-24.
28. Ben Yaala, M.; Saeedi, A.; Scherrer, D.-F.; Moser, L.; Steiner, R.; Zutter, M.; Oberkofler, M.; De Temmerman, G.; Marot, L.; Meyer, E., Plasma-assisted catalytic formation of ammonia in N2–H2 plasma on a tungsten surface. *Physical Chemistry Chemical Physics* **2019,** *21* (30), 16623-16633.
29. Shah, J.; Wang, W.; Bogaerts, A.; Carreon, M. L., Ammonia Synthesis by Radio Frequency Plasma Catalysis: Revealing the Underlying Mechanisms. *ACS Applied Energy Materials* **2018,** *1* (9), 4824-4839.
30. Gómez-Ramírez, A.; Montoro-Damas, A. M.; Cotrino, J.; Lambert, R. M.; González-Elipe, A. R., About the enhancement of chemical yield during the atmospheric plasma synthesis of ammonia in a ferroelectric packed bed reactor. *Plasma Processes and Polymers* **2017,** *14* (6), 1600081.